\documentclass[aps,prd,superscriptaddress]{revtex4}
\usepackage{amssymb}
\usepackage{amsmath}
\usepackage{amsfonts}
\usepackage{hyperref}
\usepackage{epsfig}
\usepackage{graphicx}
\usepackage{tabularx}
\usepackage{orcidlink}
\usepackage{color}

\newcommand{\beaa}{\begin{eqnarray*}}
\newcommand{\enaa}{\end{eqnarray*}}
\newcommand{\bea}{\begin{eqnarray}}
\newcommand{\ena}{\end{eqnarray}}
\newcommand{\seq}{\begin{subequations}}
\newcommand{\sen}{\end{subequations}}
\newcommand{\eq}{\begin{eqnarray}}
\newcommand{\en}{\end{eqnarray}}

\def\shiftdown#1{#1\llap{\lower.04ex\hbox{#1}}}

\def \be  {\begin{equation}}
\def \ee  {\end{equation}}
\def \ba  {\begin{eqnarray}}
\def \ea  {\end{eqnarray}}
\def \baa {\begin{eqnarray*}}
\def \eaa {\end{eqnarray*}}

\def \bb  {\begin{thebibliography}}
\def \eb  {\end{thebibliography}}

\def \lab #1 {\label{#1}}

\begin{document}

\title{Perturbative $T$-odd asymmetries in the Drell-Yan process revisited}

\author{Valery E. Lyubovitskij\,\orcidlink{0000-0001-7467-572X}}
\affiliation{Institut f\"ur Theoretische Physik,
Universit\"at T\"ubingen, \\
Kepler Center for Astro and Particle Physics, \\ 
Auf der Morgenstelle 14, D-72076 T\"ubingen, Germany} 
\affiliation{Departamento de F\'\i sica y Centro Cient\'\i fico 
Tecnol\'ogico de Valpara\'\i so-CCTVal, \\ 
Universidad T\'ecnica Federico Santa Mar\'\i a,
Casilla 110-V, Valpara\'\i so, Chile}
\affiliation{Millennium Institute for Subatomic Physics at 
the High-Energy Frontier (SAPHIR) of ANID, \\
Fern\'andez Concha 700, Santiago, Chile}
\author{Werner Vogelsang\,\orcidlink{0000-0002-4003-3099}}
\affiliation{Institut f\"ur Theoretische Physik, 
Universit\"at T\"ubingen, \\
Kepler Center for Astro and Particle Physics, \\ 
Auf der Morgenstelle 14, D-72076 T\"ubingen, Germany} 
\author{Fabian Wunder\,\orcidlink{0009-0007-4136-7844}}
\affiliation{Institut f\"ur Theoretische Physik,
Universit\"at T\"ubingen, \\
Kepler Center for Astro and Particle Physics, \\ 
Auf der Morgenstelle 14, D-72076 T\"ubingen, Germany} 
\author{Alexey S. Zhevlakov\,\orcidlink{0000-0002-7775-5917}}
\affiliation{Millennium Institute for Subatomic Physics at 
the High-Energy Frontier (SAPHIR) of ANID, \\
Fern\'andez Concha 700, Santiago, Chile}

\date{\today}

\begin{abstract}
  
We calculate the perturbative $T$-odd contributions to the
lepton angular distribution in the Drell-Yan process.   
Using collinear factorization, we work at the first order 
in QCD perturbation theory where these contributions appear,  
${\cal O}(\alpha_s^2)$, and address both $W^\pm$ and $\gamma/Z^0$ 
boson exchange. A major focus of our calculation is on the regime
where the boson's transverse momentum $Q_T$ is much smaller than
its mass $Q$. We carefully expand our results up to
next-to-next-to-leading power in $Q_T/Q$. Our calculation provides
a benchmark for studies of $T$-odd contributions that employ
transverse-momentum dependent parton distribution functions.
In the neutral-current case we compare our results for
the $T$-odd structure functions to available ATLAS data. 

\end{abstract}
        
\maketitle

\section{Introduction}

There is a long history of the study of $T$-odd asymmetries
in QCD hard-scattering processes, starting with the seminal
papers~\cite{DeRujula:1971nnp,DeRujula:1978bz,Fabricius:1980wg,Korner:1980np,Hagiwara:1981qn,Hagiwara:1982cq} that showed how $T$-odd effects may be generated
by absorptive parts of QCD loop diagrams. $T$-odd behavior refers
to noninvariance of observables under so-called {\it naive} time reversal,
that is, under reversal of momenta and spins without interchange of initial
and final states. As shown in the early papers, such behavior can 
occur even in theories that are manifestly invariant under {\it true}
time reversal.
Subsequent work~\cite{Pire:1983tv,Hagiwara:1984hi,Bilal:1990wi,Carlitz:1992fv,Mirkes:1992hu,Brandenburg:1995nv,Ahmed:1999ix,Korner:2000zr,Yokoya:2007xe,Hagiwara:2007sz,Frederix:2014cba,Gehrmann-DeRidder:2016cdi,Benic:2019zvg,Benic:2021gya,Benic:2024fvk,Abele:2022spu}
explored $T$-odd QCD phenomena in a wide range of scattering reactions, 
among them especially the Drell-Yan process. In Ref.~\cite{Hagiwara:1984hi}
it was proposed to study $T$-odd asymmetries appearing in the angular distributions
of the charged leptons from the decay of $W^\pm$ bosons produced with high transverse
momentum at hadron colliders. The asymmetries manifest themselves as terms proportional
to $\sin\phi$ or $\sin2\phi$ in the lepton distribution, where $\phi$ is a suitably
defined azimuthal angle between the lepton plane and the hadron plane.
The $T$-odd part of the spin-averaged differential cross section for $W^\pm$ production
was expanded in~\cite{Hagiwara:1984hi} in terms of three structure functions which were
computed to lowest order of perturbation theory. The results were later obtained
independently in Ref.~\cite{Mirkes:1992hu} and extended to the case of longitudinal 
polarization of one of the initial hadrons in~\cite{Yokoya:2007xe,Benic:2024fvk}.

In parallel developments, it was realized that $T$-odd effects in QCD may also arise
in hadronic matrix elements, especially in parton distribution functions
(PDFs)~\cite{Sivers:1989cc,Boer:1997nt,Brodsky:2002cx,Collins:2002kn},
where they are associated with correlations among three-momenta, transverse momenta,
and polarizations of partons and hadrons
and again generate azimuthal-angle dependent terms. 
This has given rise to an intensive experimental program aiming at the extraction
of such $T$-odd transverse-momentum dependent parton distributions (TMDs)
(for a recent review, see Ref.~\cite{Boussarie:2023izj}), either in semi-inclusive
lepton scattering (SIDIS) or via the Drell-Yan process. 

The precise connection between $T$-odd effects in perturbative
(collinearly factorized) 
hard scattering on the one hand and $T$-odd TMDs on the other has been
an area of active research as well. 
This issue is important both theoretically and for
phenomenology, where it is central for the ``matching'' of resummed calculations
based on TMDs to fixed-order perturbation theory. While much progress has been made for
leading-twist observables~\cite{Collins:1984kg,Collins:2011zzd,Aybat:2011zv,Ji:2006ub,Ji:2006vf,Boglione:2014oea,Collins:2016hqq,Gutierrez-Reyes:2017glx,Echevarria:2018qyi,Gutierrez-Reyes:2018iod,Scimemi:2019gge,Rein:2022odl},
it was also realized that for many of the 
azimuthal-angle dependent terms in the Drell-Yan and SIDIS cross sections --
both $T$-odd and $T$-even -- this matching is non-trivial and will involve TMD PDFs
at next-to-leading power in the hard scale~\cite{Bacchetta:2008xw,Boer:2006eq}.
Correspondingly, TMD factorization theorems at next-to-leading power were developed
in the literature~\cite{Bacchetta:2019qkv,Ebert:2021jhy,Gamberg:2022lju,Rodini:2022wic,Rodini:2023plb}. 

In the present paper, we advance this area of research by specifically exploring
the low-transverse momentum limit of the $T$-odd terms appearing in the Drell-Yan
hard-scattering calculation. The work is carried out in the spirit
of Ref.~\cite{Abele:2022spu} that addressed $T$-odd effects in SIDIS, but goes well
beyond it in terms of calculational techniques. For our purpose, we first perform
an independent new analytical calculation of the lowest-order $T$-odd terms in
the Drell-Yan cross section, recovering results at this order from the previous 
literature~\cite{Hagiwara:1984hi,Mirkes:1992hu}, but also extending them
by presenting results for pure $Z$-boson exchange and $\gamma$-$Z$ interference. 
As one application, we will compare our results to available ATLAS data~\cite{ATLAS:2016rnf}
for the $T$-odd angular terms taken around the $Z$ resonance. Our main focus,
however, is to carefully expand the results for low $Q_T^2/Q^2$, where $Q_T$ and $Q$ are 
the boson's transverse momentum and mass, respectively. We do this to first
and second power in this ratio, identifying logarithmic behavior as well. 
As a byproduct we also uncover a novel simple relation between two of the $T$-odd
structure functions valid at leading power in $Q_T^2/Q^2$ for both partonic
channels, $q\bar q$ annihilation and $q g$ Compton scattering. 

We hope that our explicit results will be useful in testing TMD factorization
at next-to-leading power and ultimately contribute to a better understanding 
of the matching between the TMD and collinearly factorized regimes. As has been 
shown in Refs.~\cite{Boer:1999mm,Lu:2011mz,Liu:2012fha}, at leading power $\gamma/Z$
interference generates a $\sin2\phi$ azimuthal 
dependence in the unpolarized Drell-Yan cross section, entering with
the Boer-Mulders function~\cite{Boer:1997nt}, while a term proportional 
to $\sin\phi$ is not generated. Effects beyond leading power have been investigated
in Ref.~\cite{Ebert:2020dfc}.

In more general terms, TMD factorization theorems make a prediction also for
the large-$Q_T$ ``tail'' of the transverse-momentum distribution they provide,
which may be confronted with the terms generated by the collinearly factorized cross section 
expanded to low $Q_T/Q$. An important issue is whether there is an overlap region of $Q_T$
where the two approaches agree. This decides whether the TMD and collinear contributions
to the cross section are manifestations of the same physical origin, or 
should be regarded as genuinely separate pieces. While such an overlap has been demonstrated in a few
important cases, notably the Sivers function~\cite{Ji:2006ub,Ji:2006vf,Scimemi:2019gge},
the situation is not clear for next-to-leading power observables~\cite{Bacchetta:2008xw}, especially for
those that arise only from loop corrections in the collinear-factorization case. 
In any case, knowledge of both the TMD and the large-$Q_T$ (collinear) parts
of the cross section is vital for phenomenology, in order to obtain a formalism
that encompasses the full range of $Q_T$. 
We will not address the potential ramifications of our results for TMDs in this paper, 
but rather view our work 
as providing a part of a ``library'' of hard-scattering functions
at low transverse momenta. We stress that
our techniques for expanding the cross sections for low $Q_T^2/Q^2$ are completely general
and may also be used in a variety of other settings, such as for $T$-even contributions,
collisions of polarized hadrons, and so forth, or perhaps even at the next order
in perturbation theory, as available from~\cite{Gehrmann-DeRidder:2016cdi}. 

Our paper is organized as follows.
In Sec.~\ref{sec: Setup} we present the definition of the structure functions
parametrizing the lepton-angular distribution for the Drell-Yan process
and the main ingredients 
for the perturbative calculation of the $T$-odd contributions.
In Sec.~\ref{sec: Results} we collect the analytic results, and subsequently
in Sec.~\ref{sec: Small QT} the small $Q_T$ expansion is performed.
In Sec.~\ref{sec: Phenomenology} we compare our results with the ATLAS data.
Section~\ref{sec: Conclusion} concludes our paper.
Some calculational details are collected
in the Appendices~\ref{app:Str_Func}--\ref{app:calc_tech}, 
and the lengthy results for the small-$Q_T$ expansion to
next-next-to-leading-power
are presented in Supplemental Material~\cite{SuppMaterial:2024sp}
to this article.

\section{$T$-odd structure of the Drell-Yan hadronic tensor}
\label{sec: Setup}

The hadronic tensor $W^{\mu\nu}$ for the Drell-Yan process can be written
in terms of nine structure functions $W_i$.
The most straightforward decomposition of this tensor 
is obtained by using the helicity formalism proposed
in Ref.~\cite{Lam:1978pu} for reactions with photon exchange and
extended in Ref.~\cite{Mirkes:1992hu} to the electroweak case. 
The results of Ref.~\cite{Mirkes:1992hu} for
the expansion of $W^{\mu\nu}$ can be conveniently
rewritten using a basis of orthogonal unit vectors
$T^\mu = q^\mu/\sqrt{Q^2} = (1,0,0,0)$, 
$X^\mu = (0,1,0,0)$, 
$Z^\mu = (0,0,0,1)$, 
$Y^\mu = \epsilon^{\mu\nu\alpha\beta}
T_\nu Z_\alpha X_\beta =(0,0,1,0)$, proposed in Ref.~\cite{Lam:1978pu}
and constructed from the hadron and virtual-boson momenta. 
This expansion reads as 
\eq
W^{\mu\nu} 
&=& 
        (X^\mu X^\nu + Y^\mu Y^\nu) W_T 
\,+\, i (X^\mu Y^\nu - Y^\mu X^\nu) W_{T_P} 
\,+\,    Z^\mu Z^\nu W_L \nonumber\\[3mm]
 &+&    (Y^\mu Y^\nu - X^\mu X^\nu) W_{\Delta\Delta}  
\,-\,   (X^\mu Y^\nu + Y^\mu X^\nu) W_{\Delta\Delta_P}\nonumber\\[3mm]  
 &-&    (X^\mu Z^\nu + Z^\mu X^\nu) W_{\Delta} 
\,-\,   (Y^\mu Z^\nu + Z^\mu Y^\nu) W_{\Delta_P}\nonumber\\[3mm]   
 &+&  i (Z^\mu X^\nu - X^\mu Z^\nu) W_{\nabla} 
\,+\, i (Y^\mu Z^\nu - Z^\mu Y^\nu) W_{\nabla_P}
\,, 
\en 
where $q$ is the momentum of the gauge boson $\gamma$, $W^\pm$, or $Z^0$, with $q^2 = Q^2$ 
its Minkowski momentum squared, and $\epsilon^{\mu\nu\alpha\beta}$ is the four-dimensional
Levi-Civita tensor defined via ${\rm tr}(\gamma^5 \gamma^\mu \gamma^\nu\gamma^\alpha
\gamma^\beta) = 4 \, i \, \epsilon^{\mu\nu\alpha\beta}$,
with $\epsilon^{0123} = - \epsilon_{0123} = -1$.

\begin{figure}[htp]
	\begin{center}
	  \includegraphics[width=0.75\textwidth,clip]{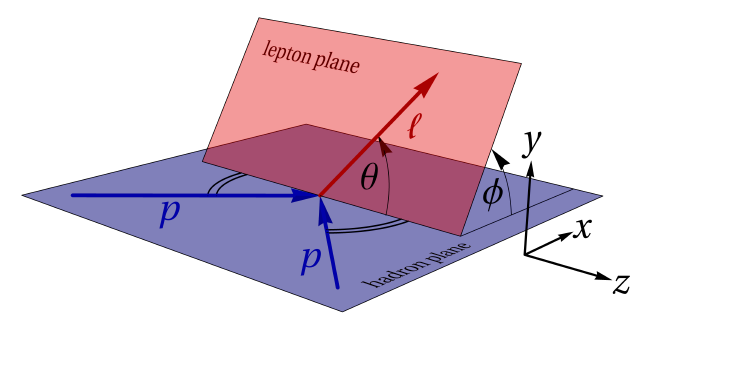}
	\end{center}
        \vspace*{-1cm}
	\caption{Definition of the polar and the
          azimuthal angles for the Drell-Yan process
          in the Collins-Soper frame. The hadron plane is depicted in blue,
          the lepton plane in red.}        	
	\label{fig:CS_frame}
\end{figure}

The number of structure functions, $9 = 3 \times 3$, is determined by the number of
possible helicity settings of the gauge boson in the amplitude and its complex conjugate. 
In the case of the purely weak Drell-Yan reactions or for $\gamma$-$Z^0$ interference we have 
nine functions, while in the case of the purely electromagnetic Drell-Yan we have only
four $T$-even structure functions. In general, the
Drell-Yan hadronic structure functions may be classified as:
\begin{itemize}
\item[(a)] two transverse functions, the $P$-even $W_T$ 
and the $P$-odd $W_{T_P}$; 
\item[(b)] one longitudinal function $W_L$ which is $P$-even;  
\item[(c)] two transverse-transverse interference (double-spin-flip)
functions, the $P$-even $W_{\Delta\Delta}$ and the
$P$-odd $W_{\Delta\Delta_P}$;  
\item[(d)] four transverse-longitudinal interference (single-spin-flip) 
functions, the $P$-even $W_{\Delta}$, $W_{\nabla}$ and 
the $P$-odd $W_{\Delta_P}$, $W_{\nabla_P}$. 
\end{itemize}
The lepton angular distribution $dN/d\Omega$ is expanded in terms of the hadronic
structure functions as 
\eq 
\frac{dN}{d\Omega} &=& \frac{3}{8 \pi (2 W_T + W_L)} \,
\biggl[ 
      g_T             \, W_T
\,+\, g_L             \, W_L 
\,+\, g_\Delta         \, W_\Delta
\,+\, g_{\Delta\Delta}  \, W_{\Delta\Delta}
\nonumber\\
&+&   g_{T_P}          \, W_{T_P}
\,+\, g_{\nabla_P}      \, W_{\nabla_P}
\,+\, g_{\nabla}        \, W_{\nabla}
\,+\, g_{\Delta\Delta_P} \, W_{\Delta\Delta_P} \,
\,+\, g_{\Delta_P}      \, W_{\Delta_P} \biggr] \,,
\en
where
$g_i = g_i(\theta,\phi)$ denote the angular coefficients 
\eq \label{gcoeffs}
g_{_T} & = &  1 + \cos^2\theta\,,  \quad \hspace*{.6cm}
g_{_L} \, = \, 1 - \cos^2\theta\,,  \quad \hspace*{.45cm}
g_{_{T_P}} \, = \, \cos\theta\,, \nonumber\\
g_{_{\Delta\Delta}} & = & \sin^2\theta \, \cos 2\phi\,, \quad \hspace*{.2cm}
g_{_{\Delta}} \, =  \, \sin 2\theta \, \cos\phi\,, \quad  \hspace*{.15cm}
g_{_{\nabla_P}} \, =  \, \sin\theta \, \cos\phi\,,\nonumber \\ 
g_{_{\Delta\Delta_P}} & = & \sin^2\theta \, \sin 2\phi\,, 
\quad  \hspace*{.04cm}
g_{_{\Delta_P}} \, = \, \sin 2\theta \, \sin\phi\,, \quad 
\hspace*{.4cm}
g_{_{\nabla}} \, = \,\sin\theta \, \sin\phi \,,
\en
with $\theta$ and $\phi$ being the polar and azimuthal angles of one of the decay leptons
in the center-of-mass system (c.m.s.) of the lepton pair. The angle $\phi$ may be taken
to define the orientation of the lepton plane with respect to the hadron plane. 
In Fig.~\ref{fig:CS_frame} we show the polar and the azimuthal angles for the Drell-Yan
process in the Collins-Soper frame.

The six angular coefficients $g_i$ ($i=T$, $L$, $\Delta\Delta$, $\Delta$,
$\Delta\Delta_P$, $\Delta_P$) in~(\ref{gcoeffs}) are invariant under the 
$P$-parity transformation $\theta \to \pi - \theta$ and $\phi \to \pi + \phi$,
while the other three coefficients $g_{i}$ ($i=T_P$, $\nabla$, $\nabla_P$) change
their sign in that case. 
Therefore, the six partial lepton angular distributions
$dN_i/d\Omega$ ($i$=$T$, $L$, $\Delta\Delta$, $\Delta$, $T_P$, $\nabla_P$) are also 
$P$ invariant, whereas the other three distributions $dN_i/d\Omega$ ($i$=$\Delta\Delta_P$,
$\Delta_P$, $\nabla$) are $P$ odd and also $T$ odd. As can be seen from
Eq.~(\ref{gcoeffs}), the latter distributions are all proportional to either
$\sin\phi$ or $\sin2\phi$. 

We note in passing that two other commonly employed, and equivalent,
parametrizations 
of the lepton angular distribution
are~\cite{Lam:1978pu,Collins:1977iv,Mirkes:1992hu,Boer:2006eq,Berger:2007jw}
\eq \label{eq4-sf}
\frac{dN}{d\Omega} &=& \frac{3}{16\pi} \, 
\biggl( 
1 + \cos^2\theta 
+ \frac{A_0}{2} (1-3\cos^2\theta) 
+ A_1 \sin 2\theta  \cos\phi 
+ \frac{A_2}{2} \sin^2\theta  \cos 2\phi \nonumber\\[2mm] 
&+& A_3 \sin\theta  \cos\phi 
+ A_4 \cos\theta  
+ A_5 \sin^2\theta \sin 2\phi 
+ A_6 \sin 2\theta \sin\phi 
+ A_7 \sin\theta  \sin\phi  
\biggr)\,,
\en
and 
\eq 
\frac{dN}{d\Omega} &=& \frac{3}{4\pi} \, \frac{1}{\lambda+3} \, 
\biggl( 
1 + \lambda\cos^2\theta 
+ \mu \sin 2\theta  \cos\phi 
+ \frac{\nu}{2} \sin^2\theta  \cos 2\phi \nonumber\\[2mm]    
&+& \tau \sin\theta  \cos\phi
+ \eta \cos\theta  
+ \xi \sin^2\theta \sin 2\phi 
+ \zeta \sin 2\theta \sin\phi 
+ \chi \sin\theta  \sin\phi  
\biggr) \,. 
\en 
The relations between the three sets of structure functions are recalled
in Appendix~\ref{app:Str_Func}.

\begin{figure}[htp]
  \begin{center}
  	  \includegraphics[scale=0.75]{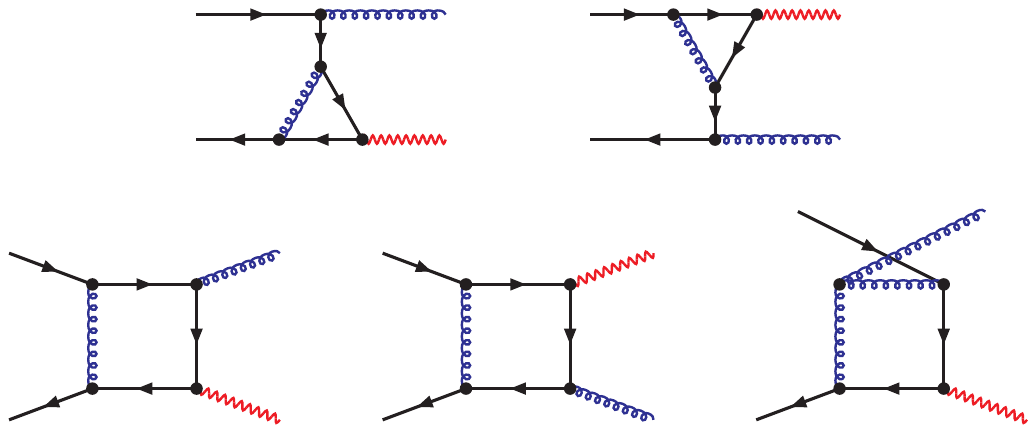}

          \vspace*{-16cm} 
		\end{center}
	\caption{One-loop diagrams for $q \bar q \to g \gamma$
          that produce an absorptive part.} 
          	\label{fig:loop1}
	\begin{center}
            	  \includegraphics[scale=0.75]{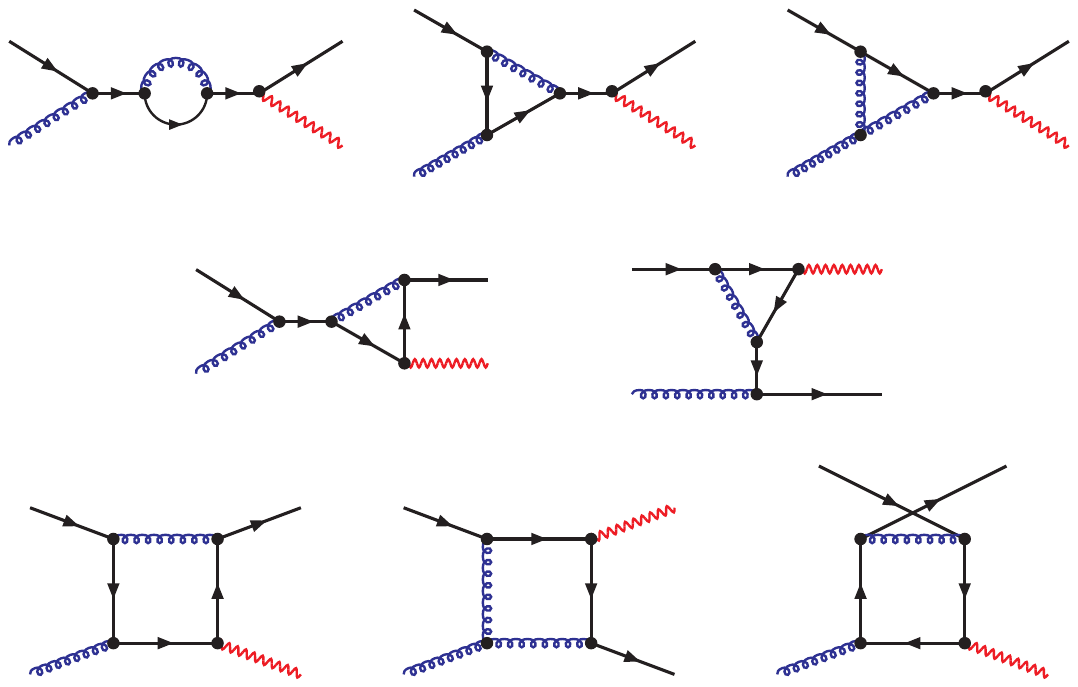}

          \vspace*{-12cm} 
        \end{center}
	\caption{One-loop diagrams for $q g \to q \gamma$
          that produce an absorptive part.} 
	\label{fig:loop2}
\end{figure}

The $T$-odd structure functions 
$W_{\nabla}$, $W_{\Delta\Delta_P}$, and $W_{\Delta_P}$ are generated at ${\cal O}(\alpha_s^2)$
in the strong coupling constant $\alpha_s$ by the absorptive parts of parton scattering
amplitudes.
The leading contributions arise from the interference of one-loop and tree-level diagrams.
The relevant channels are quark-antiquark annihilation and quark-gluon Compton scattering.
Their one-loop diagrams providing an absorptive part for photon exchange in Drell-Yan are shown 
in Figs.~\ref{fig:loop1} and~\ref{fig:loop2}; the diagrams with $Z^0$ and $W^\pm$ bosons
are generated analogously. 
The ensuing $T$-odd effects were first studied in Ref.~\cite{Hagiwara:1984hi} and
later recalculated in~\cite{Mirkes:1992hu}. Here we will
present an independent derivation that will allow us to explore the low-$Q_T$ limit
of the results.

For our calculation of the $T$-odd structure functions we use a convenient orthogonal
basis of vectors $P,R,K$~\cite{Lyubovitskij:2021ges}, defined by
\eq\label{basis_PRK}
P^\mu &=& (p_1 + p_2)^\mu \,, 
\nonumber\\
R^\mu &=& (p_1 - p_2)^\mu \,,
\nonumber\\
K^\mu &=& k_1^\mu
      - P^\mu  \, \frac{P \cdot k_1}{P^2}
      - R^\mu  \, \frac{R \cdot k_1}{R^2}
      = - q^\mu
      + P^\mu  \, \frac{P \cdot q}{P^2}
      + R^\mu  \, \frac{R \cdot q}{R^2}\,,
\en
which obey the conditions 
\eq
P^2 = - R^2 = \hat{s}\,,
\quad
K^2 = - \frac{\hat{u} \hat{t}}{\hat{s}}\,,
\quad 
P \cdot R = P \cdot K = R \cdot K = 0  \,. 
\en
Here $p_1$, $p_2$, and $k_1$ are the momenta of the two initial partons and
the final-state parton, respectively, satisfying the momentum conservation relation
$p_1 + p_2 = k_1 + q$. Furthermore, 
$\hat{s} = (p_1+p_2)^2$,
 $\hat{t} = (p_1-q)^2$,
$\hat{u} = (p_2-q)^2$, with $\hat{s} + \hat{t} + \hat{u} = Q^2$ the parton-level 
Mandelstam variables. 

The $(P,R,K)$ and $(T,X,Y,Z)$ bases are related by 
\eq 
X^\mu &=& \frac{T^\mu \, \sqrt{1+\rho^2}}{\rho}
- \frac{P^\mu z_{12}^{+}  + R^\mu z_{12}^{-}}
{2 Q \rho \sqrt{1+\rho^2}} 
\,=\, \frac{\rho \Big(P^\mu z_{12}^{+}  + R^\mu z_{12}^{-}\Big)}
    {2 Q \sqrt{1+\rho^2}} - \frac{K^\mu \, \sqrt{1+\rho^2}}{Q \rho}
\,, \nonumber\\[2mm]   
Z^\mu &=& \frac{P^\mu z_{12}^{-}  + R^\mu z_{12}^{+}}
{2 Q \sqrt{1+\rho^2}}
\,, \nonumber\\[2mm]      
Y^\mu &=& - \, \epsilon^{\mu PRK}  \frac{z_1 z_2}{Q^3 \rho (1+\rho^2)} \,,
\en
where $z_{12}^{\pm} = z_1 \pm z_2$, $Q = \sqrt{Q^2}$, and
$\epsilon^{\mu PRK} = \epsilon^{\mu\nu\alpha\beta} 
\, P_\nu \, R_\alpha \, K_\beta$.  
We have $z_i = x_i/\xi_i$, with the momentum fractions $\xi_i$ of the partons
defined by $p_i = \xi_i P_i$, and with the momentum fractions of the light-cone
components of the gauge boson, 
\eq\label{x12}
x_{1,2} = e^{\pm y} \sqrt{\frac{Q^2+Q_T^2}{s}}
\en
at nonzero $Q_T$. We also introduce the variables
\eq\label{x120}
x_{1,2}^0 = e^{\pm y} \, \frac{Q}{\sqrt{s}}
\en 
relevant in the $Q_T = 0$ limit.

The hadronic structure functions $W(x_1,x_2,\rho^2)$ for the Drell-Yan process
with colliding hadrons $H_1$ and $H_2$ are related to the parton-level structure
functions $w^{ab}(x_1,x_2,\rho^2)$ by the QCD collinear factorization formula
\eq\label{factorization}
W(x_1,x_2,\rho^2) =
\frac{1}{x_1 x_2} \, \sum\limits_{a,b}
\, \int\limits_{x_1}^1 \, dz_1
\, \int\limits_{x_2}^1 \, dz_2
\ w^{ab}(z_1,z_2,\rho^2) 
\, f_{a/H_1}\Big(\frac{x_1}{z_1}\Big) 
\, f_{b/H_2}\Big(\frac{x_2}{z_2}\Big) \,,
\en
where $f_{i/H}(\xi)$ is the PDF describing the
$\xi$ distribution of partons of type $i$ in hadron $H$.
We are suppressing here the scale dependence of the PDFs. 

We may project onto the parton-level $T$-odd structure functions
in the following way (here we drop parton labels):
\eq
w_{\Delta\Delta_P} &=& - \frac{1}{2} (X^\mu Y^\nu + X^\nu Y^\mu) \, w_{\mu\nu}
\nonumber\\
&=& - \frac{z_1z_2}{4 Q^4 \, \rho^2 \, (1+\rho^2)^{3/2}}  \,
\biggl[ \epsilon^{\mu PRK} \Big(P^\nu z_{12}^{+} + R^\nu z_{12}^{-} \Big)
      + \epsilon^{\nu PRK} \Big(P^\mu z_{12}^{+} + R^\mu z_{12}^{-} \Big)
  \biggr]  \, w_{\mu\nu} \,,
\label{wdd}\\[5mm]  
w_{\Delta_P} &=& - \frac{1}{2} (Y^\mu Z^\nu + Y^\nu Z^\mu) \, w_{\mu\nu}
\nonumber\\  
&=& \frac{z_1z_2}{4 Q^4 \, \rho \, (1+\rho^2)^{3/2}}  \, 
\biggl[ \epsilon^{\mu PRK}
  \Big(P^\nu z_{12}^{-} + R^\nu z_{12}^{+} \Big) 
  + \epsilon^{\nu PRK}
  \Big(P^\mu z_{12}^{-} + R^\mu z_{12}^{+} \Big) 
\biggr]  \, w_{\mu\nu} \,,
\label{wd}\\[5mm]  
w_{\nabla} &=& \frac{i}{2} (X^\mu Z^\nu - X^\nu Z^\mu) \, w_{\mu\nu}
= \frac{i \, z_1 z_2}{2 Q^2 \, \rho \, (1+\rho^2)}  \, 
\Big(P^\nu R^\mu - P^\mu R^\nu\Big) \, w_{\mu\nu} \,.
\label{wnabla} 
\en
In the evaluation of the absorptive parts of the one loop diagrams we use
the following set of imaginary parts of scalar one-loop 
integrals~\cite{Lyubovitskij:2021ges,Haug:2022hkr}:

\eq
& &
   {\rm Im}B_0(Q^2) = {\rm Im}B_0(\hat{s}) = \pi 
    \,, \ \ 
   {\rm Im}B_0(\hat{u}) = {\rm Im}B_0(\hat{t}) = 0
   \,,
   \nonumber\\
& &
  {\rm Im}C_0(\hat{s},0) =  \frac{\pi}{\hat{s}}
  \biggl( \frac{1}{\bar\epsilon} - \log\frac{\hat{s}}{\mu^2}
  \biggr)
   \,, \ \ 
   {\rm Im}C_0(\hat{u},0) =  {\rm Im}C_0(\hat{t},0) =  0  \,,
   \nonumber\\   
& &
   {\rm Im}C_0(Q^2,0) =  \frac{\pi}{Q^2}
   \biggl( \frac{1}{\bar\epsilon} - \log\frac{Q^2}{\mu^2}
   \biggr)
   \,, \ \ 
     {\rm Im}C_0(Q^2,\hat{s}) = - \frac{\pi}{Q^2-\hat{s}} \, 
   \log\frac{Q^2}{\hat{s}} \,,
   \nonumber\\
& &       
   {\rm Im}C_0(Q^2,\hat{u}) =  \frac{\pi}{Q^2-\hat{u}}
   \biggl( \frac{1}{\bar\epsilon} - \log\frac{Q^2}{\mu^2}
   \biggr)
   \,, \ \
   {\rm Im}C_0(Q^2,\hat{t}) =  \frac{\pi}{Q^2-\hat{t}}
   \biggl( \frac{1}{\bar\epsilon} - \log\frac{Q^2}{\mu^2} \biggr) \,,
   \nonumber\\
& &
   {\rm Im}C_0(\hat{s},\hat{u}) =  \frac{\pi}{\hat{s}-\hat{u}}
   \biggl( \frac{1}{\bar\epsilon} - \log\frac{\hat{s}}{\mu^2} \biggr) 
   \,,
   \ \
   {\rm Im}C_0(\hat{s},\hat{t}) =  \frac{\pi}{\hat{s}-\hat{t}}
   \biggl( \frac{1}{\bar\epsilon} - \log\frac{\hat{s}}{\mu^2} \biggr) \,,
   \nonumber\\
& & 
   {\rm Im}D_0(Q^2,\hat{s},\hat{u}) =  - \frac{2 \pi}{\hat{s} \hat{u}}
   \log\frac{Q^2-\hat{u}}{Q^2} \,,
   \ \
   {\rm Im}D_0(Q^2,\hat{s},\hat{t}) =  - \frac{2 \pi}{\hat{s} \hat{t}}
   \log\frac{Q^2-\hat{t}}{Q^2} \,,
   \nonumber\\
& & 
   {\rm Im}D_0(Q^2,\hat{t},\hat{u}) =  \frac{2 \pi}{\hat{u} \hat{t}}
   \biggl(-  \frac{1}{\bar\epsilon} + \log\frac{Q^2}{\mu^2} - \log\frac{(Q^2-\hat{u})
   (Q^2-\hat{t})}{\hat{u} \hat{t}} 
   \biggr) \,,     
\en
where $B_0$ denotes the two-propagator (bubble) diagram, $C_0$ the three-propagator
(triangle) diagram, and $D_0$ the four-propagator (box) diagram.
We have used dimensional regularization with $D=4-2\epsilon$ space-time dimensions;
as 
usual $1/\bar\epsilon = 1/\epsilon + \log(4 \pi) + \gamma_\text{E}$ with
the Euler-Mascheroni constant $\gamma_E$.

We note that in the cases of the electroweak contributions to $w_{\Delta\Delta_P}$ and
$w_{\Delta_P}$ we have to deal with an odd number of $\gamma^5$ matrices in the relevant
Dirac traces. We adopt the Larin scheme for treating $\gamma_5$; details are discussed
in Appendix~\ref{app:gamma5}.

\section{Analytical results for the partonic $T$-odd structure functions}
\label{sec: Results}

In this section we present our analytical results for the parton-level $T$-odd structure
functions $w_{\Delta\Delta_{P}}$, $w_{\Delta_{P}}$, and $w_\nabla$.
We first introduce some notation.
We use the QCD color factors $C_F = (N_c^2-1)/(2 N_c) = 4/3$, $C_A = N_c = 3$, $T_F = 1/2$,
$C_1 = C_F - N_c/2 = - 1/(2 N_c) = - 1/6$, which at large $N_c$ scale 
as ${\cal  O}(N_c)$, ${\cal  O}(N_c)$, ${\cal  O}(1)$, ${\cal  O}(1/N_c)$, respectively.
Specifically, the color factors for $q\bar q$ annihilation and $qg$ scattering are
$C_{q\bar q} = C_F/N_c = (N_c^2-1)/(2 N_c^2) = 4/9$ and $C_{qg} = T_F/N_c = 1/(2 N_c) = 1/6$.
Furthermore, it is convenient to introduce the coupling factors $g_{q\bar q; i} =
C_{q\bar q} \, g_{{\rm EW}; i}^{Z\gamma/W}
\, e_q^2 \, \alpha_s^2/(4\pi)$ and
$g_{qg; i} =  C_{qg} \,  g_{{\rm EW}; i}^{Z\gamma/W} \, e_q^2 \, \alpha_s^2/(4\pi)$, 
where $i=1,2$.
Here, $e_q$ is the electric charge of a quark of flavor $q$. 
The electroweak couplings $g_{{\rm EW}; 1}$ and $g_{{\rm EW}; 2}$,
which incorporate the products of couplings of the gauge bosons
($W^\pm$, $Z^0$, $\gamma$) with quarks and leptons, are given by
\eq
g_{\rm EW; 1}^{Z\gamma} &=& 2 \, g_{Zq}^{A} \, \Big[
g_{Zq}^{V}  \, \Big((g_{Z\ell}^{V})^2 + (g_{Z\ell}^{A})^2\Big) \, |D_Z(Q^2)|^2
+ g_{Z\ell}^{V} \, \mathrm{Re}[D_Z(Q^2)]\Big]  
\,,
\nonumber\\ 
g_{\rm EW; 2}^{Z\gamma} &=& 2 \, g_{Z\ell}^{A} \, \Big[
g_{Z\ell}^{V} \, \Big((g_{Zq}^{V})^2 + (g_{Zq}^{A})^2\Big) \ |D_Z(Q^2)|^2
+ g_{Zq}^{V} \, \mathrm{Re}[D_Z(Q^2)]\Big]  \,,
\label{ccouplingZ}
\en
in the case of electrically neutral gauge bosons ($Z^0$, $\gamma$) and
\eq
g_{\rm EW; 1}^W &=& 
2 \, g_{Wqq'}^{V} \, g_{Wqq'}^{A}
\, \Big((g_{W\ell}^{V})^2 + (g_{W\ell}^{A})^2\Big)
|V_{qq'}|^2 \, |D_W(Q^2)|^2 
\,,
\nonumber\\ 
g_{\rm EW; 2}^W
&=& 2 \, g_{W\ell}^{V} \, g_{W\ell}^{A}
\, \Big((g_{Wqq'}^{V})^2 + (g_{Wqq'}^{A})^2\Big) \,
|V_{qq'}|^2 \, |D_W(Q^2)|^2\,,
\label{ccouplingW} 
\en
in the case of the $W^\pm$ gauge bosons, where 
\eq
g_{W\ell}^{V} &=& g_{W\ell}^{A}
\, = \, g_{Wqq'}^{V} \, = \, g_{Wqq'}^{A}
\, = \,  \frac{1}{2 \sin\theta_W \sqrt{2}}
\,,
\nonumber\\
g_{Z\ell}^{V} &=& - \frac{1 - 4 \sin^2\theta_W}{2 \sin 2\theta_W}
\,,
\quad \hspace*{.1cm} 
g_{Z\ell}^{A} \, = \, - \frac{1}{2 \sin 2\theta_W}
\,,
\nonumber\\
g_{Zu}^{V} &=& \frac{1 - 8/3 \sin^2\theta_W}{2 e_q \sin 2\theta_W}
\,,
\quad 
g_{Zd}^{V} \, = \, - \frac{1 - 4/3 \sin^2\theta_W}{2 e_q \sin 2\theta_W}
\,,
\nonumber\\
g_{Zu}^{A} &=&   \frac{1}{2 e_q \sin 2\theta_W}
\,,
\quad \hspace*{.6cm} 
g_{Zd}^{A} \, = \, - \frac{1}{2 e_q \sin 2\theta_W}    \,.
\en
In the above expressions, $V_{qq'}$ is the relevant element of
the Cabibbo-Kabayashi-Maskawa (CKM) matrix and 
$\theta_W$ is the Weinberg angle measured to be
$\sin^2\theta_W = 0.23121$~\cite{ParticleDataGroup:2022pth}.
Furthermore, $D_G(Q^2)$  ($G=W^\pm, Z^0$) denotes the product of the Breit-Wigner
propagator of a weak gauge boson and $Q^2$. 
Its real and imaginary part are given by
\eq
\mathrm{Re}[D_G(Q^2)] &=& 
\frac{(Q^2-M_G^2) Q^2}{(Q^2-M_G^2)^2 + M_G^2 \Gamma_G^2} \,, 
\nonumber\\
\mathrm{Im}[D_G(Q^2)] &=& 
\frac{M_G \Gamma_G Q^2}{(Q^2-M_G^2)^2 + M_G^2 \Gamma_G^2} \,.
\en
The masses $M_{G}$ and total widths $\Gamma_{G}$ of the bosons,
taken from the Particle Data Group~\cite{ParticleDataGroup:2022pth},
are
$M_{W^\pm} = 80.377 \pm 0.012$ GeV,
$M_{Z^0} = 91.1876 \pm 0.0021$ GeV,
$\Gamma_{W^\pm} = 2.085 \pm 0.042$ GeV, and
$\Gamma_{Z^0} = 2.4955 \pm 0.0023$ GeV.
Note that in Eqs.~(\ref{ccouplingZ}) and (\ref{ccouplingW}) the terms proportional
to the squares of the Breit-Wigner propagators ($|D_Z(Q^2)|^2$ and $|D_W(Q^2)|^2$)
correspond to the purely weak $ZZ$ and $WW$ contributions to the couplings,
while the terms proportional to the real part of the Breit-Wigner $Z$ boson
propagator $\mathrm{Re}[D_Z(Q^2)]$ correspond to $\gamma$-$Z$ interference.  

As a final ingredient, we note that all one-loop partonic structure functions
contain a factor $\delta\big((\hat{s}+\hat{t}+\hat{u}-Q^2)/\hat{s}\big)$ 
arising from phase space and corresponding to the fact that the recoil
in the final state consists of a single massless parton.
It is convenient to write
\eq
w^{ab}(z_1,z_2,\rho^2)\,=\,\tilde{w}^{ab}(z_1,z_2,\rho^2)
\,\delta\big((\hat{s}+\hat{t}+\hat{u}-Q^2)/\hat{s}\big)\,.
\en
With this notation in place, we obtain the following partonic $T$-odd structure
functions. For the $q \bar q$ annihilation subprocess, we find 
\eq
\tilde{w}_{\Delta\Delta_{P}}^{q \bar q}
&=& \frac{g_{q\bar q; 1}}{2} \,
\sqrt{
   \frac{Q^2 \hat{s}}{(Q^2-\hat{u}) (Q^2-\hat{t})}} \,
   \biggl[- \frac{C_F}{2}
   \biggl(\frac{Q^2-\hat{t}}{Q^2-\hat{u}}
 + \frac{Q^2-\hat{u}}{Q^2-\hat{t}} 
   \biggr)
   \nonumber\\
  &+& C_1 \biggl(
   \frac{Q^2-\hat{t}}{\hat{t}}
   \biggl(1 - \frac{\hat{s}}{\hat{t}}
   \log\frac{Q^2-\hat{u}}{\hat{s}} \biggr)
+  \frac{Q^2-\hat{u}}{\hat{u}}
   \biggl(1 - \frac{\hat{s}}{\hat{u}}
   \log\frac{Q^2-\hat{t}}{\hat{s}}
\biggr)   
\biggr] \,,
\\
\tilde{w}_{\Delta_{P}}^{q \bar q}
&=& \frac{g_{q\bar q; 1}}{2} \,
\, 
\frac{Q^2 \hat{s}}{\sqrt{(Q^2-\hat{u}) (Q^2-\hat{t}\,) \hat{u} \hat{t}}} \,
\biggl[ C_F
\biggl(\frac{Q^2-\hat{t}}{Q^2-\hat{u}} - \frac{Q^2-\hat{u}}{Q^2-\hat{t}} 
\biggr)
\nonumber\\
  &+& C_1 \biggl(
       \frac{Q^2-\hat{u}}{\hat{u}}
   \log\frac{Q^2-\hat{t}}{\hat{s}} 
    -  \frac{Q^2-\hat{t}}{\hat{t}}
   \log\frac{Q^2-\hat{u}}{\hat{s}} \biggr)   
   \biggr] \,,
\\
\tilde{w}_\nabla^{q \bar q}
&=& g_{q\bar q; 2} \, \sqrt{\frac{Q^2 \hat{s}}{\hat{u} \hat{t}}} \,
\biggl[ \frac{C_F}{2} \frac{(2 Q^2 \hat{s} + \hat{u} \hat{t}) (Q^2+\hat{s})
    (\hat{u}-\hat{t})}{(Q^2-\hat{u})^2 (Q^2-\hat{t})^2}
  \nonumber\\
  &+& C_1 \biggl( -  \frac{Q^2 (\hat{u}-\hat{t})}{(Q^2-\hat{u}) (Q^2-\hat{t})}
  + \frac{\hat{s}}{\hat{u}} \log\frac{Q^2-\hat{t}}{\hat{s}}
  - \frac{\hat{s}}{\hat{t}} \log\frac{Q^2-\hat{u}}{\hat{s}} 
  \biggr) \biggr] \,.
\en

For $q g$ scattering we have
\eq
\tilde{w}_{\Delta\Delta_{P}}^{q g}
&=& \frac{g_{q g; 1}}{2} \, \frac{\hat{u}}{\hat{s}} \,
\sqrt{\frac{Q^2 \hat{s}}{(Q^2-\hat{u}) (Q^2-\hat{t})}}  \,
\biggl[ \frac{C_F}{2}
  \frac{2Q^2 + \hat{s}}{Q^2-\hat{t}} 
  \nonumber\\
  &-& C_1 \biggl(
   \frac{Q^2}{Q^2-\hat{u}}
   + \frac{Q^2-\hat{u}}{\hat{s}}
   \log\frac{(Q^2-\hat{u}) (Q^2-\hat{t})}{\hat{u} \hat{t}}
+  \frac{Q^2-\hat{t}}{\hat{t}}
\biggl(1 - \frac{Q^2-\hat{u}}{\hat{t}}  \log\frac{Q^2-\hat{u}}{\hat{s}}
\biggr) \biggr)   
\biggr] \,,
\\
\tilde{w}_{\Delta_{P}}^{q g}
&=& \frac{g_{q g; 1}}{2} \,
\frac{Q^2 \hat{u}}
{\sqrt{(Q^2-\hat{u}) (Q^2-\hat{t}) \hat{u} \hat{t}}}  \,
\biggl[ \frac{C_F}{2} \frac{\hat{t} - 2 \hat{u}}{Q^2-\hat{t}}
 \nonumber\\
  &-& C_1 \biggl(
  \frac{\hat{u}}{Q^2-\hat{u}}
+ \frac{Q^2}{\hat{s}}
- \frac{(Q^2 - \hat{u}) \hat{u}}{\hat{s}^2}
  \log\frac{(Q^2-\hat{u}) (Q^2-\hat{t})}
     {\hat{u} \hat{t}}\biggr)
\biggr] \,,
\\
\tilde{w}_\nabla^{q g} &=& g_{qg; 2} \, \frac{\hat{u}}{\hat{s}} \, 
  \sqrt{\frac{Q^2 \hat{s}}{\hat{u} \hat{t}}}  \,
  \biggl[- \frac{C_F}{2} \frac{2 Q^2\hat{u}
+ \hat{s} \hat{t}}{(Q^2-\hat{t})^2}
  \nonumber\\
  &-& C_1 \biggl( \frac{2 Q^2 \hat{s}}{(Q^2-\hat{u})^2}
  - \frac{Q^4 - \hat{u} \hat{t}}{(Q^2-\hat{u}) (Q^2-\hat{t})}
  + \frac{\hat{u}}{\hat{t}} \log\frac{Q^2-\hat{u}}{\hat{s}}
  - \frac{\hat{u}}{\hat{s}} \log\frac{(Q^2-\hat{u})
    (Q^2-\hat{t})}{\hat{u}\hat{t}} 
  \biggr) \biggr] \,. 
\en
We note that in the large $N_c$ limit the terms proportional to $C_1$
in the structure functions are suppressed by a factor  $1/N_c^2$ relative
to those proportional to $C_F$ and $C_A$.
The results for $g \bar q$ scattering are obtained from the 
ones for the $q g$ process by interchange of momenta, $p_1 \leftrightarrow p_2$, 
which corresponds to an interchange of Mandelstam variables $u \leftrightarrow t$.

For calculating the hadronic structure functions via the factorization
formula~\eqref{factorization} and subsequently investigating their small
$Q_T$ behavior, it is convenient to express our results in terms of the
variables $z_1 = x_1/\xi_1$, $z_2 = x_2/\xi_2$, and $\rho^2=Q^2/Q_T^2$.
Using
\eq
& &\hat{s} = \frac{Q^2+Q_T^2}{z_1 z_2} \,,
\quad 
Q^2 - \hat{t} = \frac{Q^2+Q_T^2}{z_1} \,,
\quad 
Q^2 - \hat{u} = \frac{Q^2+Q_T^2}{z_2} \,,
\nonumber\\
& &\frac{\hat{u} \hat{t}}{Q^2 \hat{s}} = \rho^2 \,,
\quad 
\frac{(Q^2-\hat{u}) (Q^2-\hat{t})}{Q^2 \hat{s}} = 1 + \rho^2 \,,
\en
one gets for $q \bar q$ annihilation  
\eq
\tilde{w}_{\Delta\Delta_{P}}^{q \bar q}
&=& - \frac{g_{q\bar q; 1}}{4 z_1 z_2} \,
\, \frac{1}{\sqrt{1+\rho^2}} \,
\biggl[ C_A \frac{z_1^2+z_2^2}{2} + C_1 \Big(z_1^2 F_1(z_2)
  + z_2^2 F_1(z_1) \Big)
\biggr] \,,
\\
\tilde{w}_{\Delta_{P}}^{q \bar q}
&=& - \frac{g_{q\bar q; 1}}{2 z_1 z_2} \,
\, \frac{1}{\rho \sqrt{1+\rho^2}}
\biggl[ C_A \frac{z_1^2-z_2^2}{2} + C_1 \Big(z_1^2 F_2(z_2)
  - z_2^2 F_2(z_1) \Big)
\biggr] \,,
\\
\tilde{w}_\nabla^{q \bar q}
&=& - \frac{g_{q\bar q; 2}}{z_1 z_2}
\, \frac{1}{\rho}
\biggl[ \Big(C_A -  \frac{\rho^2}{1+\rho^2} C_F\Big)
    \frac{z_1^2-z_2^2}{2}
  + C_1 \Big(z_1 F_2(z_2) - z_2 F_2(z_1)\Big)
  \biggr] \,,
  \label{qq_full}
\en
and for $q g$ scattering 
\eq
\tilde{w}_{\Delta\Delta_{P}}^{q g}
&=& - \frac{g_{q g; 1}}{2} \, \frac{1-z_2}{z_1 z_2}
\, \frac{1}{\sqrt{1+\rho^2}}
\, 
\biggl[ \frac{C_F}{2} z_1
\biggl(1 + \frac{2 z_1z_2}{1 + \rho^2}\biggr) 
  \nonumber\\
  &+& C_1 z_1 z_2 \biggl( \Big(F_1(z_1) - \frac{1-\rho^2}{1+\rho^2}\Big)
  \frac{z_2}{2} 
  + z_1 \log\frac{\rho^2}{1+\rho^2}
 \biggr)\biggr]
 \,,
\\
\tilde{w}_{\Delta_{P}}^{q g}
&=& - \frac{g_{q g; 1}}{2} \, \frac{1-z_2}{z_1 z_2}
\, \frac{1}{\rho \sqrt{1+\rho^2}}
\, 
\biggl[ \frac{C_F}{2} z_1 (1 + z_1 - 2 z_2) 
  \nonumber\\
  &+& C_1 z_2 \biggl(1-z_2 - \frac{z_1^2 z_2}{1+\rho^2}
   +  z_1^2 (1-z_2) \log\frac{\rho^2}{1+\rho^2} 
 \biggr)\biggr]
 \,,
 \\
\tilde{w}_\nabla^{q g}
&=& - g_{q g; 2}  \frac{1-z_2}{z_1 z_2}
\, \frac{1}{\rho} 
\biggl[C_F z_1 \biggl(\frac{z_1 (1-z_2)}{1+\rho^2}
  + \frac{1-z_1}{2 z_2} \biggr) 
  \nonumber\\
  &+& C_1 z_2 \biggl(\frac{z_1^2 - 2 z_2}{1+\rho^2}
  + z_1 (1-z_2) \log\frac{\rho^2}{1+\rho^2} 
  - (1-z_2) F_2(z_1) \biggr)
  \biggr]
\,.
\en 
Here, the functions $F_1$ and $F_2$ are defined as:
\eq\label{Fdef}
F_1(z) &\equiv& \frac{1+z}{1-z}
+ \frac{2 z \log(z)}{(1-z)^2}
= 2 \sum\limits_{N=1}^{\infty} \, \frac{(1-z)^N}{(N+1) (N+2)} 
= {\cal O}(1-z)
\,, \nonumber\\ 
F_2(z) &\equiv& 1 + \frac{z \log(z)}{1-z}
=  \frac{1-z}{2} \, \Big(1 + F_1(z) \Big)
= \sum\limits_{N=1}^{\infty} \, \frac{(1-z)^N}{N (N+1)} 
= {\cal O}(1-z)
\,. 
\en
They obey $F_1(1) = F_2(1) = 0$ and $F_1(0) = F_2(0) = 1$. For later reference,
we have also given their expansions around $z=1$.

We now have
\eq\label{factorizationdelta}
W(x_1,x_2,\rho^2) =
\frac{1}{x_1 x_2} \, \sum\limits_{a,b}
\, \int\limits_{x_1}^1 \, dz_1
\, \int\limits_{x_2}^1 \, dz_2
\ \tilde{w}^{ab}(z_1,z_2,\rho^2)\,\delta\left((1-z_1)(1-z_2)
-\frac{\rho^2}{1+\rho^2}z_1 z_2\right) 
\, f_{a/H_1}\Big(\frac{x_1}{z_1}\Big) 
\, f_{b/H_2}\Big(\frac{x_2}{z_2}\Big) \,.\quad
\en
This factorization is formally valid when $Q_T$ is of order $Q$, that is,
for $\Lambda_\text{QCD}\ll Q\sim Q_T$. 
For $Q_T\ll Q$ the appropriate factorization formalism is TMD factorization.
Here, we will take the collinear factorization and extrapolate to small values
of $Q_T$ by formally expanding the result about $Q_T=0$.
This will result in an expansion in powers of $\rho^2$ which can be matched
to TMD results in the region of intermediate $Q_T$ for a smooth transition
from the TMD to the collinear regime. 
We will perform the expansion in $\rho^2$ beyond the leading power,
which has been the main focus in the existing literature, to provide information
about which higher-power corrections are accounted for in the collinear formalism.

\section{Small-$Q_T$ expansion}
\label{sec: Small QT}

In the expansion of hadronic structure functions of the form
in Eq.~\eqref{factorizationdelta} we have three contributions.
First, there is the direct dependence of the partonic structure function on $Q_T$.
Second, the phase space delta function has nontrivial $Q_T$ dependence.
Third, the variables $x_1$, $x_2$ have implicit $Q_T$ dependence.
The first type of contribution may be straightforwardly taken into account
by simple expansion of the partonic structure functions.
The second and third contributions require more discussion.

The phase space delta function in~(\ref{factorizationdelta}) is well known
in the literature, see, e.g., Refs.~\cite{Meng:1995yn,Boer:2006eq,Berger:2007jw}.
Its expansion to leading power in $\rho^2 = Q_T^2/Q^2$ was also given in that
reference and reads as 
\begin{align}\label{delta_QT}
\delta\!\left((1-z_1)(1-z_2)-\frac{\rho^2}{1+\rho^2}z_1 z_2 \right)
=\frac{\delta(1-z_1)}{(1-z_1)_+}+\frac{\delta(1-z_2)}{(1-z_2)_+}
-\delta(1-z_1)\delta(1-z_2)\log\rho^2+\mathcal{O}(\rho^2)&\,.
\end{align}
Here the ``plus'' distribution is defined by
\begin{equation}
\int_0^1 dz\,\frac{f(z)}{(1-z)_+}\equiv\int_0^1 dz\,\frac{f(z)-f(1)}{1-z}\,,
\end{equation}
for a function $f$ that is regular at $z=1$. We note in passing that
in Ref.~\cite{Gelfand} a general method for the expansion of 
distributions was developed, based on Mellin integral techniques.
Building on these ideas, we recently formulated~\cite{LVWZ:2023} an algorithm 
for the small-$Q_T$ expansion of singular functions valid to arbitrary order
of $\rho^2$ and arbitrary number of radiated partons. 
This will be presented in a separate publication.
Here we are only concerned with the expansion of {\it integrals} containing the phase 
space delta function in \eqref{delta_QT}.
For a general regular function $\varphi(z_1,z_2)$ such an integral can be expanded
for small $Q_T$ including $\mathcal{O}(\rho^4,\rho^4\log\rho^2)$ terms
in the following way:
\eq\label{small_QT_I}
I_0 &\equiv& \int\limits_{x_1}^1 dz_1 \int\limits_{x_2}^1 dz_2
\, \delta\Big((1-z_1) (1-z_2)
- \frac{\rho^2}{1+\rho^2} z_1 z_2\Big) \, \varphi(z_1,z_2)
\nonumber\\
&=& \int\limits_{x_1}^1 dz_1 \int\limits_{x_2}^1 dz_2
\, \biggl( \delta(1-z_2) G_1(z_1,z_2)
+ \delta(1-z_1) G_1(z_2,z_1)
\nonumber\\
&+& \delta(1-z_1) \delta(1-z_2) 
G_2(z_1,z_2)
\biggr) \, \varphi(z_1,z_2)
+ {\cal O}(\rho^6,\rho^6\log\rho^2)
\,,
\en
with
\eq
G_1(z_1,z_2) &=& \frac{(1+\rho^2) (1+\rho^2\partial_{z_2}) +
  \rho^4\partial^2_{z_2}/2}{(1-z_1)_{+}}
\nonumber\\
&-& \frac{\rho^2 (1+\rho^2 + (1+3\rho^2) \partial_{z_2} +
  \rho^2 \partial^2_{z_2})}{(1-z_1)^2_{+,1}}
+ \frac{\rho^4 (1+2 \partial_{z_2}+\partial^2_{z_2}/2)}{(1-z_1)^3_{+,2}}
\,, \nonumber\\
G_2(z_1,z_2) &=&
\rho^2 \biggl(1 + \rho^2 \Big(\frac{1}{2}
+\partial_{z_1}+\partial_{z_2}+\partial_{z_1z_2}^2\Big)\biggr)
- \log\rho^2 \biggl((1+\rho^2) (1+\rho^2 (\partial_{z_1}+\partial_{z_2}
+\partial^2_{z_1z_2}))
\nonumber\\
&+& \frac{\rho^4}{2}
\, (\partial_{z_1} + \partial_{z_2})^2 
+ \rho^4 \partial^2_{z_1z_2} \,
\Big(1 + \partial_{z_1} + \partial_{z_2} 
+ \frac{\partial^2_{z_1z_2}}{4}\Big)
\biggr)
\,. 
\en
Here  $1/(1-z)^m_{+,m-1}$ is a generalized plus distribution of power $m$, defined by
\begin{equation}\label{genplus}
\int_0^1 dz \frac{f(z)}{(1-z)^m_{+,m-1}}\equiv
\int_0^1 dz\frac{f(z)-\mathcal{T}^{m-1}_{z=1}f(z)}{(1-z)^m}\,,
\end{equation}
where $f(z)$ is again a sufficiently regular test function and
$\mathcal{T}^{m-1}_{z=1}f(z)$ denotes the Taylor polynomial of $f(z)$
about $z=1$ to order $m-1$,
\begin{equation}
\mathcal{T}^{m-1}_{z=1}f(z)=\sum_{k=0}^{m-1}\frac{(-1)^k f^{(k)}(1)}{k!}\,(1-z)^k.
\end{equation}
A lower integration bound of $x$ instead of zero introduces
additional boundary terms of the form
\eq\label{PDF_d}
\int\limits_{x}^1 dz \, 
\frac{f(z)}{(1-z)^m_{+,m-1}}
&=& 
\int\limits_{x}^1 dz \, \biggl[\frac{1}{(1-z)^m_{+x,m-1}}
+ \delta(1-z) \, \log(1-x_1) \, \frac{(-1)^{m-1}}{(m-1)!} 
\, \partial_{z}^{m-1}
\nonumber\\
&-& \delta(1-z) \, 
\sum\limits_{j=2}^m
\, \frac{(-1)^{m-j}}{(j-1) \, (m-j)!} \, 
\biggl( \frac{1}{(1-x_1)^{j-1}} - 1 \biggr)
\, \partial_{z}^{m-j} 
\biggr] 
f(z_1) \,, 
\en
where $f(z)/(1-z)^m_{+x,m-1}$ is the generalized plus distribution defined
for an integral starting at a finite lower limit $x$, i.e.
\begin{equation}
\int_x^1 dz \frac{f(z)}{(1-z)^m_{+x,m-1}}\equiv\int_x^1 dz\frac{f(z)
-\mathcal{T}^{m-1}_{z=1}f(z)}{(1-z)^m}\,.
\end{equation}
Comparing Eq.~(\ref{small_QT_I}) with \eqref{delta_QT} one can see that we reproduce
the known leading terms, while the terms of order $\rho^2$, $\rho^2 \log\rho^2$, $\rho^4$,
and $\rho^4 \log\rho^2$ are new.

Substituting the small-$Q_T$ expansion of the parton-level structure functions
$w^{ab}(z_1,z_2,\rho^2)$ for the various partonic channels into Eq.~(\ref{factorization})
we get for the contributions to the small-$Q_T$ expansion of
the hadronic structure function $W(x_1,x_2,\rho^2)$:   
\eq\label{small_QT_II}
W_{\delta-\textrm{fct.}}(x_1, x_2,\rho^2)
= W_0(x_1, x_2,L_\rho) + \rho^2 \, W_1(x_1, x_2,L_\rho) +
\rho^4 \, W_2(x_1, x_2,L_\rho) + {\cal O}(\rho^6) \,,
\en
were we have abbreviated
\eq
L_\rho\equiv \log\rho^2\,.
\en
The expansion coefficients with $i=1,2,3$ have the structure
\eq\label{Wi_general}
W_i(x_1, x_2,L_\rho) &=& \frac{1}{x_1x_2}\sum_{a,b}
\, \biggl[
   R_{ab,i}(x_1,x_2,L_\rho)
\, f_{a/H_1}(x_1)
\, f_{b/H_2}(x_2)
\nonumber\\
&+& \Big(P_{ba,i} \otimes f_{b/H_2}\Big)(x_2,x_1,L_\rho)  \ f_{a/H_1}(x_1)  
 +  \Big(P_{ab,i} \otimes f_{a/H_1}\Big)(x_1,x_2,L_\rho)  \ f_{b/H_2}(x_2) 
\biggr] 
\,,
\en 
where
\eq\label{convolution_general}
\big({\cal P} \otimes f\big)(x,y,L_\rho) = \int_x^1 
\, \frac{dz}{z} \, {\cal P}(z,y,L_\rho) \, f\Big(\frac{x}{z}\Big) 
\en  
denotes a generalized convolution, $R_i(x_1,x_2,L_\rho)$,
$P_{ba,i}(z_2,x_1,L_\rho)$, and $P_{ab,i}(z_1,x_2,L_\rho)$
are perturbative coefficient
functions containing differential operators acting on
the PDFs $f_{a/H_1}(x_1)$ and $f_{b/H_2}(x_2)$. 
We note that the generalized convolution~(\ref{convolution_general})
reverts to the ordinary one,
\eq\label{convolution_usual}
\big({\cal P} \otimes f\big)(x) = \int_x^1 
\, \frac{dz}{z} \, {\cal P}(z) \, f\Big(\frac{x}{z}\Big) \,,
\en 
when ${\cal P}(z,y,L_\rho)$ does not depend on $y$ and $L_\rho$.
Details are given in Appendix~\ref{app:calc_tech}.
We stress that, as indicated in Eq.~(\ref{small_QT_II}),
the functions $W_i$ may carry dependence on $\log\rho^2$,
on top of the overall power of $\rho$ that they multiply.

However, Eq.~(\ref{small_QT_II}) is not yet the complete expansion.
As mentioned above, we need to take into account that $x_1$ and $x_2$ are
defined at finite $Q_T$ (see Eq.~(\ref{x12})) and hence must also be
expanded about their respective values at $Q_T = 0$, $x_1^0$ and
$x_2^0$ in~(\ref{x120}).  
Therefore, we substitute $x_i = x_i^0 \sqrt{1+\rho^2}$ as arguments of
the structure functions $W_i$ and perform the $\rho^2$ expansions of
the latter. We now present our final result for the full small-$Q_T$
expansion of the hadronic structure functions, including the
leading-power (LP) term $W^{\rm LP}(x_1^0, x_2^0,L_\rho)$, the
next-to-leading-power (NLP) term $W^{\rm NLP}(x_1^0, x_2^0,L_\rho)$, and the
next-next-to-leading-power (NNLP)term $W^{\rm NNLP}(x_1^0, x_2^0,L_\rho)$:
\eq
W(x_1, x_2,\rho^2) = W^{\rm LP}(x_1^0, x_2^0,L_\rho)
+ \rho^2 \, W^{\rm NLP}(x_1^0, x_2^0,L_\rho)
+ \rho^4 \, W^{\rm NNLP}(x_1^0, x_2^0,L_\rho)             
+ {\cal O}(\rho^6) \,,
\en
where 
\eq
W^{\rm LP}(x_1^0, x_2^0,L_\rho) &=& W_0(x_1^0, x_2^0,L_\rho) \,,
\\
W^{\rm NLP}(x_1^0, x_2^0,L_\rho) &=& W_1(x_1^0, x_2^0,L_\rho)
+   \frac{1}{2} \biggl(
x_1^0 \ \partial_{x_1^0} W_0(x_1^0, x_2^0,L_\rho) 
    +   x_2^0 \ \partial_{x_2^0} W_{0}(x_1^0, x_2^0,L_\rho) 
    \biggr) \,, 
\\
W^{\rm NNLP}(x_1^0, x_2^0,L_\rho) &=& W_2(x_1^0, x_2^0,L_\rho)
+  \frac{1}{4} x_1^0 x_2^0
\ \partial_{x_1^0} \partial_{x_2^0} W_0(x_1^0, x_2^0,L_\rho) 
\nonumber\\
&-&   \frac{1}{8} \biggl(
    x_1^0    \ \partial_{x_1^0} W_0(x_1^0, x_2^0,L_\rho)
- 4 x_1^0    \ \partial_{x_1^0} W_1(x_1^0, x_2^0,L_\rho) 
-  (x_1^0)^2 \ \partial^2_{x_1^0} W_0(x_1^0, x_2^0,L_\rho) 
\biggr)
\nonumber\\  
&-&   \frac{1}{8} \biggl(
    x_2^0    \ \partial_{x_2^0}  W_0(x_1^0, x_2^0,L_\rho)
- 4 x_2^0    \ \partial_{x_2^0}  W_1(x_1^0, x_2^0,L_\rho) 
-  (x_2^0)^2 \ \partial^2_{x_2^0} W_0(x_1^0, x_2^0,L_\rho)
\biggr)  \,.
\en
Here $\partial^{m}_{x_1} \partial^{n}_{x_2} W_i(x_1,x_2,L_\rho)$ denotes
the $m$th partial derivative with respect to $x_1$ and the $n$th
partial derivative with respect to $x_2$. 
The calculational techniques for taking these derivatives are discussed
in Appendix~\ref{app:calc_tech}.

Explicitly we obtain the following analytical results for the LP contributions
$W_{J}^{{\rm LP}; a b}(x_1^0,x_2^0,L_\rho)$ to the $T$-odd hadronic structure functions
(here $a b = q \bar q, q g$ and $J = \Delta\Delta_P, \Delta_P, \nabla$): 
\eq
\label{qq_LP}
W^{{\rm LP}; q \bar q}_{\Delta\Delta_P}(x_1^0,x_2^0,L_\rho)
&=& \frac{g_{q\bar q; 1}}{4 x_1^0 x_2^0}  
\, C_A \, \left(L_\rho + \frac{3}{2}\right)
\, q_1(x_1^0)  \, \bar q_2(x_2^0)
\nonumber\\
&-& \frac{g_{q\bar q; 1}}{4 x_1^0 x_2^0} \, \frac{C_A}{2 C_F} \,  
\Big[
  q_1(x_1^0) \, \Big(P_{qq} \otimes \bar q_2\Big)(x_2^0) 
+ \Big(P_{qq} \otimes q_1\Big)(x_1^0) \, \bar q_2(x_2^0)  
\Big]  
\nonumber\\
&-& \frac{g_{q\bar q; 1}}{4 x_1^0 x_2^0} \, C_1 \,  
\Big[
  q_1(x_1^0) \, \Big(f_1 \otimes \bar q_2\Big)(x_2^0)
+ \Big(f_1 \otimes q_1\Big)(x_1^0) \, \bar q_2(x_2^0)
\Big]  \,, 
\\
W^{{\rm LP}; q \bar q}_{\nabla}(x_1^0,x_2^0,L_\rho)  &=&  
2 \, \beta \, 
W^{{\rm LP}; q \bar q}_{\Delta_P}(x_1^0,x_2^0) 
\nonumber\\
&=& - \frac{g_{q\bar q; 1}}{\rho \, x_1^0 x_2^0}  
\, \frac{C_A}{2 C_F} \,  
\Big[
  q_1(x_1^0) \, \Big(\tilde P_{qq} \otimes \bar q_2\Big)(x_2^0) 
- \Big(\tilde P_{qq} \otimes q_1\Big)(x_1^0) \, \bar q_2(x_2^0)  
\Big]  
\nonumber\\
&-& \frac{g_{q\bar q; 1}}{\rho \, x_1^0 x_2^0} \, C_1 \,  
\Big[
  q_1(x_1^0) \, \Big(f_2 \otimes \bar q_2\Big)(x_2^0)
- \Big(f_2 \otimes q_1\Big)(x_1^0) \, \bar q_2(x_2^0)
\Big]  
\,,
\en
\eq
W^{{\rm LP}; q g}_{\Delta\Delta_P}(x_1^0,x_2^0,L_\rho)
&=& - \frac{g_{q g; 1}}{4 x_1^0 x_2^0} \,   
  q(x_1^0) \, \Big(P'_{qg} \otimes g\Big)(x_2^0)  
\,,
\\ 
W^{{\rm LP}; q g}_{\nabla}(x_1^0,x_2^0,L_\rho) &=& 
2 \, \beta \, 
W^{{\rm LP}; q g}_{\Delta_P}(x_1^0,x_2^0,L_\rho)
\, = \, - \frac{g_{q g; 1}}{\rho \, x_1^0 x_2^0} \,
q(x_1^0) \, \Big(P''_{qg} \otimes g\Big)(x_2^0) 
\,,\label{discovery}
\en
where
\eq\label{Pqq}
\beta &=& \frac{g_{\rm EW; 1}}{g_{\rm EW; 2}} \,, \nonumber\\ 
P_{qq}(z) &=& C_F \biggl[ \frac{1+z^2}{(1-z)_+} 
+ \frac{3}{2} \delta(1-z)\biggr] \,, \nonumber\\
\tilde P_{qq}(z) &=& C_F (1+z) \,, \nonumber\\
P'_{qg}(z,L_\rho)
&=& C_F (1+2 z) + C_1 z \, (2 L_\rho - z) \,,
\nonumber\\
P''_{qg}(z,L_\rho)
&=& C_F (1-z) + C_1 z \, (L_\rho (1 - z) + 1 - 2 z)   
\,,
\en
and 
\eq
f_i(z) = \frac{F_i(z)}{1-z},
\en 
with $F_i$ as defined in Eq.~(\ref{Fdef}).
Note that the $f_i$ are regular functions
with $f_1(1) =1/3$, $f_2(1) =1/2$.

As shown in~(\ref{discovery}), there is an interesting relation between
the structure functions 
$W^{{\rm LP}; a b}_{\nabla}(x_1^0,x_2^0,L_\rho)$ and
$W^{{\rm LP}; a b}_{\Delta_P}(x_1^0,x_2^0,L_\rho)$:
\eq
W^{{\rm LP}; a b}_{\nabla}(x_1^0,x_2^0,L_\rho) =  
2 \beta \, W^{{\rm LP}; a b}_{\Delta_P}(x_1^0,x_2^0,L_\rho)\,,
\label{ALA_LamTung_T_odd}
\en 
valid both for the $q \bar q$ and the $q g$ subprocess at leading power.
The NLP and NNLP contributions to the $T$-odd hadronic structure function
are listed in the Supplemental Material
to this paper~\cite{SuppMaterial:2024sp}. 

\begin{figure}[t]
	\includegraphics[height=5.3cm]{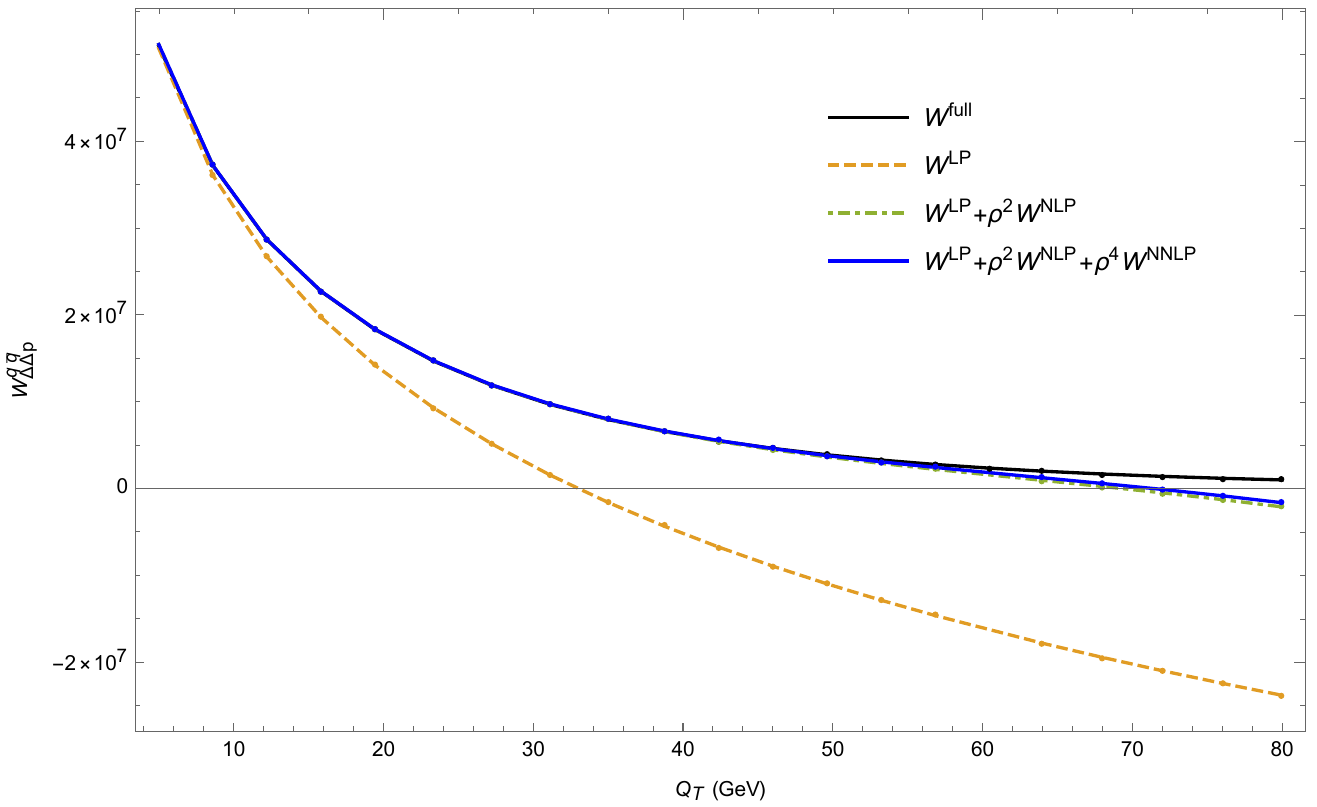} 
        \caption{Comparison of the full analytical result for 
        the quark-channel contribution to $W_{\Delta\Delta_P}$  
        (black solid line, taken from Eq.~(\ref{qq_full}) 
        with expansions to LP (dashed), NLP (dot-dashed),
        NNLP (blue solid) as given in the
        Supplemental Material~\cite{SuppMaterial:2024sp}. 
\label{NLP_NNLP}}
\end{figure}

To illustrate the numerical behavior of these expansions, 
we consider the $q\bar{q}$ contribution to the hadronic double-flip
structure function, $W^{q \bar q}_{\Delta\Delta_P}(x_1,x_2)$, as an example.
In Fig.~\ref{NLP_NNLP} we compare the full expression without $Q_T$ expansion 
with the LP, NLP, and NNLP results. Here we use the CTEQ 6.1M PDFs of
Ref.~\cite{Stump:2003yu}, taken from LHAPDF~\cite{Buckley:2014ana},
along with their ManeParse~\cite{Clark:2016jgm}
Mathematica implementation. We choose $\sqrt{s}=8$~TeV, $Q=100$~GeV, 
as representative of the kinematics in the ATLAS measurements \cite{ATLAS:2016rnf},
and the renormalization and factorization scales in the calculations are set to $\mu=\sqrt{Q^2+Q_T^2}$.
As one can see, the LP piece describes the full result only at low $Q_T$ and rapidly 
departs from it for $Q_T>10$~GeV or $\rho^2 > 0.01$.
By contrast, already inclusion of the NLP term leads to excellent
agreement with the full result out to $Q_T=40$~GeV ($\rho^2 = 0.16$),
only marginally further improved by the NNLP contribution. 
In particular, for $Q_T=20$~GeV, the LP result deviates from the full one by 
about 20\%, whereas at NNLP the relative deviation is only $\sim 0.4\%$.

\section{Comparison to ATLAS data}
\label{sec: Phenomenology}

As shown in Eq.~(\ref{eq4-sf}), the Drell-Yan cross section can be 
expressed in terms of eight angular coefficients $A_{i=0,\ldots,7}$.  
The relations of these coefficients to the hadronic helicity structure
functions are recalled in Appendix~\ref{app:Str_Func}.
Previous experimental and phenomenological studies mostly focused on
the first five $A_i$
coefficients~\cite{CMS:2015cyj,ATLAS:2016rnf,NA10:1986fgk,LHCb:2022tbc,CDF:2011ksg,Lambertsen:2016wgj,Gehrmann-DeRidder:2016cdi,Richter-Was:2016avq,Mirkes:1990vn} 
which are related to the $T$-even structure functions.
The $T$-odd structure functions have received less attention.
Experimentally, it has not yet been possible to measure the $T$-odd
angular coefficients in $W$ boson production, but results in
neutral-current scattering in the vicinity of the $Z$-boson mass 
peak are available from the ATLAS Collaboration~\cite{ATLAS:2016rnf}.
Specifically, the measurement was performed in the $Z$-boson invariant
mass window of 80--100 GeV, as a function of $Q_T$, and also in
three bins of rapidity $y$:
(a) $|y|<1, (b) 1<|y|<2 $, and (c) $2<|y|<3.5$. We note that near $Q=m_Z$
the contribution by $\gamma$-$Z^0$ interference is suppressed relative
to that for pure $Z^0$ exchange.
In the following, we compare our results for the angular coefficients
$A_5, A_6$, and $A_7$ to the ATLAS data. Here we use the full expressions
at ${\cal O}(\alpha_s^2)$ for the helicity structure functions
$W_{\Delta\Delta_P}, W_{\Delta_P}$, and $W_{\nabla}$.
In the denominator of the coefficients, we use the ${\cal O}(\alpha_s)$
expressions for the transverse and longitudinal structure functions $W_T$
and $W_L$
(see details in Refs.~\cite{Altarelli:1984pt,Boer:2006eq,Berger:2007jw}).
This approach thus consistently gives the leading contribution to
$A_5, A_6$, and $A_7$, which is of order $\alpha_s$. 

\begin{figure}[t]
	\includegraphics[height=5cm]{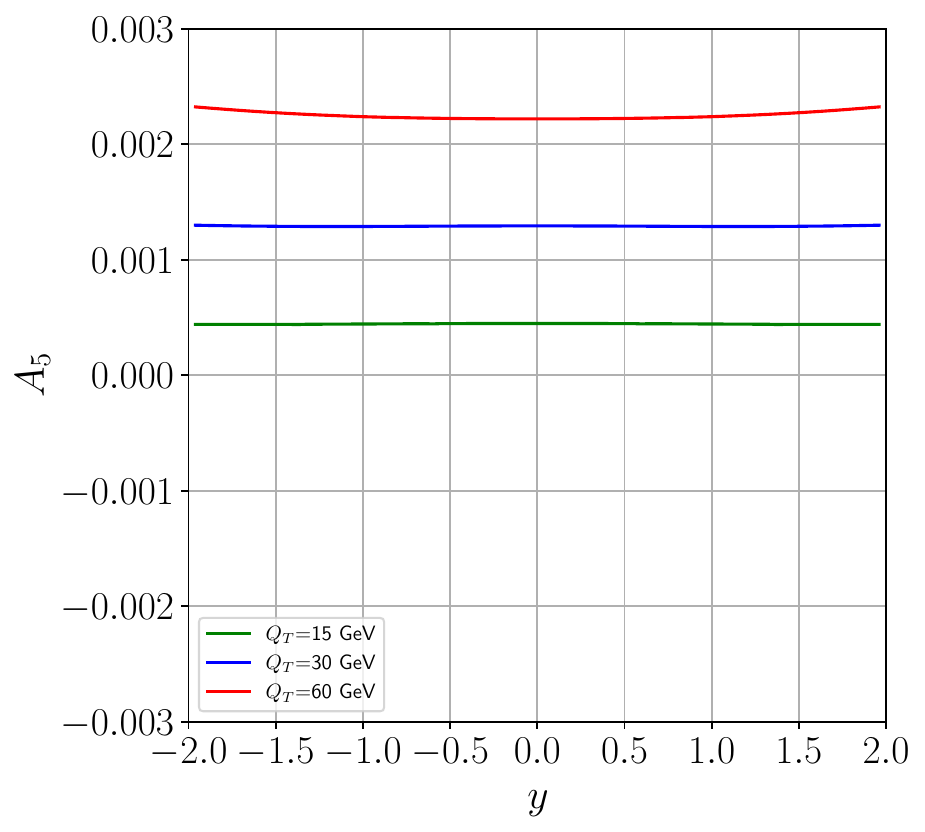}
	\includegraphics[height=5cm]{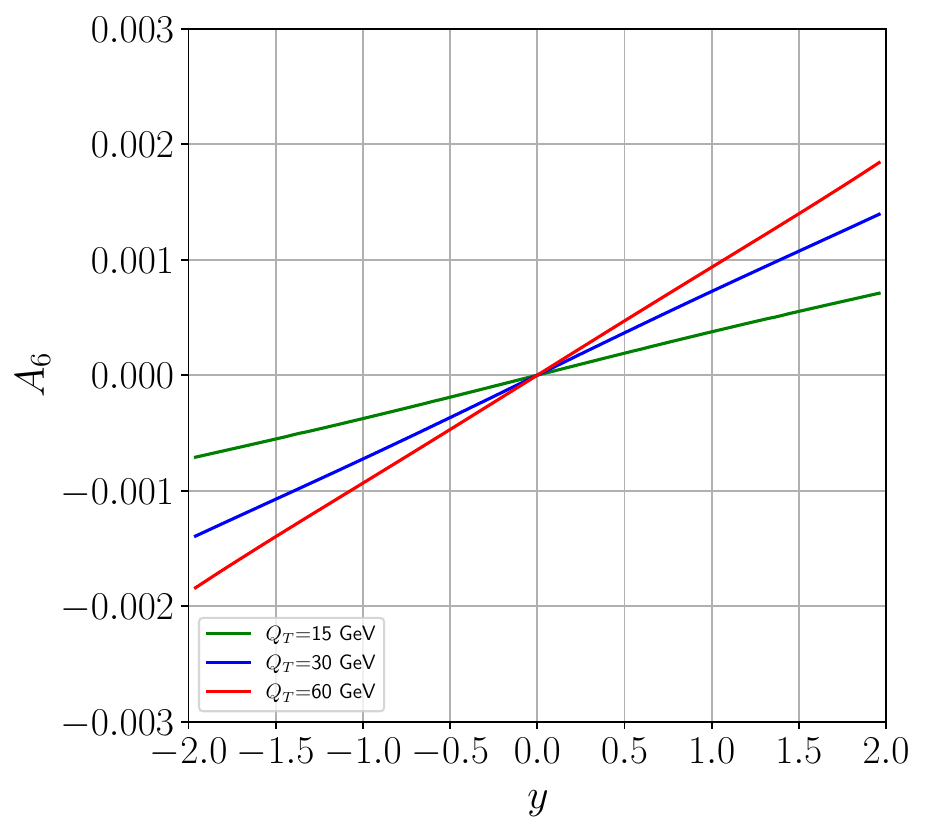}
	\includegraphics[height=5cm]{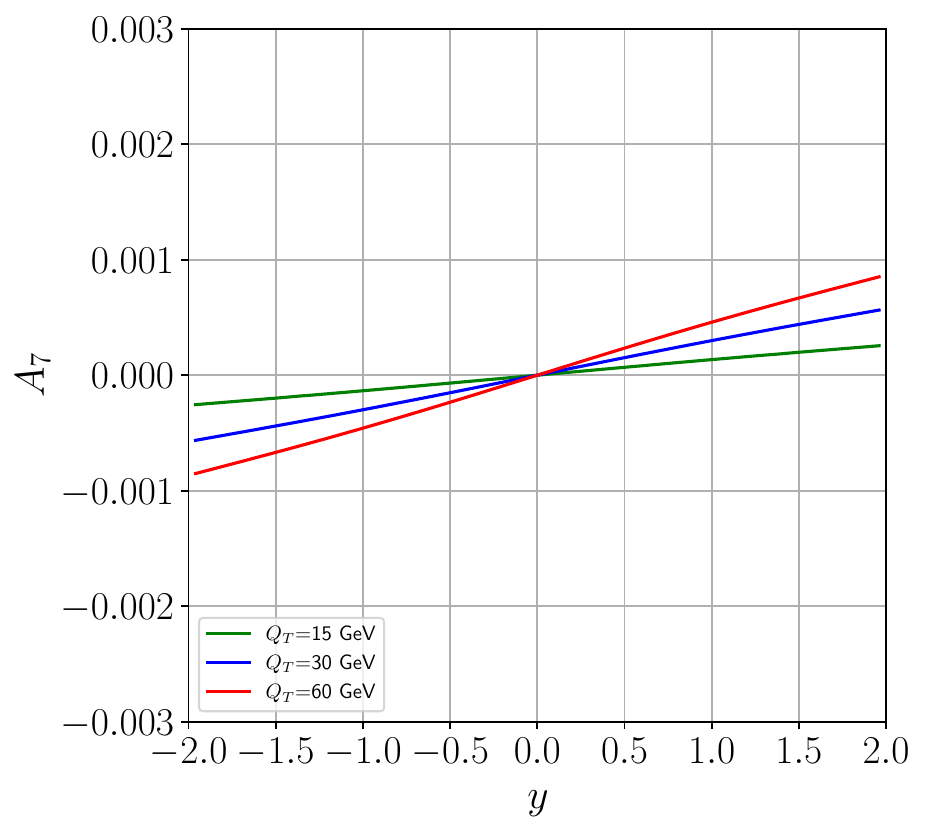}        
	\caption{$T$-odd angular coefficients $A_5$, $A_6$ and $A_7$
          for various transverse momenta $Q_T$ of the lepton pair,
          as functions of pair
	rapidity at $Q\sim m_Z$ and $\sqrt{s}=8$ TeV.}
	\label{YrapidityDependence}	
\end{figure}

\begin{figure}[t]
\includegraphics[height=4cm]{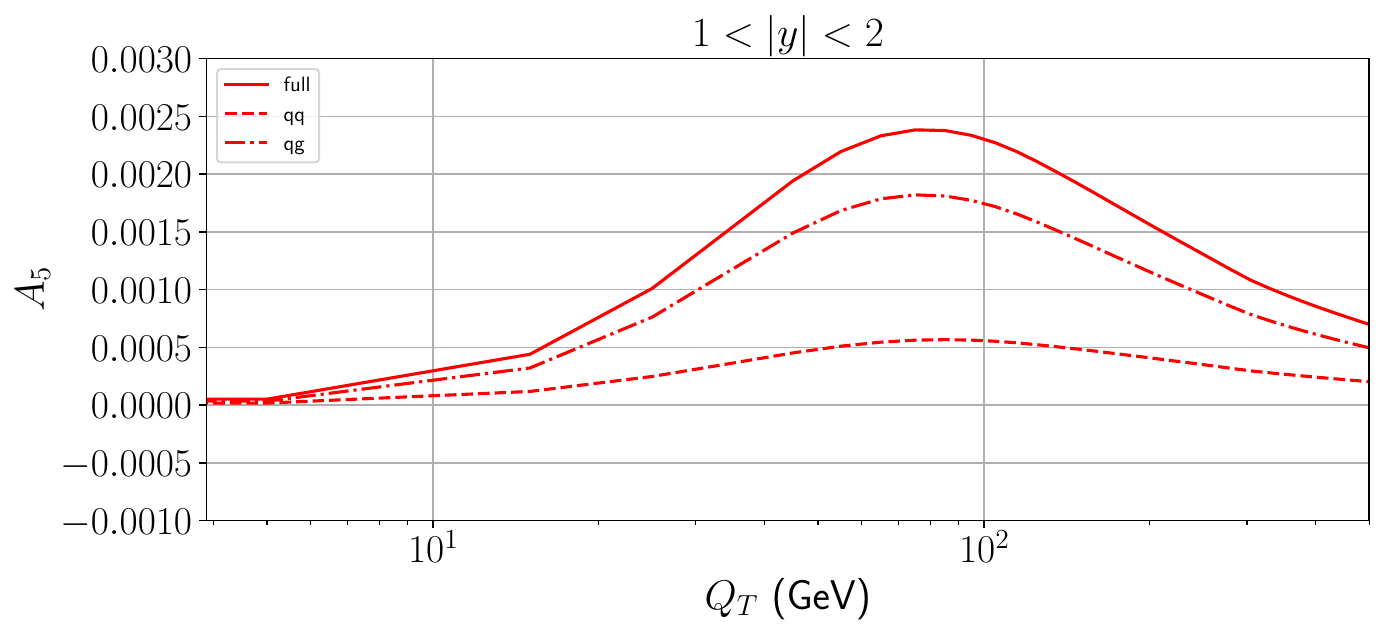}
\includegraphics[height=4cm]{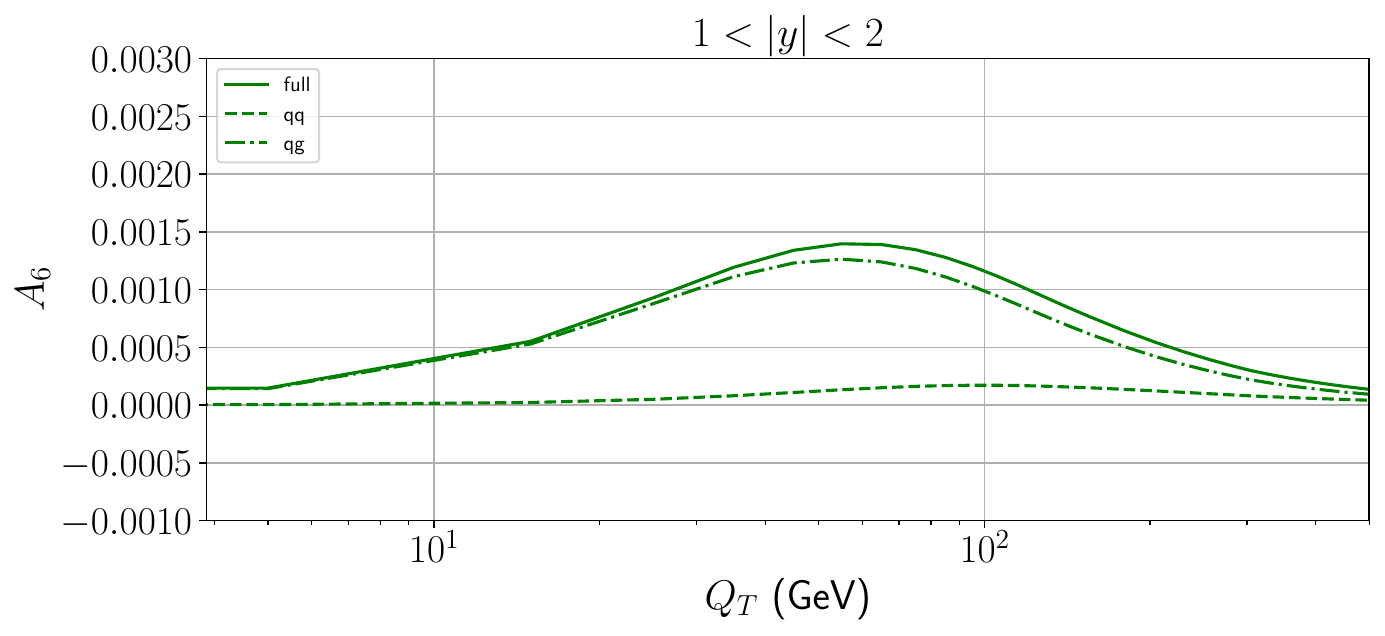}
\includegraphics[height=4cm]{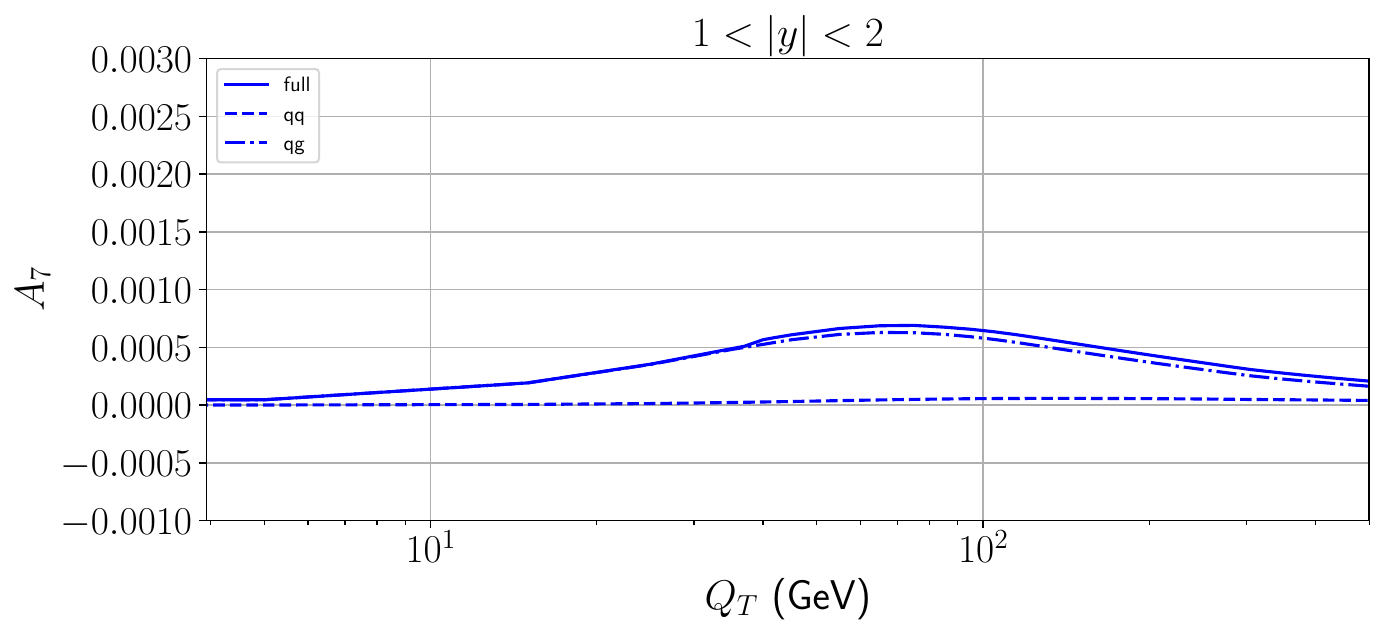}
\caption{$Q_T$ dependence of the angular coefficients $A_5$, $A_6$, and $A_7$
for the rapidity interval $1<|y|<2$ at $Q\sim m_Z$
and $\sqrt{s}=8$ TeV. We also show the individual $q\bar{q}$
and $qg$ contributions.}
\label{qq_qg_A567}

\vspace*{.5cm}

\includegraphics[height=4cm]{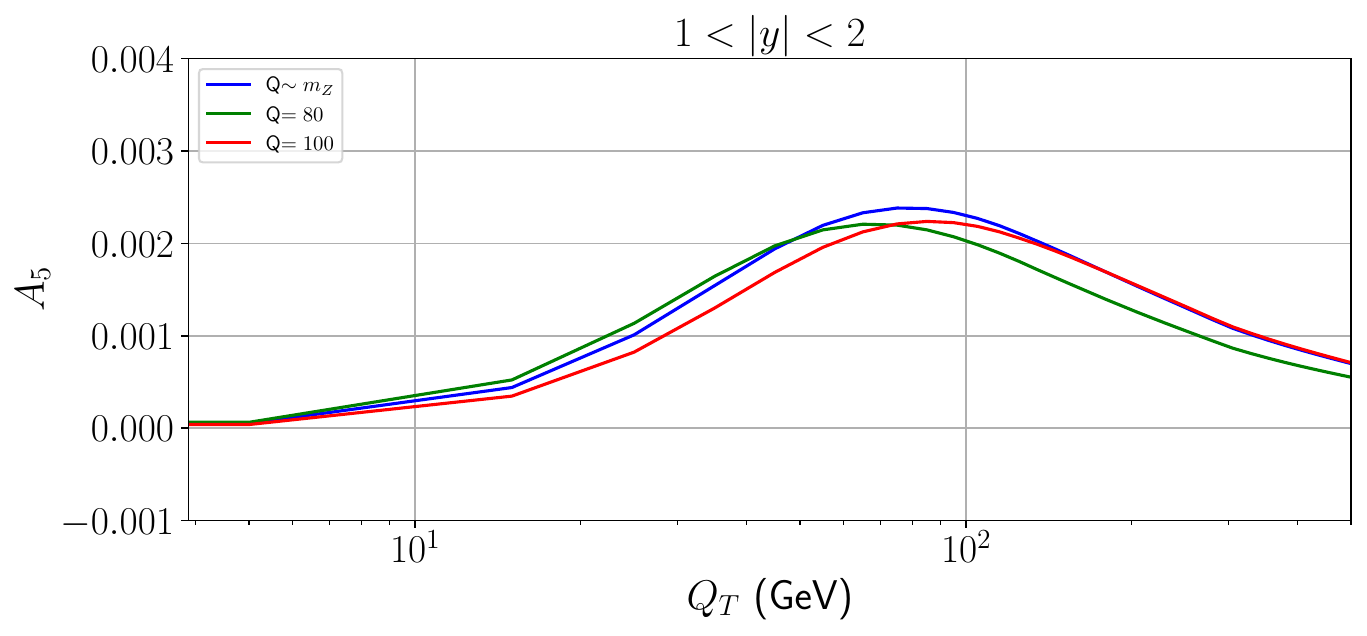}
\includegraphics[height=4cm]{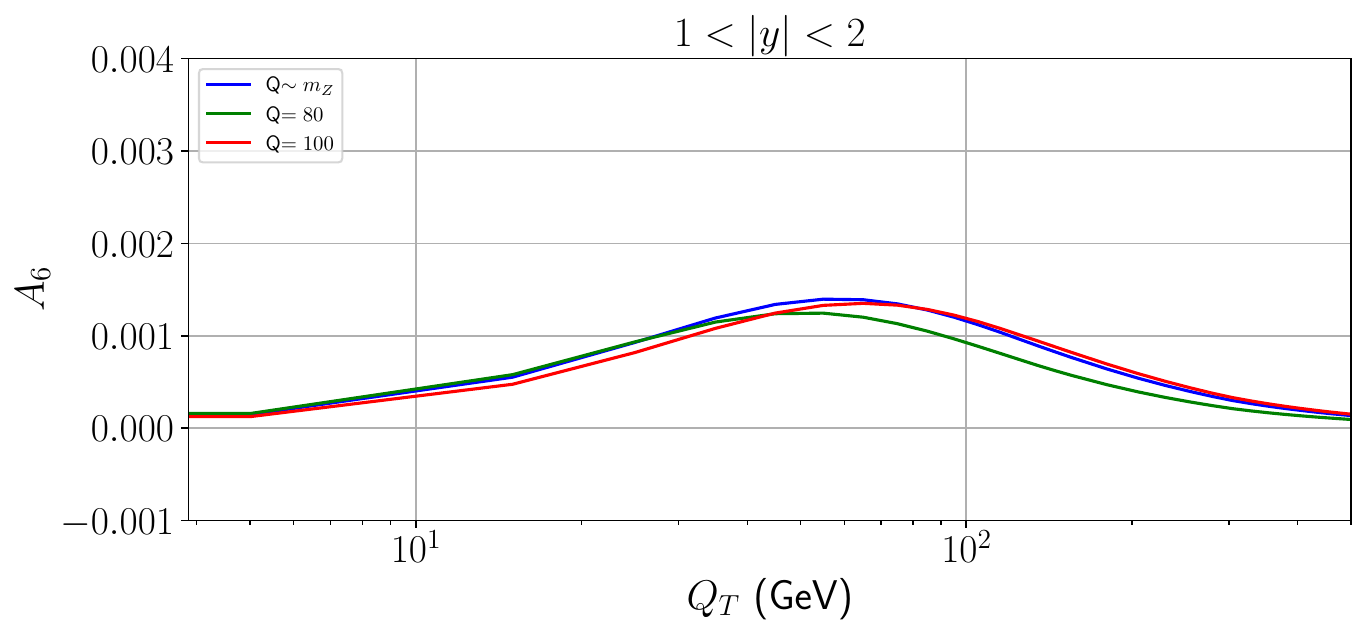}
\includegraphics[height=4cm]{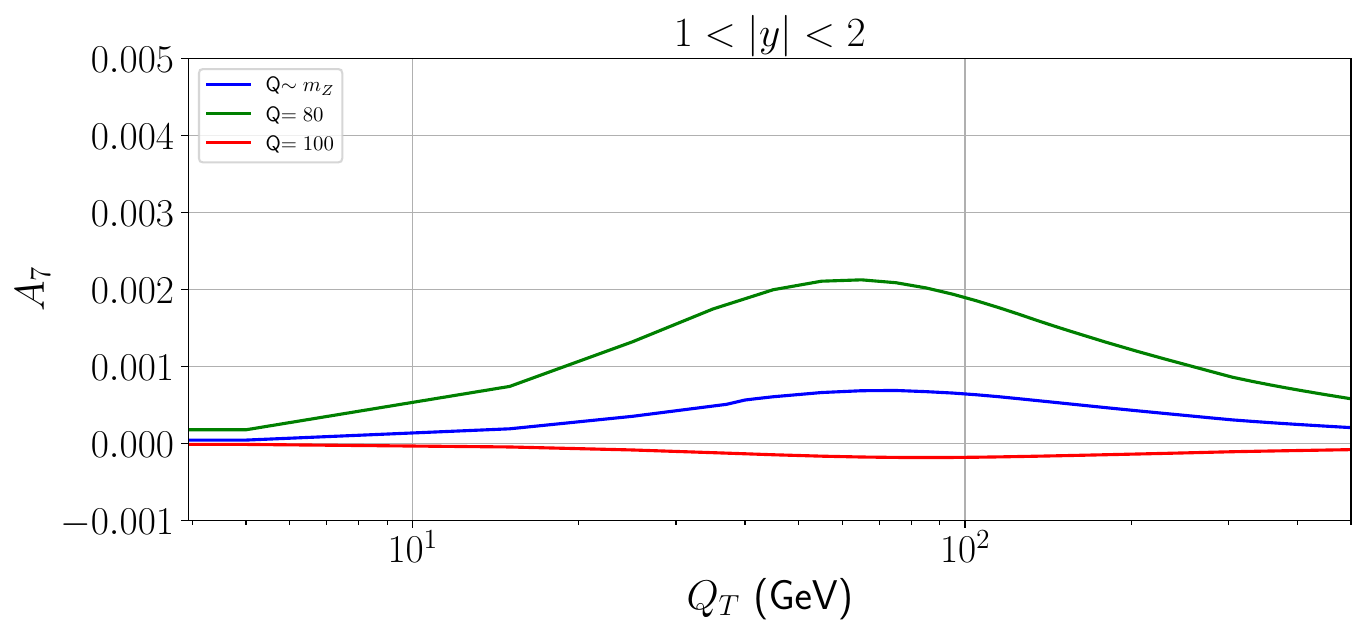}
\caption{$Q_T$ dependence of the angular coefficients $A_5$, $A_6$, and $A_7$
  for rapidity $1<|y|<2$ at $\sqrt{s}=8$ TeV and for different $Q$.}
\label{A567_diffQ}
\end{figure}

To begin with, we investigate the rapidity and $Q_T$ dependences of
the angular coefficients $A_5$, $A_6$, $A_7$ near the $Z^0$ pole. As in the previous section 
the calculation is done using $\sqrt{s}=8 $ TeV and $\mu=\sqrt{Q^2+Q_T^2}$.
Figure~\ref{YrapidityDependence} shows
the rapidity distribution for fixed $Q_T$,
while Fig.~\ref{qq_qg_A567} presents the results as functions of $Q_T$
in one of the three rapidity bins accessed by ATLAS. As the plots show,
the coefficients are overall small, reaching at most 0.1--0.2\% near
$Q_T\sim m_Z$ or toward larger $|y|$. This finding does not really come as a surprise:
Small values of the $T$-odd $A_{5,6,7}$ coefficients have been predicted
in Refs.~\cite{Mirkes:1992hu,Hagiwara:1984hi} also for $W$ boson production.
In the case of $Z$ boson production $A_{5,6,7}$ are further suppressed because of
the smallness of the corresponding weak couplings, 
relative to the couplings appearing in $W_T$ and $W_L$.  

The increase of $A_{5,6,7}$ at larger values of rapidity
in Fig.~\ref{YrapidityDependence} 
-- which is consistent with the leading-power relation between
$W_{\nabla}$ and $W_{\Delta_P}$
we found in Eq.~(\ref{ALA_LamTung_T_odd}) -- 
follows the trend observed in Ref.~\cite{Hagiwara:1984hi} 
for $W$ boson production in $p\bar{p}$ collisions at $\sqrt{s}=540$ GeV.
Likewise, a similar dependence on rapidity was found for the angular
coefficients in Refs.~\cite{ATLAS:2016rnf,CMS:2015cyj}.
At fixed rapidity, the $T$-odd coefficients are small for small $Q_T$ and 
then increase, peaking when $Q_T$ is near the $Z$ mass
(see Fig.~\ref{qq_qg_A567}).
We also show in the figure the individual contributions
by $q\bar{q}$ and $qg$ scattering, the latter dominating for all kinematics.
Among the $T$-odd structure functions, $W_{\Delta\Delta_{P}}$, being 
symmetric under interchange $z_1 \leftrightarrow z_2$, has the largest
contribution from quark-antiquark annihilation.

\begin{figure}
  \includegraphics[width=0.49\textwidth,
    trim={5cm 12.5cm 2cm 4cm},clip]{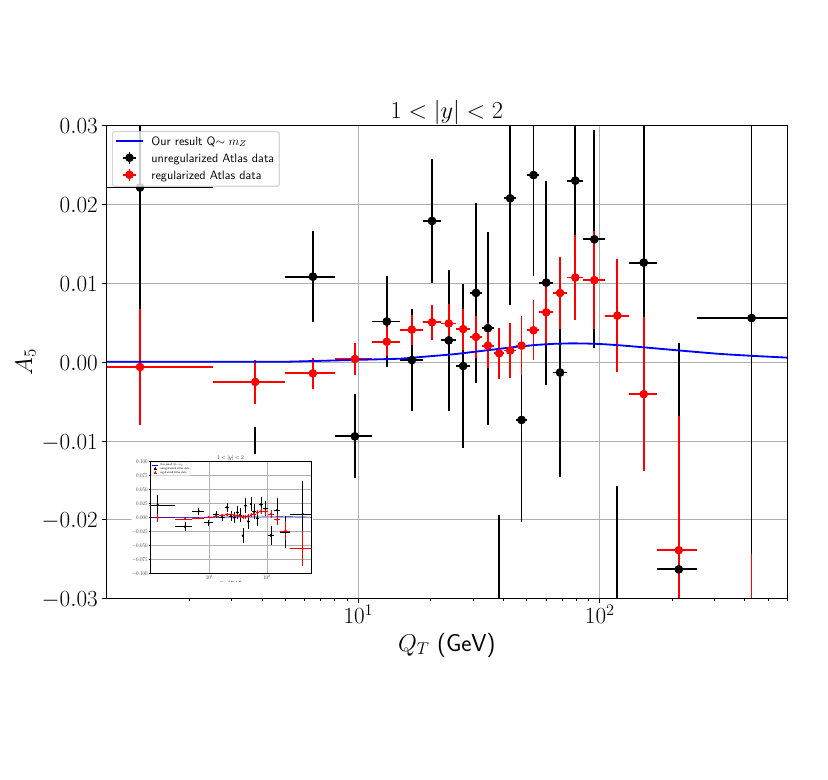}
  \includegraphics[width=0.49\textwidth,
    trim={5cm 12.5cm 2cm 4cm},clip]{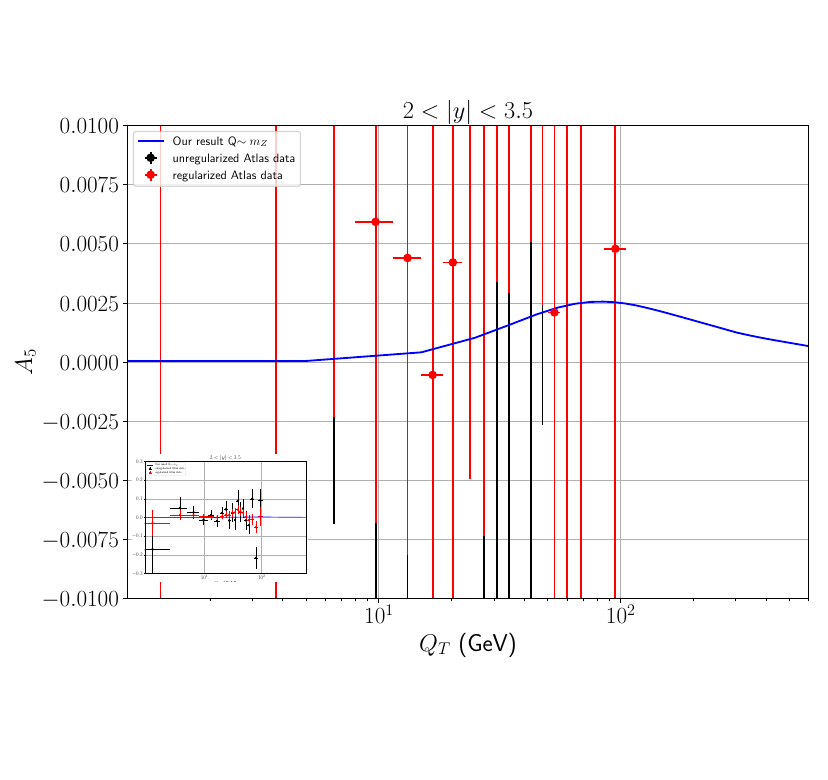}
  \vspace*{-.5cm}
  \caption{Comparison of our result for $A_5$
  to ATLAS data~\cite{ATLAS:2016rnf} at $Q\sim m_Z$.
  The black and red experimental points denote the unregularized
  and regularized data,
  respectively, and show their statistical error.
  The left panel shows the results
  for $1<|y|<2$, while the right is for $2<|y|<3.5$.
  (The scale on the $y$ axis 
  has been chosen for better visibility of the small values
  of the coefficient;
  as a result, some data points fall outside the plot range.
  The full dataset is shown
  in the left lower inset in each plot.)}
  \label{A5_withAtlas}

\includegraphics[width=0.49\textwidth,
  trim={5cm 12.5cm 2cm 4cm},clip]{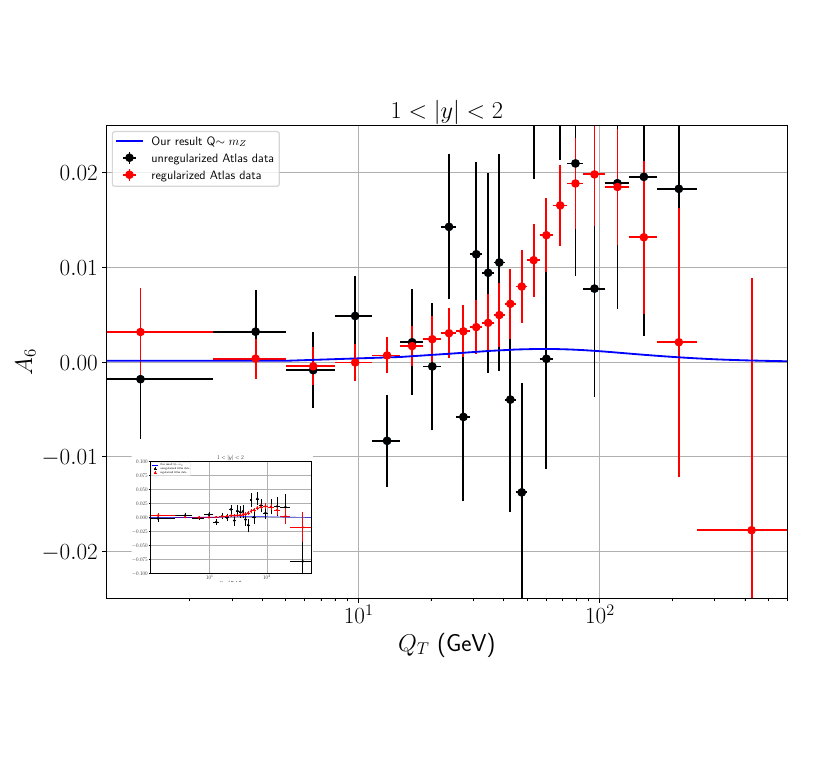}
\vspace*{-.5cm}
\caption{Same as Fig.~\ref{A5_withAtlas},
  but for the coefficient $A_6$ for rapidity $1<|y|<2$.}
\label{A6_withAtlas}

\includegraphics[width=0.49\textwidth,
  trim={5cm 12.5cm 2cm 4cm},clip]{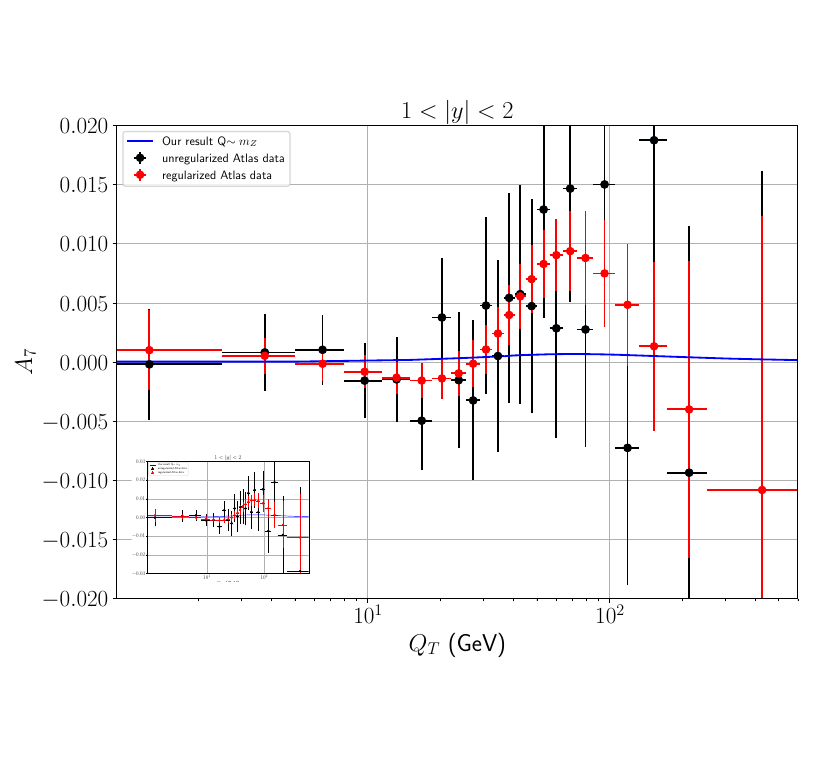}
\includegraphics[width=0.49\textwidth,
  trim={5cm 12.5cm 2cm 4cm},clip]{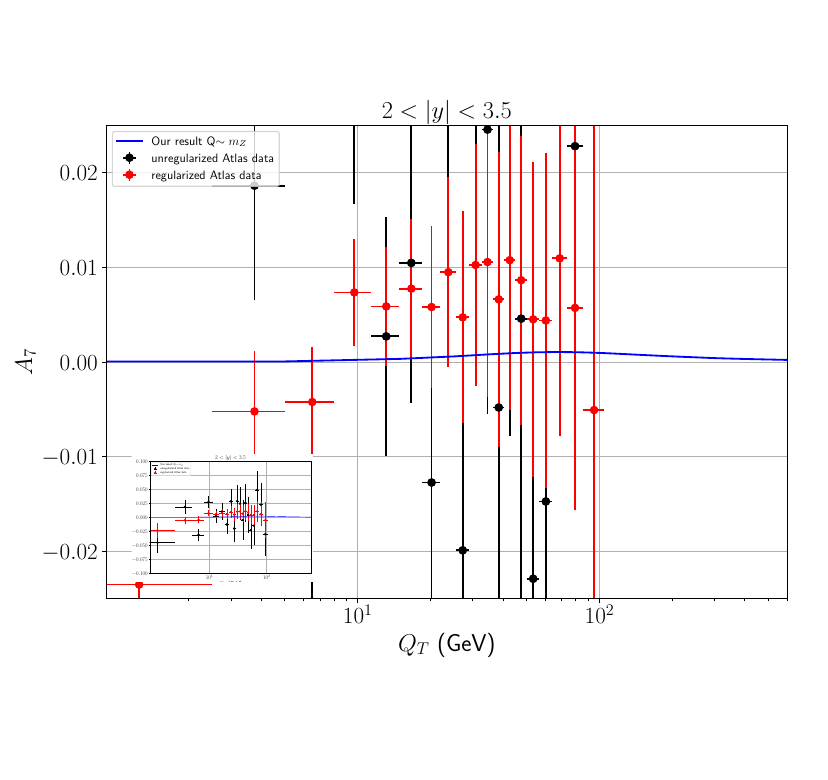}
\vspace*{-.35cm}
\caption{Same as Fig.~\ref{A5_withAtlas}, but for the coefficient $A_7$.}
\label{A7_withAtlas}
\end{figure}

Figure~\ref{A567_diffQ} explores the role played by the pair mass $Q$
for the $Q_T$ distribution. We show results for $Q=80$~GeV,
$Q=m_Z$, and $Q=100$~GeV,
which span the range of $Q$ used for the ATLAS measurements. As one can see, 
$A_5$ and $A_6$ are rather insensitive to $Q$, whereas $A_7$ exhibits
a strong dependence, even turning negative at high $Q$. This suggests
that $A_7$ will be quite sensitive to smearing effects if data are sampled
over a sizable range in $Q$. 

We now turn to the actual comparison with the ATLAS data~\cite{ATLAS:2016rnf}.
Of their three rapidity bins ($|y|<1, 1<|y|<2$, and $2<|y|<3.5$)
we only use the two with higher $|y|$ since, as we saw above, the angular 
coefficients are very small for $|y|<1$. We note that ATLAS presents the data
in two ways, as an ``unregularized'' and a ``regularized'' set. 
The regularization smoothes the data by correcting for bin migration.
This procedure involves the use of Monte-Carlo pseudo data which are
at lowest-order accuracy. Details about the data regularization method are
presented in Appendixes C and E of Ref.~\cite{ATLAS:2016rnf}.

Figures~\ref{A5_withAtlas}-\ref{A7_withAtlas} show the comparison.
Neither the regularized nor the unregularized data are in 
particularly good agreement with our theoretical predictions.
At best, there is qualitative agreement in that
the theoretical results show positive values for all three $T$-odd
angular coefficients, with a similar trend in the data. 
In particular, for the bin $1<|y|<2$ one observes a rise of $A_{5,6,7}$
with $Q_T$ up to about $Q_T\sim m_Z$, exactly as predicted theoretically.
Quantitatively, however, the regularized data -- which is the set primarily
to be used for comparisons -- shows overall much higher coefficients than
obtained in our calculation.
We note that ATLAS used the DYNNLO package~\cite{Catani:2009sm} to obtain
theoretical results for $A_{5,6,7}$.
DYNNLO predicts values of up to $0.005$ for the coefficients $A_{5,6,7}$.
However, as stated in \cite{ATLAS:2016rnf}, the prediction of nonzero 
values is at the limit of sensitivity
of both the theoretical calculation and the data. 
Hence it seems preliminary to delve into a detailed analysis of the visible discrepancies between our result and the existing data. 

More robust conclusions regarding the agreement of data and theory for the $T$-odd structure
functions will only become possible after significant improvements of both the experimental
results and the theoretical description. A natural question to ask is whether higher-order
(say, NLO) QCD corrections to the angular coefficients could lead to a better
agreement with the data. In this context it is important to keep 
in mind that the $T$-odd coefficients effectively carry an overall factor $\alpha_s$. 
This immediately means that they will be more susceptible to QCD corrections than 
ratios of cross sections would normally be. Just to give a simple estimate: 
For the values of $Q_T$ relevant here, varying the scale in $\alpha_s$ from
$Q_T/2$ to $2Q_T$ easily generates differences of $\pm 15\%$ or more 
in the calculated coefficients. On top of this, there will be smaller uncertainties 
associated with the scale dependence and uncertainties of the PDFs. 
One would thus expect the NLO corrections to $A_{5,6,7}$ to be overall
non-negligible, even though judging by the sizeable discrepancy in magnitude
and shape between the existing data and our result it would come as
a surprise if higher-order corrections were to account for the entire
difference. We note that NLO corrections could in fact be obtained
from Ref.~\cite{Gehrmann-DeRidder:2016cdi}. 
Clearly, a phenomenological study of $A_{5,6,7}$ at NLO will be an 
interesting project for the future.
Along with hopefully improved future data
it would open the door to careful assessments of the validity
of fixed-order perturbation theory for the $T$-odd angular coefficients.

\section{Conclusion}
\label{sec: Conclusion}

We have performed a detailed analysis of perturbative $T$-odd effects in charged-
and neutral-current Drell-Yan processes, taking into account $W^\pm$ and $Z^0$ exchange,
as well as $\gamma$-$Z^0$ interference. To this end, we have computed
the relevant $T$-odd structure functions for the $q\bar{q}$ annihilation and $qg$
Compton channels at order ${\cal O}(\alpha_s^2)$,
where they become nonvanishing thanks to absorptive contributions to loop amplitudes.
While the corresponding results are not new, we have used them in novel ways.
Foremost, we have presented a new formalism to expand the results for low transverse
momentum, or low $\rho=Q_T/Q$, with the goal of facilitating comparisons to frameworks
that analyze $T$-odd effects in terms of TMDs, especially at  nonleading power. 
Our new formalism is completely general and can in principle be used to obtain expansions
to arbitrary order in $Q_T/Q$. 
As a proof of concept, we have applied it to the $T$-odd structure functions and expanded
them to order ${\cal O}(\rho^4)$. In doing so, we uncovered a new relation between two of
the $T$-odd structure functions, $W^\text{LP}_{\nabla}(x_1,x_2)$ and
$W^\text{LP}_{\Delta_P}(x_1,x_2)$, valid at leading power in the small-$Q_T$ expansion.
Although in the present paper we have not attempted to connect our results to calculations
based on TMD factorization, we think that our paper has much to offer for such comparisons
in the future.

We have also presented numerical results for the validity of the expansion
in $\rho^2$, and we have compared our full results for the $T$-odd structure
functions to available data from the ATLAS experiment. We found that in the
present situation it is impossible to draw any quantitative conclusions from
this comparison.

In the present paper we have restricted our analysis to the case of
the $T$-odd effects in the Drell-Yan process with unpolarized beams.
Extensions to the $T$-even sector and to polarized scattering will be
natural extensions of our work. 

\begin{acknowledgments} 

We are grateful to Marc Schlegel for helpful discussions. 
A.S.Z. thanks T\"{u}bingen University for warm hospitality during
his visit during which the main part of this paper was completed. 
This work was funded
by BMBF (Germany) ``Verbundprojekt 05P2021 (ErUM-FSP T01) -- 
Run 3 von ALICE am LHC:
Perturbative Berechnungen von Wirkungsquerschnitten f\"ur ALICE''
(F\"orderkennzeichen: 05P21VTCAA),
by ANID PIA/APOYO AFB220003 (Chile),
by FONDECYT (Chile) under Grants No. 1230160 and No. 1240066, and
by ANID$-$Millen\-nium Program$-$ICN2019\_044 (Chile). 

\end{acknowledgments}

\appendix
\section{Relations among different sets of the structure functions}
\label{app:Str_Func}

The three sets of structure functions $\{A_i\}$,
$\{W_i\}$ and $\{\lambda,\mu,\nu,\ldots\}$ are related
as~\cite{Lam:1978pu,Collins:1977iv,Mirkes:1992hu,Boer:2006eq,Berger:2007jw}  
\eq 
\lambda &=& \frac{W_T-W_L}{W_T+W_L} = \frac{2-3A_0}{2+A_0}\,, 
\quad 
\mu \, = \, \frac{W_\Delta}{W_T+W_L} = \frac{2A_1}{2+A_0}\,, 
\quad 
\nu \, = \, \frac{2 W_{\Delta\Delta}}{W_T+W_L} = \frac{2A_2}{2+A_0}\,, 
\nonumber\\[2mm] 
\tau &=& \frac{W_{\nabla_P}}{W_T+W_L} = \frac{2A_3}{2+A_0}\,, 
\quad \hspace*{.2cm}
\eta \, = \, \frac{W_{T_P}}{W_T+W_L} = \frac{2A_4}{2+A_0}\,, 
\quad \hspace*{.1cm}
\xi \, = \, \frac{W_{\Delta\Delta_P}}{W_T+W_L} = \frac{2A_5}{2+A_0}\,, 
\nonumber\\[2mm] 
\zeta &=& \frac{W_{\Delta_P}}{W_T+W_L} = \frac{2A_6}{2+A_0}\,, 
\quad \hspace*{.15cm}
\chi \, = \, \frac{W_{\nabla}}{W_T+W_L} = \frac{2A_7}{2+A_0} \,,
\en 
and
\eq 
A_0 &=& \frac{2W_L}{2W_T+W_L} = \frac{2 (1-\lambda)}{3+\lambda} \,, 
\quad 
A_1 \, = \, \frac{2W_\Delta}{2W_T+W_L} = \frac{4 \mu}{3+\lambda} \,, 
\quad 
A_2 \, = \, \frac{4W_{\Delta\Delta}}{2W_T+W_L} = \frac{4 \nu}{3+\lambda} \,, 
\nonumber\\[2mm] 
A_3 &=& \frac{2W_{\nabla_P}}{2W_T+W_L} = \frac{4 \tau}{3+\lambda} \,, 
\quad \hspace*{.4cm}
A_4 \, = \, 
\frac{2W_{T_P}}{2W_T+W_L} = \frac{4 \eta}{3+\lambda} \,, 
\quad \hspace*{.05cm}
A_5 \, = \,  \frac{2W_{\Delta\Delta_P}}{2W_T+W_L} = \frac{4 \xi}{3+\lambda} \,, 
\nonumber\\[2mm] 
A_6 &=& \frac{2W_{\Delta_P}}{2W_T+W_L} = \frac{4 \zeta}{3+\lambda} \,, 
\quad \hspace*{.4cm}
A_7 \, = \, \frac{2W_{\nabla}}{2W_T+W_L} = \frac{4 \chi}{3+\lambda} \,. 
\en

\section{Treatment of $\gamma^5$ matrix in calculation
  of structure functions}
\label{app:gamma5}

In the calculation of the $T$-odd structure functions we have to deal with
the $\gamma^5$ matrix in dimensional regularization. We encounter two different cases: 
\begin{itemize}
\item[(a)] Dirac traces with an even number (in practice, two) of $\gamma^5$ matrices 
in case of $w_{\nabla}$;  
\item[(b)] Dirac traces with an odd  number (in practice, one) of $\gamma^5$ matrices 
in case of $w_{\Delta\Delta_P}$ and $w_{\Delta_P}$. 
\end{itemize}
For case (a), it is permissible even in dimensional regularization
to anticommute the two $\gamma^5$ matrices toward each other and 
to use $(\gamma^5)^2 = 1$. As a result, the contributions of [vector $\otimes$ vector]
and [axial-vector $\otimes$ axial-vector] couplings to 
$w_{\nabla}$ are identical.  

In case (b) we use techniques established in the literature for the treatment of
$\gamma^5$ in dimensional regularization.
In particular, for the axial-vector spin matrix we use
the Larin prescription~\cite{Akyeampong:1973xi,Larin:1991tj,Larin:1993tq,Zijlstra:1992kj},
expressing $\gamma^5$ as the product of the four-dimensional Levi-Civita tensor
$\epsilon^{\mu\nu\alpha\beta}$ and three gamma matrices:
\eq\label{Larin_sub}
\gamma^\mu \gamma^5 = \frac{i}{6} \,
\epsilon^{\mu\nu\alpha\beta} \, \gamma_\nu
\, \gamma_\alpha \, \gamma_\beta 
\,. 
\en
Next, in order to evaluate the structure functions $w_{\Delta\Delta_P}$ and $w_{\Delta_P}$,
we apply the method proposed in Ref.~\cite{Zijlstra:1992kj} for the 
contraction of two Levi-Civita tensors. 
One Levi-Civita tensor occurs in the definition of the structure functions
$w_{\Delta\Delta_P}$ and $w_{\Delta_P}$ [see Eqs.~(\ref{wdd}) and~(\ref{wd})] 
and the other appears because of the Larin substitution~(\ref{Larin_sub}).
In general, the product of two Levi-Civita tensors must be evaluated in terms of
the $D$-dimensional Kronecker tensor $\delta^\mu_\nu$ in order to preserve Lorentz
invariance~\cite{Zijlstra:1992kj}:
\eq\label{det}
\epsilon^{\mu_1\mu_2\mu_3\mu_4} \, \epsilon_{\nu_1\nu_2\nu_3\nu_4}
= - {\rm det}
\left|
\begin{array}{cccc}
  \delta^{\mu_1}_{\nu_1} & \delta^{\mu_1}_{\nu_2} &
  \delta^{\mu_1}_{\nu_3} & \delta^{\mu_1}_{\nu_4} \\[3mm]
  \delta^{\mu_2}_{\nu_1} & \delta^{\mu_2}_{\nu_2} &
  \delta^{\mu_2}_{\nu_3} & \delta^{\mu_2}_{\nu_4} \\[3mm]
  \delta^{\mu_3}_{\nu_1} & \delta^{\mu_3}_{\nu_2} &
  \delta^{\mu_3}_{\nu_3} & \delta^{\mu_3}_{\nu_4} \\[3mm]
  \delta^{\mu_4}_{\nu_1} & \delta^{\mu_4}_{\nu_2} &
  \delta^{\mu_4}_{\nu_3} & \delta^{\mu_4}_{\nu_4} \\    
\end{array}
\right|\,.
\en 
For our purposes we need to evaluate two types of contractions [see Eqs.~(\ref{wdd})
  and~(\ref{wd})],
\eq\label{IDT1}
i \epsilon^{\mu P R K} \, \gamma_\mu \gamma_5 
= - \frac{1}{6} \, \epsilon^{\mu P R K} \,
\epsilon_{\mu\rho\alpha\beta} \,
\gamma^\rho \, \gamma^\alpha \, \gamma^\beta \,,
\en  
and
\eq\label{IDT2}
i \epsilon^{\mu P R K} \, \gamma_\nu \gamma_5 
= - \frac{1}{6} \, \epsilon^{\mu P R K} \,
\epsilon_{\nu\rho\alpha\beta} \,
\gamma^\rho \, \gamma^\alpha \, \gamma^\beta \,. 
\en
Using Eq.~(\ref{det}) we get
\eq\label{ID1}
i \epsilon^{\mu P R K} \, \gamma_\mu \gamma_5 =
(D-3) \, \not\! P \not\! R \not\! K \,,
\en 
and
\eq\label{ID2}
i \epsilon^{\mu P R K} \, \gamma_\nu \gamma_5 =
\delta^\mu_{\nu; \perp} \, \not\! P \not\! R \not\! K
+ \gamma^\mu_{\perp} \,
\Big(P_\nu \not\! K \not\! R
   + R_\nu \not\! P \not\! K
   + K_\nu \not\! R \not\! P\Big)
\,,
\en 
where $D = 4 - 2 \epsilon$ and 
$\delta^\mu_{\nu; \perp}$ and 
$\gamma^\mu_{\perp}$ are 
the perpendicular $D$-dimensional Kronecker tensor and gamma matrices, respectively,
defined as
\eq\label{kinematics1}  
\delta^\mu_{\nu; \perp} 
&\equiv& \delta^\mu_{\nu}
- \frac{P^\mu P_\nu}{P^2}
- \frac{R^\mu R_\nu}{R^2}
- \frac{K^\mu K_\nu}{K^2}
\,, \nonumber\\[1mm]
\gamma^\mu_{\perp} &\equiv&
\delta^\mu_{\nu; \perp} \, \gamma^\nu 
= \gamma^\mu
- \frac{P^\mu \!\! \not\! P}{P^2}
- \frac{R^\mu \!\! \not\! R}{R^2}
- \frac{K^\mu \!\! \not\! K}{K^2}  \,,
\en
which are manifestly orthogonal to all momenta of the basis $(P,R,K)$.
The obey the following conditions:
\eq
& &  \delta^\mu_{\nu; \perp} \, \delta^\nu_{\mu; \perp}
    = \delta^\mu_{\nu; \perp} \, \delta^\nu_{\mu}
    = \delta^\mu_{\mu; \perp} 
    = D - 3 
\,, \nonumber\\[1mm]
& &\delta^\mu_{\nu; \perp} \, P_\mu = 
   \delta^\mu_{\nu; \perp} \, R_\mu = 
   \delta^\mu_{\nu; \perp} \, K_\mu = 0 
\,. 
\en
Similarly one can define the perpendicular $D$-dimensional
metric tensor $g^{\mu\nu}_\perp$ introduced in Ref.~\cite{Lyubovitskij:2021ges},
\eq\label{kinematics2}  
g^{\mu\nu}_\perp
= g^{\mu\nu} 
- \frac{P^\mu P^\nu}{P^2}
- \frac{R^\mu R^\nu}{R^2}
- \frac{K^\mu K^\nu}{K^2}  \,. 
\en
One should mention that the identities~(\ref{IDT2}) and~(\ref{ID2})
are generalizations of~(\ref{IDT1}) and~(\ref{ID1}), respectively. 
In particular, Eq.~(\ref{ID1}) follows from~(\ref{ID2}) for $\nu \to \mu$.
In this limit, $\delta^\mu_{\mu; \perp} = D-3$, the second term
on the rhs of Eq.~(\ref{ID2}) vanishes, and we arrive at the identity~(\ref{ID1}).  

We also stress that using the orthogonal basis $(P,R,K)$ along with the orthogonal
metric tensors and gamma matrices turns out to be very useful and economical in our 
analytical calculations. In addition, we note that the use of identity~(\ref{ID2})
is further simplified in the case of the evaluation of $w_{\Delta\Delta_P}$ and $w_{\Delta_P}$. 
The reason is that the Levi-Civita tensor  $\epsilon^{\mu P R K}$ on the lhs 
of Eq.~(\ref{ID2}) is accompanied by the basis vector $P^\nu$ or $R^\nu$
[see Eqs.~(\ref{wdd}) and~(\ref{wd})].
Therefore, the first Lorentz structure
$\delta^\mu_{\nu; \perp} \not\!\! P \not\!\! R \not\!\! K$
on the rhs of Eq.~(\ref{ID2}) vanishes thanks to
$P^\nu \, \delta^\mu_{\nu; \perp} = R^\nu \, \delta^\mu_{\nu; \perp} = 0$.
The second Lorentz structure 
$\gamma^\mu_{\perp} \,
\Big(P_\nu \not\!\! K \not\!\! R + R_\nu \not\!\! P \not\!\! K  
+ K_\nu \not\!\! R \not\!\! P\Big)$
on the rhs of the identity~(\ref{ID2}) is also simplified
because of the orthogonality of the $(P,R,K)$ basis.
In particular, depending on the accompanying momentum $P^\nu$ or $R^\nu$, we deduce
from the identity~(\ref{ID2}) two simplified identities useful 
for the calculation of $w_{\Delta\Delta_P}$ and $w_{\Delta_P}$: 
\eq\label{ID2_simple}
i \epsilon^{\mu P R K} \, \not\! P \gamma_5 &=&
\gamma^\mu_{\perp} \not\! K \not\! R \ P^2
\,, \nonumber\\[1mm]
i \epsilon^{\mu P R K} \, \not\! R \gamma_5 &=&
\gamma^\mu_{\perp} \not\! P \not\! K \ R^2
\,.
\en

\section{Calculational technique for the partial derivatives of
  the structure functions $W_i(x_1,x_2)$}
\label{app:calc_tech}

In this appendix we discuss the calculational technique for the partial
derivatives of the structure functions $W_i(x_1,x_2)$ defined
in Eq.~(\ref{Wi_general}), which have the following form 
\eq\label{Wi_general_Appendix}
W_i(x_1, x_2,L_\rho) &=& \frac{1}{x_1x_2}\sum_{a,b}
\, \biggl[
   R_{ab,i}(x_1,x_2,L_\rho)
\, f_{a/H_1}(x_1)
\, f_{b/H_2}(x_2)
\nonumber\\
&+& \Big(P_{ba,i} \otimes f_{b/H_2}\Big)(x_2,x_1,L_\rho)  \ f_{a/H_1}(x_1)  
 +  \Big(P_{ab,i} \otimes f_{a/H_1}\Big)(x_1,x_2,L_\rho)  \ f_{b/H_2}(x_2) 
\biggr] 
\,,
\en
where $R_{ab,i}(x_1,x_2,L_\rho)$, $P_{ab,i}(z_1,x_2,L_\rho)$,
and $P_{ba,i}(z_2,x_1,L_\rho)$ are
perturbative coefficient functions containing differential operators
acting on the PDFs $f_{a/H_1}(x_1)$ and $f_{b/H_2}(x_2)$.
In particular, the functions $R_{ab,i}(x_1,x_2,L_\rho)$,
$P_{ab,i}(z_1,x_2,L_\rho)$,
and $P_{ba,i}(z_2,x_1,L_\rho)$ can be expanded using the set of the differential
operators to factorize the dependence on the variables $x_1$ and $x_2$,
$z_1$ and $x_2$, $z_2$ and $x_1$, respectively (see
detailed discussion in the Supplemental Material
to this paper~\cite{SuppMaterial:2024sp}). 
One can see that $W_i(x_1,x_2,L_\rho)$ is composed of three main terms,
one term proportional to the perturbative function $R_i(x_1,x_2,L_\rho)$
times the PDFs, and two terms each containing a single integral representation
of the convolution of the perturbative functions $P_{a,i}(z_1,L_\rho)$,
and $P_{b,i}(z_2,L_\rho)$ with the PDFs.
Also the latter terms depend on the second variable $x_i = x_1$ or $x_2$
via the simple form of a product of $f(x_i)$ with some polynomial in $x_i$.  

The first term can be trivially differentiated with respect to $x_1$ and $x_2$
to the desired order.
In the case of the other two terms one can straightforwardly
differentiate the nonintegral terms. 
It remains to specify how to take the partial derivative of the terms with
integrals. Let us discuss the treatment of such terms by considering
the following integral:
\eq
J(x) &=& \frac{I(x)}{x}\,, \qquad 
I(x) \, = \, (P \otimes f)(x) = 
\int\limits_{x}^1 \frac{dz}{z} \, P(z)
\, f\Big(\frac{x}{z}\Big)
\,. 
\en
The $n$th derivative of the integral $J(x)$ can
be taken using the binomial formula 
\eq
\frac{\partial^n J(x)}{\partial x^n} &=& \sum\limits_{m=0}^{n}
\, \frac{n!}{m!}
\, \frac{(-1)^{n-m}}{x^{n-m+1}} \,
\frac{\partial^m I(x)}{\partial x^m} \,. 
\en
Therefore, we only need to derive an analytical formula
for the $n$th derivative
of the integral $I(x)$, where $n$ is an arbitrary natural number.
In our derivation we will consider two possible choices for
the function $P(z)$:
\begin{itemize}
\item[(a)] $P(z) = R(z)$ is a regular function of  the variable $z$,
\item[(b)] $P(z) = [1/(1-z)^m]_{+,m-1}$ is the generalized plus distribution 
of power $m$, defined in Eq.~(\ref{genplus}) in the main text.
\end{itemize}

We first consider the simpler case (a). Here, the first-order derivative
of $I(x)$ reads as 
\eq
\frac{\partial I(x)}{\partial x} &=&
\frac{\partial}{\partial x} \,
\int\limits_{x}^1 \frac{dz}{z} \, R(z)
\, f\Big(\frac{x}{z}\Big)
= 
\frac{\partial}{\partial x} \,
\int\limits_{x}^1 \frac{d\xi}{\xi}
\, R(x/\xi) \, f(\xi)
\nonumber\\
&=&
- \int\limits_{x}^1 \frac{d\xi}{\xi}
\, \delta(\xi-x) \, R(x/\xi) \, f(\xi)
+ \int\limits_{x}^1 \frac{d\xi}{\xi}
\, \frac{\partial R(x/\xi)}{\partial x}\, f(\xi)
\nonumber\\
&=&
- \int\limits_{x}^1 \frac{d\xi}{\xi^2}
\, \delta(1-x/\xi) \, R(x/\xi) \, f(\xi)
+ \int\limits_{x}^1 \frac{d\xi}{\xi^2}
\, \frac{\partial R(x/\xi)}{\partial (x/\xi)}\, f(\xi)
\nonumber\\
&=&
- \frac{1}{x} \int\limits_{x}^1 dz 
\, \delta(1-z) \, R(z) \, f\Big(\frac{x}{z}\Big)
+ \frac{1}{x}
\int\limits_{x}^1 dz  
\, \frac{\partial R(z)}{\partial z}\, f\Big(\frac{x}{z}\Big)
\nonumber\\
&=&
\frac{1}{x} \biggl[- R(1) \, f(x) 
+ \int\limits_{x}^1 dz \, R'(z) \, f\Big(\frac{x}{z}\Big)\biggr]
\,,
\en
where $R'(z) = \partial R(z)/\partial z$. 

One can prove by induction that the $n$th derivative of $I(x)$ is given by
\eq
\frac{\partial^n I(x)}{\partial x^n} =
- \sum\limits_{k=1}^{n} 
\, \biggl(\frac{f(x)}{x^{k}}\biggr)^{(n-k)}_x \, R^{(k-1)}(1) 
+ \frac{1}{x^n} \,
\int\limits_{x}^1 dz  
\, z^{n-1} \, R^{(n)}(z) \, f\Big(\frac{x}{z}\Big)
\,,
\en
where
\eq
R^{(k)}(z) =  \partial^k R(z)/\partial z^k\,,  \qquad 
(\cdots)^{(k)}_x =  \partial^{k} (\cdots)/\partial x^{k} \,. 
\en

For case (b) we recall the definition for
the generalized plus distribution when applied
to a PDF $f(x/z)$:
\eq\label{PDF_dist}
\int\limits_x^1 dz \, 
\frac{f\Big(\frac{x}{z}\Big)}{(1-z)^m_{+,m-1}}
&=& 
\int\limits_x^1 dz \, \biggl[\frac{1}{(1-z)^m_{+x,m-1}}
+ \delta(1-z) \, \log(1-x) \, \frac{(-1)^{m-1}}{(m-1)!} 
\, \partial_z^{m-1}
\nonumber\\
&-& \delta(1-z) \, 
\sum\limits_{j=2}^m
\, \frac{(-1)^{m-j}}{(j-1) \, (m-j)!} \, 
\biggl( \frac{1}{(1-x)^{j-1}} - 1 \biggr) \, \partial_z^{m-j} 
\biggr] 
f\Big(\frac{x}{z}\Big) \,.
\en
The $x$ dependence of $f(x/z)/(1-z)^m_{+x,m-1}$ induces a subtraction of
the $(m-1)$-th order Taylor polynomial $T[f(x/z)]^{m-1}_{z=1}$ evaluated
at $z=1$,
\eq
\frac{f\Big(\frac{x}{z}\Big)}{(1-z)^m_{+x,m-1}} &=& 
\frac{f\Big(\frac{x}{z}\Big)
  -T\Big[f\Big(\frac{x}{z}\Big)\Big]^{m-1}_{z=1}}{(1-z)^m}
\,, \nonumber\\
T\Big[f\Big(\frac{x}{z}\Big)\Big]^{m-1}_{z=1}
&=& \sum\limits_{k=0}^{m-1} \frac{(z-1)^k}{k!} \, 
\partial_{z}^{k} f\Big(\frac{x}{z}\Big)\Big|_{z=1}
\nonumber\\
&=& f(x) + \sum\limits_{k=1}^{m-1} \sum\limits_{\ell = 1}^{k}
\, C_{k-1}^{\ell-1} \, (1-z)^k \,
\frac{x^\ell}{\ell !} \, f^{(\ell)}(x) \,, 
\en
where $C_m^\ell = m!/(\ell! (m-\ell)!)$ is the binomial coefficient and
$f^{(\ell)}(x) = \partial^\ell f(x)/\partial x^\ell$.  
Note that the partial derivatives $\delta(1-z) \partial_z^{m-j} f(x/z)$ with respect
to $z$ in Eq.~(\ref{PDF_dist}) can be simplified and reduced to derivatives with respect
to $x$ as
\eq
\delta(1-z) \, \partial_z^{m} f\Big(\frac{x}{z}\Big)
= \delta(1-z) \, 
\sum\limits_{\ell=1}^{m} (-1)^m C_m^\ell \, \frac{(m-1)!}{(\ell-1)!} 
\, x^\ell  \, f^{(\ell)}(x) \,. 
\en
In all cases the derivatives $\partial^n I(x)/\partial x^n$ can be easily taken 
by changing the integration variable $z \to x/\xi$ and, after some simplifications,
returning back to the integration over $z$.
We stress that the choice $P(z) = R(z) [1/(1-z)^m]_{+,m-1}$, where $R(z)$
is a regular function, can be reduced to case (b) by carrying out 
a Taylor expansion of $R(z)$ around $z=1$:
\eq
R(z) =  \sum\limits_{k=0}^{\infty} \, \frac{(z-1)^k}{k!} \, R^{(k)}(1) \, 
\en
where $R^{(k)}(1) = (\partial^k R(z)/\partial z^k)_{z=1}$, which results
in the cancellation
of the respective powers of $(1-z)$ between the numerator and the denominator of
the integrand of $I(x)$.
In particular, the distribution $P_{qq}(z)$ defined in Eq.~(\ref{Pqq}) contains
the term $(1+z^2)/(1-z)_+$, which can be represented as the sum of a regular term
corresponding to case (a) and a single distribution $2/(1-z)_+$ corresponding to case (b), 
\eq
\frac{1+z^2}{(1-z)_+} =  - (1+z) + \frac{2}{(1-z)_+} \,.
\en
In case (b) the $n$th derivative of the integral $I(x)$ reads as 
\eq
\frac{\partial^n I(x)}{\partial x^n} &=&
- \lim\limits_{z \to 1} \, \sum\limits_{k=1}^{n} \,
\left(\frac{1}{x^{k}} \,
\frac{\partial^{k-1}}{\partial z^{k-1}}
\, \biggl[\frac{1}{(1-z)^m_{+x,m-1}}\biggr]
\, z^{k-2}
f\Big(\frac{x}{z} \Big)
\right)^{(n-k)}_x
\nonumber\\
&+& \frac{1}{x^n} \, \int\limits_{x}^1 dz \,
\frac{\partial^n}{\partial z^n}
\, \biggl[\frac{1}{(1-z)^m_{+,m-1}}\biggr]
\, z^{n-1} f\Big(\frac{x}{z}\Big)
\nonumber\\
&=&
- \lim\limits_{z \to 1} \, \sum\limits_{k=1}^{n} \,
\left(\frac{(m+k-2)!}{x^k \, (m-1)!} \,
\frac{z^{k-2} f\Big(\frac{x}{z}\Big)}{(1-z)^{m+k-1}_{+x,m+k-2}}
\right)^{(n-k)}_x
\nonumber\\
&+& \frac{(m+n-1)!}{x^n \, (m-1)!}
\, \int\limits_{x}^1 dz \,
\, \frac{z^{n-1} f\Big(\frac{x}{z}\Big)}{(1-z)^{m+n}_{+,m+n-1}}
\nonumber\\
&=&
\lim\limits_{z \to 1} \, \sum\limits_{k=1}^{n} \,
\left(\frac{(-1)^{m+k}}{x^k \, (m-1)!  \, (m+k-1)} \,
\frac{\partial^{m+k-1}}{\partial z^{m+k-1}}
\, \biggl[z^{k-2} \, f\Big(\frac{x}{z}\Big)\biggr]
\right)^{(n-k)}_x
\nonumber\\
&+& \frac{(m+n-1)!}{x^n \, (m-1)!}
\, \int\limits_{x}^1 dz \,
\, \frac{z^{n-1} f\Big(\frac{x}{z}\Big)}{(1-z)^{m+n}_{+,m+n-1}}
\,.
\en

\clearpage

\renewcommand{\thesection}{S\arabic{section}}  
\renewcommand{\thetable}{S\arabic{table}}  
\renewcommand{\thefigure}{S\arabic{figure}}
\renewcommand{\theequation}{S\arabic{equation}}
\setcounter{equation}{0}

\section*{Supplementary Material: NLP and NNLP contributions to
  the $T$-odd hadronic structure functions
  $W_i(x_1,x_2)$~\cite{SuppMaterial:2024sp}.}

Here we list the higher-order terms $W^{\rm NLP}(x_1^0, x_2^0,L_\rho)$
and $W^{\rm NNLP}(x_1^0, x_2^0,,L_\rho)$ in the small-$Q_T$ expansion,
\eq 
W^{\rm NLP}(x_1^0, x_2^0,L_\rho) &=& W_1(x_1^0, x_2^0,L_\rho)
+   \frac{1}{2} \biggl(
x_1^0 \ \partial_{x_1^0} W_0(x_1^0, x_2^0,L_\rho) 
    +   x_2^0 \ \partial_{x_2^0} W_{0}(x_1^0, x_2^0,L_\rho) 
    \biggr) \,, 
\\
W^{\rm NNLP}(x_1^0, x_2^0,L_\rho) &=& W_2(x_1^0, x_2^0,L_\rho)
+  \frac{1}{4} x_1^0 x_2^0
\ \partial_{x_1^0} \partial_{x_2^0} W_0(x_1^0, x_2^0,L_\rho) 
\nonumber\\
&-&   \frac{1}{8} \biggl(
    x_1^0    \ \partial_{x_1^0} W_0(x_1^0, x_2^0,L_\rho)
- 4 x_1^0    \ \partial_{x_1^0} W_1(x_1^0, x_2^0,L_\rho) 
-  (x_1^0)^2 \ \partial^2_{x_1^0} W_0(x_1^0, x_2^0,L_\rho) 
\biggr)
\nonumber\\  
&-&   \frac{1}{8} \biggl(
    x_2^0    \ \partial_{x_2^0}  W_0(x_1^0, x_2^0,L_\rho)
- 4 x_2^0    \ \partial_{x_2^0}  W_1(x_1^0, x_2^0,L_\rho) 
-  (x_2^0)^2 \ \partial^2_{x_2^0} W_0(x_1^0, x_2^0,L_\rho)
\biggr)  \,.
\en
Also we have abbreviated $L_\rho\equiv \log\rho^2$. 

The higher-order terms $W_{1; J}^{a b}(x_1,x_2,L_\rho)$
and $W_{2; J}^{ab}(x_1,x_2,L_\rho)$ are given by 
\eq\label{W_exp}
W^{ab}_{k; J}(x_1,x_2,L_\rho)
&=& - \frac{g_{a b}}{4 x_1 x_2} \biggl[ 
R_{ab, k}^{J}(x_1,x_2,L_\rho)  \, f_{a/H_1}(x_1) \,  f_{b/H_2}(x_2) 
\nonumber\\
&+& \Big(P_{ba,k}^{J} \otimes f_{b/H_2}\Big)(x_2,x_1,L_\rho) \  f_{a/H_1}(x_1) 
 +  \Big(P_{ab,k}^{J} \otimes f_{a/H_1}\Big)(x_1,x_2,L_\rho) \  f_{b/H_2}(x_2)
\biggr]  \,,
\en
where $k = 1, 2$ and $J = \Delta\Delta_P, \Delta_P,\nabla$.

In the following, for convenience, we introduce functions $F_i(z)$
with $i=1,2,3,4,5,6$. Functions $F_i(z)$ with $i=3,4,5,6$ 
are obtained from functions $F_1(z)$ and $F_2(z)$ 
by subtraction of the leading and sub-leading terms in an expansion
in $(1-z)$. The main idea of such substracted functions is to make
the ratios $F_i(z)/(1-z)^n$ regular at $z=1$.  
All functions $F_i(z)$ with $i=1,\ldots,6$ are defined as 
\eq
F_1(z) &=& \frac{1+z}{1-z}
+ \frac{2 z \log(z)}{(1-z)^2}
= 2 \sum\limits_{N=1}^{\infty} \, \frac{(1-z)^N}{(N+1) (N+2)} 
= {\cal O}(1-z)
\,, \nonumber\\ 
F_2(z) &=& 1 + \frac{z \log(z)}{1-z}
=  \frac{1-z}{2} \, \Big(1 + F_1(z) \Big)
= \sum\limits_{N=1}^{\infty} \, \frac{(1-z)^N}{N (N+1)} 
= {\cal O}(1-z)
\,, \nonumber\\ 
F_3(z) &=& F_1(z) - \frac{1-z}{3}
= 2 \sum\limits_{N=2}^{\infty} \, \frac{(1-z)^N}{(N+1) (N+2)} 
= {\cal O}\Big((1-z)^2\Big) \,,
\nonumber\\
F_4(z) &=& F_1(z) - \frac{1-z}{3} - \frac{(1-z)^2}{6}
= 2 \sum\limits_{N=3}^{\infty} \, \frac{(1-z)^N}{(N+1) (N+2)} 
= {\cal O}\Big((1-z)^3\Big) \,,
\nonumber\\
F_5(z) &=& F_2(z) - \frac{1-z}{2} =
\frac{1-z}{2} F_1(z) 
= \sum\limits_{N=2}^{\infty} \, \frac{(1-z)^N}{N (N+1)} 
= {\cal O}\Big((1-z)^2\Big) \,,
\nonumber\\
F_6(z) &=& F_2(z) - \frac{1-z}{2}  - \frac{(1-z)^2}{6} =
\frac{1-z}{2} \Big(F_1(z) - \frac{1-z}{3}\Big) =
\frac{1-z}{2} F_3(z)
\nonumber\\
&=& \sum\limits_{N=3}^{\infty} \, \frac{(1-z)^N}{N (N+1)} 
= {\cal O}\Big((1-z)^3\Big)\,.
\en
They obey the following normalization conditions:
\eq
F_i(1) = 0\,, \quad 
F_1(0) = F_2(0) = 1\,, \quad
F_3(0) = \frac{2}{3}\,, \quad
F_4(0) = F_5(0) = \frac{1}{2}\,, \quad
F_6(0) = \frac{1}{3} \,.
\en
It is clear that the ratios
\eq
f_{i_1}(z)  = \frac{F_{i_1}(z)}{1-z}    \,, \ i_1 = 1,2;  \qquad
f_{i_2}(z) = \frac{F_{i_2}(z)}{(1-z)^2}\,, \ i_2 = 3,5;  \qquad
f_{i_3}(z) = \frac{F_{i_3}(z)}{(1-z)^3}\,, \ i_3 = 4,6\,,
\en
are regular at $z=1$.
In the following, we express the NLP and NNLP contributions to
the hadronic structure functions in terms of the functions $f_i(z)$.

The perturbative functions $R_{ab,k}^{J}(x_1,x_2,L_\rho)$,
$P_{ab,k}^{J}(z_1,x_2,L_\rho)$, and
$P_{ba,k}^{J}(z_2,x_1,L_\rho)$ 
with $k=1,2$
can be expanded as
\eq\label{RP_exp}
R_{ab,k}^{J}(x_1,x_2,L_\rho) &=& \sum\limits_{N=1}^9
\, R_{ab; N}^{k; J}(L_\rho) \, T^R_N(x_1,x_2) \,,
\nonumber\\
P_{ab,k}^{J}(z_1,x_2,L_\rho) 
&=& \sum\limits_{N=1}^3
\, P_{a, 1; N}^{k; J}(z_1,L_\rho) \, T^P_N(x_2) \,,
\nonumber\\
P_{ba,k}^{J}(z_2,x_1,L_\rho) 
&=& \sum\limits_{N=1}^3
\, P_{ab, 2; N}^{k; J}(z_2,L_\rho) \, T^P_N(x_1) \,,
\en
where
\eq\label{RP_exp2}
& &T^R_1(x_1,x_2) = 1\,,   \hspace*{2.5cm}  
   T^R_2(x_1,x_2) = x_1 \, \partial_{x_1}\,, \hspace*{1.6cm}  
   T^R_3(x_1,x_2) = x_2 \, \partial_{x_2}\,, 
\nonumber\\
& &T^R_4(x_1,x_2) = x_1 x_2 \, \partial_{x_1x_2}^2\,, \hspace*{1.1cm}
   T^R_5(x_1,x_2) = x_1^2   \, \partial_{x_1}^2\,,    \hspace*{1.6cm}
   T^R_6(x_1,x_2) = x_2^2   \, \partial_{x_2}^2\,,   
\nonumber\\
& &T^R_7(x_1,x_2) = x_1^2 x_2   \, \partial_{x_1}^2 \partial_{x_2}
\,, \hspace*{.9cm}
    T^R_8(x_1,x_2) = x_1   x_2^2 \, \partial_{x_1}   \partial_{x_2}^2
\,, \hspace*{0.75cm}
   T^R_9(x_1,x_2) = x_1^2 x_2^2 \, \partial_{x_1}^2 \partial_{x_2}^2
\en
and
\eq 
T^P_1(x) = 1\,, \qquad
T^P_2(x) = x \, \partial_{x}\,, \qquad
T^P_3(x) = x^2 \, \partial_{x}^2 
\,.
\en

Below we list only the nonvanishing terms 
$R_{a b; N}^{k; J}$,
$P_{a b, 1; N}^{k; J}(z_1)$,
$P_{a b, 2; N}^{k; J}(z_2)$ for the various structure functions.

For the $\Delta\Delta_P$ structure function we have,
in the $q \bar q$ annihilation channel:
\eq
R_{q q; 1}^{1; \Delta\Delta_P} &=&
C_A + \Big(\frac{C_A}{2} + \frac{4 C_1}{3}\Big) \, L_\rho 
\,, \nonumber\\
R_{q q; 2}^{1; \Delta\Delta_P} &=& R_{q q; 3}^{1; \Delta\Delta_P} \, = \,
\Big(C_A -  \frac{C_1}{3}\Big) \, L_\rho
\,, \nonumber\\
R_{q q; 4}^{1; \Delta\Delta_P} &=&
- C_A \, L_\rho
\,.
\en

\eq
P_{q q, 1; 1}^{1; \Delta\Delta_P}(z) &=&
P_{q q, 2; 1}^{1; \Delta\Delta_P}(z) \, = \, 
 - \frac{C_A}{4} \, 
 \frac{1 + 3 z + z^2 - z^3}{(1-z)^2_{+,1}} 
- \frac{C_1}{6} \, \frac{1 + 2 z - 3 z^2 - 2 z^3}{(1-z)^2_{+,1}} 
\nonumber\\
&-&
   \frac{C_1}{2} \, f_3(z) (1 + 3 z) 
\,, \nonumber\\
P_{q q, 1; 2}^{1; \Delta\Delta_P}(z) &=&
P_{q q, 2; 2}^{1; \Delta\Delta_P}(z) \, = \, 
 \frac{C_A}{2} \, \frac{z (1+z^2)}{(1-z)^2_{+,1}} 
+ \frac{C_1}{3} \, 
\, \frac{z}{(1-z)_+} 
\nonumber\\
  &+&  C_1 \, f_3(z) \, z 
\,.
\en

\eq
R_{q q; 1}^{2; \Delta\Delta_P} &=&
- \Big(C_A + \frac{4 C_1}{3}\Big)
- \Big(\frac{3 C_A}{8} + C_1\Big) \, L_\rho 
\,,\nonumber\\
R_{q q; 2}^{2; \Delta\Delta_P} &=&
R_{q q; 3}^{2; \Delta\Delta_P} \, = \,
- \Big(C_A - \frac{C_1}{3}\Big)
- \Big(\frac{3 C_A}{2} + \frac{5 C_1}{6}\Big) L_\rho   
\,,\nonumber\\
R_{q q; 4}^{2; \Delta\Delta_P} &=&  
  C_A 
  + \Big(\frac{3 C_A}{2} + \frac{4 C_1}{3}\Big) \, L_\rho 
\,,\nonumber\\
R_{q q; 5}^{2; \Delta\Delta_P} &=&
R_{q q; 6}^{2; \Delta\Delta_P} \, = \,
- \Big(\frac{3 C_A}{4} - \frac{C_1}{12}\Big) \, L_\rho
\,,\nonumber\\
R_{q q; 7}^{2; \Delta\Delta_P} &=&
R_{q q; 8}^{2; \Delta\Delta_P} \, = \,
\Big(\frac{C_A}{2} - \frac{C_1}{6}\Big) \, L_\rho
\,,\nonumber\\ 
R_{q q; 9}^{2; \Delta\Delta_P} &=&
- \frac{C_A}{4} \, L_\rho
\,.
\en
\eq
P_{q q, 1; 1}^{2; \Delta\Delta_P}(z) &=&
P_{q q, 2; 1}^{2; \Delta\Delta_P}(z) \, = \, 
\frac{3 C_A}{16} \,
\frac{(1 + z) (1 + 5 z - 3 z^2 + z^3)}{(1-z)^3_{+,2}}
\nonumber\\
&+& \frac{C_1}{48}  \,
\frac{9 + 42 z - 60 z^2
-  18 z^3 + 19 z^4}{(1-z)^3_{+,2}}
+ \frac{3 C_1}{8}  \, f_4(z) \, (1 + 6 z + z^2) 
\,,\nonumber\\
P_{q q, 1; 2}^{2; \Delta\Delta_P}(z) &=&
P_{q q, 2; 2}^{2; \Delta\Delta_P}(z) \, = \, 
- \frac{C_A}{4} \,
\frac{z (3 + z + 3 z^2 - 3 z^3)}{(1-z)^3_{+,2}}
- \frac{C_1}{12} \frac{z (9 - 9 z - z^2 - 3 z^3)}
{(1-z)^3_{+,2}}
\nonumber\\
&-& \frac{C_1}{2} \, f_4(z) \, z (3 + z)
\,,\nonumber\\
P_{q q, 1; 3}^{2; \Delta\Delta_P}(z) &=&
P_{q q, 2; 3}^{2; \Delta\Delta_P}(z) \, = \, 
  \frac{C_A}{4} \,
  \frac{z^2 (1 + z^2)}{(1-z)^3_{+,2}}
+ \frac{C_1}{12} \frac{z^2 (3 - z)}
{(1-z)^2_{+,1}}
+ \frac{C_1}{2} \, f_4(z) \, z^2
\,.
\en

For the $\Delta_P$ structure function we have,
in the $q \bar q$ annihilation channel:
\eq
R_{q q; 2}^{1; \Delta_P} &=& - R_{q q; 3}^{1; \Delta_P}
\, = \, - \frac{L_\rho}{\rho} \,
(2 C_A + C_1)
\,.
\en

\eq
  P_{q q, 1; 1}^{1; \Delta_P}(z) &=&
- P_{q q, 2; 1}^{1; \Delta_P}(z) \, = \, 
\frac{C_A}{2 \rho} 
\, \frac{1 + 3 z - z^2 + z^3}{(1-z)^2_{+,1}}
+ \frac{C_1}{2 \rho} \, 
\frac{1 + 2 z - 3 z^2 + 2 z^3}{(1-z)^2_{+,1}}
\nonumber\\
&+& \frac{C_1}{2 \rho} \, f_1(z) \, (1 + 3 z)
\,, \nonumber\\
  P_{q q, 1; 2}^{1; \Delta_P}(z) &=&
- P_{q q, 2; 2}^{1; \Delta_P}(z) \, = \, 
- \frac{C_A}{\rho} \, \frac{z (1+z)}{(1-z)_+}
- \frac{C_1}{\rho} \, \frac{z}{(1-z)_+} 
- \frac{C_1}{\rho} \, f_1(z) \, z
\,.
\en

\eq
R_{q q; 2}^{2; \Delta_P} &=& - R_{q q; 3}^{2; \Delta_P} \, = \,
\frac{2 C_A + C_1}{\rho} \, + 
\frac{3}{\rho} \, \Big(C_A + \frac{19 C_1}{18} \Big) \, L_\rho
\,, \nonumber\\
R_{q q; 5}^{2; \Delta_P} &=& - R_{q q; 6}^{2; \Delta_P} \, = \,
\frac{3}{2 \rho} \, \Big(C_A + \frac{2 C_1}{9}\Big) \, L_\rho
\,, \nonumber\\
R_{q q; 7}^{2; \Delta_P} &=& - R_{q q; 8}^{2; \Delta_P} \, = \,
- \frac{2 C_A + C_1}{2 \rho} \, L_\rho
\,.
\en

\eq
  P_{q q, 1; 1}^{2; \Delta_P}(z) &=&
- P_{q q, 2; 1}^{2; \Delta_P}(z) \, = \, 
- \frac{3 C_A}{8 \rho} \, 
\frac{1 + 6 z + 2 z^3 - z^4}{(1-z)^3_{+,2}}
\nonumber\\
&-& \frac{C_1}{24 \rho}  
\frac{12 + 57 z - 75 z^2 + 39 z^3 - 17 z^4}{(1-z)^3_{+,2}}
  -  \frac{3 C_1}{8 \rho} \, f_3(z) \, (1+6z+z^2) 
\,, \nonumber\\
  P_{q q, 1; 2}^{2; \Delta_P}(z) &=&
- P_{q q, 2; 2}^{2; \Delta_P}(z) \, = \,  
\frac{C_A}{2 \rho} \, 
\frac{z (3 + z - 3 (1 - z) z^2)}{(1-z)^3_{+,2}}
\nonumber\\
&+& \frac{C_1}{6 \rho}  
\frac{z (12 - 11 z - 2 z^2 + 7 z^3)}{(1-z)^3_{+,2}}
+ \frac{C_1}{2 \rho} \, f_3(z) \, z (3+z)
\,, \nonumber\\
  P_{q q, 1; 3}^{2; \Delta_P}(z) &=&
- P_{q q, 2; 3}^{2; \Delta_P}(z) \, = \,  
- \frac{C_A}{2 \rho} \, 
\frac{z^2 (1 + z)}{(1-z)^2_{+,1}}
- \frac{C_1}{6 \rho}  
\frac{z^2 (4 - z)}{(1-z)^3_{+,2}}
- \frac{C_1}{2 \rho} \, f_3(z) \, z^2
\,.
\en

For the $\nabla$ structure function we have,
in the $q \bar q$ annihilation channel:
\eq
R_{q q; N}^{1; \nabla} = 2 R_{q q; N}^{1; \Delta_P}
\,, \quad N=2,3  
\,.
\en

\eq
P_{q q, 1; 1}^{1; \nabla}(z) &=& 2 P_{q q, 1; 1}^{1; \Delta_P}(z) 
+ \frac{C_1}{\rho} \,
\Big(1 - f_1(z) \, (1+z)\Big)
\,, \nonumber\\
P_{q q, 1; 2}^{1; \nabla}(z) &=& 2 P_{q q, 1; 2}^{1; \Delta_P}(z)
\,, \nonumber\\
P_{q q, 2; N}^{1; \nabla}(z) &=&
- P_{q q, 1; N}^{1; \nabla}(z)
\,, \quad N=1,2  
\,. 
\en

\eq
R_{q q; 2}^{2; \nabla} &=& - R_{q q; 3}^{2; \nabla}
= 
2 R_{q q; 2}^{2; \Delta_P} 
+ \frac{C_1}{3 \rho} \, L_\rho
\,, \nonumber\\
R_{q q; 5}^{2; \nabla} &=& - R_{q q; 6}^{2; \nabla}
\, = \,   2 R_{q q; 5}^{2; \Delta_P}
\, = \, - 2 R_{q q; 6}^{2; \Delta_P}
\,, \nonumber\\
  R_{q q; 7}^{2; \nabla} &=&
- R_{q q; 8}^{2; \nabla}
\, = \,   2 R_{q q; 7}^{2; \Delta_P}
\, = \, - 2 R_{q q; 8}^{2; \Delta_P}
\,.
\en

\eq
  P_{q q, 1; 1}^{2; \nabla}(z) &=&
2 P_{q q, 1; 1}^{2; \Delta_P}(z)
\nonumber\\
&-&
\frac{C_A}{4 \rho} \, (1+z) 
- \frac{C_1}{12 \rho} \,
\frac{12 - 13 z - 2 z^2 + 7 z^3}{(1-z)^2_{+,1}}
\nonumber\\
&+& \frac{C_1}{4 \rho} \,
f_3(z) \, (1+3z) (3+z)
\,,\nonumber\\
    P_{q q, 1; 2}^{2; \nabla}(z) &=&
  2 P_{q q, 1; 2}^{2; \Delta_P}(z)
\nonumber\\
&+&
\frac{C_1}{3 \rho} \, 
\frac{z (2 - 3 z + z^2)}{(1-z)^2_{+,1}}
- \frac{C_1}{\rho} \, f_3(z) \, z (1+z) 
\,,\nonumber\\
    P_{q q, 2; N}^{2; \nabla}(z) &=&
  - P_{q q, 1; N}^{2; \nabla}(z)
\,, \quad N=1,2 
\,. 
\en

We now present the results for the $q g$ channel.
For the $\Delta\Delta_P$ structure function, we have:
\eq
R_{q g; 1}^{1; \Delta\Delta_P} &=&
\Big(\frac{5 C_A}{2} + 4 C_1 \Big(\frac{11}{12} + L_\rho\Big)\Big)
\, L_\rho
\,, \nonumber\\
R_{q g; 2}^{1; \Delta\Delta_P} &=&
- \Big(\frac{3 C_A}{2} + 2 C_1 (1 + L_\rho)\Big)
\, L_\rho
\,.
\en

\eq 
P_{q g, 1; 1}^{1; \Delta\Delta_P}(z) &=&
    \frac{C_A}{2} \frac{z^2 (1 + 2 z)}{(1-z)^2_{+,1}}
 + C_1 \Big(\frac{1 + 5 z}{3}
 + 2 z L_\rho\Big)
 \, \frac{z^2}{(1-z)^2_{+,1}}
\nonumber\\
&+& C_1 \, f_3(z)  \, z^2
\,, \nonumber\\
P_{q g, 2; 1}^{1; \Delta\Delta_P}(z) &=& 
- \frac{C_A}{4} \, \frac{1 + 7 z + 2 z^2}{(1-z)_+}
- \frac{C_1}{6} \, \frac{3 + 33 z - 21 z^2 + 7 z^3  
+ 6 z (1 + 3 z) L_\rho}{(1-z)_+}
\,, \nonumber\\
P_{q g, 2; 2}^{1; \Delta\Delta_P}(z) &=& 
  \frac{C_A}{2}  \, \frac{z (1 + 2z)}{(1-z)_+}
+ C_1 \, \frac{z (1 + 2 z - z^2 +  2 z L_\rho)}{(1-z)_+}
\,.
\en

\eq
R_{q g; 1}^{2; \Delta\Delta_P} &=&
- \frac{C_A}{4} \, (10 - 3 L_\rho)
- \frac{C_1}{3} \, (11 + 21 L_\rho -18 L_\rho^2) 
\,,\nonumber\\
R_{q g; 2}^{2; \Delta\Delta_P} &=&
  \frac{C_A}{4} \, (6 - 5 L_\rho)
+ C_1 \, (2 + 2 L_\rho - 5 L_\rho^2) 
\,,\nonumber\\
R_{q g; 3}^{2; \Delta\Delta_P} &=& 
-  \Big(\frac{7 C_A}{2} + \frac{C_1}{2}  \, (11 + 12 L_\rho)\Big)
\, L_\rho 
\,,\nonumber\\
R_{q g; 4}^{2; \Delta\Delta_P} &=&
\Big(\frac{5 C_A}{2}  + \frac{C_1}{3} \, (11 + 12 L_\rho)\Big)
\, L_\rho
\,,\nonumber\\
R_{q g; 5}^{2; \Delta\Delta_P} &=&
\Big(\frac{5 C_A}{4} 
+ C_1  \,  (1 + 2 L_\rho)\Big) \, L_\rho
\,,\nonumber\\
R_{q g; 7}^{2; \Delta\Delta_P} &=&
- \Big(\frac{3 C_A}{4} 
+ C_1 \, (1 + L_\rho)\Big) \, L_\rho
\,.
\en

\eq
P_{q g, 1; 1}^{2; \Delta\Delta_P}(z) &=&
- \frac{C_A}{4}
\frac{z^2 (3 + 9 z - 2 z^2)}{(1-z)^3_{+,2}}
\nonumber\\
&+& \frac{C_1}{4} \Big(5 - 27 z + 15 z^2 - z^3 
- 4 z (3 + z) L_\rho\Big) \, 
\frac{z^2}{(1-z)^3_{+,2}}
\nonumber\\
&-&\frac{3 C_1}{2} \, f_4(z) \, z^2 (1+z)
\,,\nonumber
\en
\eq
P_{q g, 1; 2}^{2; \Delta\Delta_P}(z) &=&
 \frac{C_A}{2}
\, \frac{z^3 (1+2z)}{(1-z)^3_{+,2}}
+ \frac{C_1}{6}
\frac{z^3 (3 + 8 z + z^2 + 12 z L_\rho)}{(1-z)^3_{+,2}} 
+ C_1 \, f_4(z) \, z^3
\,,\nonumber\\
P_{q g, 2; 1}^{2; \Delta\Delta_P}(z) &=&
\frac{C_A}{16} \, 
\frac{3 + 36 z + 19 z^2 - 2 z^3}{(1-z)^2_{+,1}}
\nonumber\\
&+& \frac{C_1}{8} \,
\Big(3 + 52 z - 8 z^2 + 4 z^3 - 7 z^4
+ 6 z (1 + 6 z + z^2) L_\rho\Big) 
\frac{1}{(1-z)^2_{+,1}}
\,,\nonumber\\
P_{q g, 2; 2}^{2; \Delta\Delta_P}(z) &=&
- \frac{C_A}{4} \, 
\frac{z (3 + 9 z - 2 z^2)}{(1-z)^2_{+,1}}
\nonumber\\
&-& \frac{C_1}{6} \, \Big(9 + 39 z - 39 z^2 + 13 z^3  
+ 6 z (3 + z) L_\rho\Big) \, 
\frac{z}{(1-z)^2_{+,1}}
\,,\nonumber\\
P_{q g, 2; 3}^{2; \Delta\Delta_P}(z) &=&
  \frac{C_A}{4} \, \frac{z^2 (1+2z)}{(1-z)^2_{+,1}}
+ \frac{C_1}{2} \, \frac{z^2 (1 + 2 z - z^2
  + 2 z L_\rho)}{(1-z)^2_{+,1}}
\,. 
\en

For the $\Delta_P$ structure function we have, in the $qg$ channel:
\eq
R_{q g; 1}^{1; \Delta_P} &=& \frac{L_\rho}{2\rho}
\, (C_A - 6 C_1) 
\,, \nonumber\\
R_{q g; 2}^{1; \Delta_P} &=& \frac{2 L_\rho}{\rho}
\, C_1 
\,.
\en

\eq
P_{q g, 1; 1}^{1; \Delta_P}(z) &=&
- \frac{C_A}{2\rho}  \, \frac{z^2}{(1-z)_+}
- \frac{C_1}{\rho} \, \frac{z^2 (1+z)}{(1-z)^2_{+,1}}
\,, \nonumber\\
P_{q g, 2; 1}^{1; \Delta_P}(z) &=& - \frac{C_A}{2\rho}
\, \frac{1 + z - z^2}{(1-z)_{+}}
\nonumber\\
&-& \frac{C_1}{\rho} \, \frac{1 + 4 z - 10 z^2 + 2 z^3}{(1-z)_{+}}
 -  \frac{C_1}{\rho} \, z (1 + 3 z) L_\rho
\,, \nonumber\\
P_{q g, 2; 2}^{1; \Delta_P}(z) &=& 
\frac{C_A}{\rho} \, z 
+ \frac{2 C_1}{\rho} \, \frac{z (1 - 2 z^2)}{(1-z)_{+}} 
+ \frac{2 C_1}{\rho} \, z^2 \, L_\rho
\,. 
\en

\eq
R_{q g; 1}^{2; \Delta_P} &=& - \frac{C_A}{2 \rho}
\Big(1 + \frac{3}{2} L_\rho\Big) 
+ \frac{3C_1}{\rho} \Big(1 - \frac{19}{6} L_\rho - 2 L_\rho^2\Big)
\,, \nonumber\\
R_{q g; 2}^{2; \Delta_P} &=& \frac{C_A}{2 \rho} L_\rho
- \frac{2 C_1}{\rho} (1 - 4 L_\rho  - 2 L_\rho^2) 
\,, \nonumber\\
R_{q g; 3}^{2; \Delta_P} &=& - \frac{L_\rho}{\rho} 
(C_A - 4 C_1)
\,, \nonumber\\
R_{q g; 4}^{2; \Delta_P} &=&
\frac{L_\rho}{2 \rho} \, (C_A - 6 C_1)  
\,, \nonumber\\
R_{q g; 5}^{2; \Delta_P} &=&
-\frac{L_\rho}{2 \rho} \, (C_A + 2 C_1 (5 + L_\rho))
\,, \nonumber\\
R_{q g; 7}^{2; \Delta_P} &=&
\frac{L_\rho}{\rho} \, C_1 
\,.
\en

\eq
P_{q g, 1; 1}^{2; \Delta_P}(z) &=&
\frac{1}{4 \rho} \, \Big(
C_A (3 + z^2)
+ 2 C_1 (7 + 10 z + 3 z^2 + 4 z^2 L_\rho)\Big) 
\, \frac{z^2}{(1-z)^3_{+,2}}
\,, \nonumber\\
P_{q g, 1; 2}^{2; \Delta_P}(z) &=&
- \frac{C_A}{2\rho}  \, \frac{z^3}{(1-z)^2_{+,1}}
- \frac{C_1}{\rho}   \, \frac{z^3 (1+z)}{(1-z)^3_{+,2}}
\,, \nonumber
\en
\eq
P_{q g, 2; 1}^{2; \Delta_P}(z) &=& 
\frac{C_A}{8 \rho} \,
\frac{3 + 9 z - 5 z^2 + z^3}{(1-z)^2_{+,1}}
\nonumber\\
&+& \frac{C_1}{4 \rho} \,
\frac{3 + 20 z - 37 z^2 - 8 z^3 + 6 z^4}{(1-z)^2_{+,1}}
+ \frac{3 C_1}{4 \rho} \,
\frac{z (1 + 6 z + z^2)}{(1-z)_{+}} \, L_\rho
\, 
\,, \nonumber\\
P_{q g, 2; 2}^{2; \Delta_P}(z) &=&
- \frac{C_A}{2 \rho} \, 
\frac{z (3 - 3 z + z^2)}{(1-z)^2_{+,1}}
\nonumber\\
&-& \frac{C_1}{\rho} \,
\frac{z (3 + 2 z - 14 z^2 + 6 z^3)}{(1-z)^2_{+,1}}
- \frac{C_1}{\rho} \, \frac{z^2 (3 + z)}{(1-z)_+}  \, L_\rho
\,, \nonumber\\
P_{q g, 2; 3}^{2; \Delta_P}(z) &=&
\frac{C_A}{2 \rho} \,
\frac{z^2}{(1-z)_+}
+ \frac{C_1}{\rho} \,
\frac{z^2 (1 - 2 z^2)}{(1-z)^2_{+,1}}
+ \frac{C_1}{\rho} \,
\frac{z^3}{(1-z)_+} L_\rho 
\,. 
\en

Finally, for the $\nabla$ structure function we have,
in the $qg$ channel:
\eq
R_{q g; 1}^{1; \nabla} &=&
2 R_{q g; 1}^{1; \Delta_P}
- \frac{2 L_\rho}{\rho}
\, (C_A - 6 C_1) 
\,, \nonumber\\
R_{q g; 2}^{1; \nabla} &=&
2 R_{q g; 2}^{1; \Delta_P}
\,.
\en

\eq
  P_{q g, 1; 1}^{1; \nabla}(z) &=&
2 P_{q g, 1; 1}^{1; \Delta_P}(z) 
+ \frac{2}{\rho} \, \Big(C_A z - 2 C_1 (2 + z)\Big)
\, \frac{z}{(1-z)_{+}} 
\,, 
\nonumber\\
  P_{q g, 2; 1}^{1; \nabla}(z) &=&
2 P_{q g, 2; 1}^{1; \Delta_P}(z) 
+ \frac{1}{\rho} \, 
\frac{C_A - 6 C_1 z}{(1-z)_{+}}
+ \frac{2C_1}{\rho} \, z (1+z) \, L_\rho 
\,,
\nonumber\\
  P_{q g, 2; 2}^{1; \nabla}(z) &=&
2 P_{q g, 2; 2}^{1; \Delta_P}(z) 
\,. 
\en

\eq
  R_{q g; 1}^{2; \nabla} &=& 
2 R_{q g; 1}^{2; \Delta_P}
+ \frac{C_A}{2 \rho} (4 - 7 L_\rho)
- \frac{C_1}{3 \rho} \Big(36 - 47 L_\rho - 24 L_\rho^2\Big)
\Big) 
\,,\nonumber\\
  R_{q g; 2}^{2; \nabla} &=& 
2 R_{q g; 2}^{2; \Delta_P}
+ \frac{L_\rho}{\rho} \, \Big(3 C_A - 2 C_1 (15 + 2 L_\rho)\Big) 
\,,\nonumber\\
  R_{q g; 3}^{2; \nabla} &=& 
2 R_{q g; 3}^{2; \Delta_P}
+ \frac{4 L_\rho}{\rho} \, (C_A - 4 C_1)
\,,\nonumber\\
R_{q g; 4}^{2; \nabla} &=& 
2 R_{q g; 4}^{2; \Delta_P}
- \frac{2 L_\rho}{\rho} \, (C_A - 6 C_1)  
\,.
\en

\eq 
  P_{q g, 1; 1}^{2; \nabla}(z) &=&
2 P_{q g, 1; 1}^{2; \Delta_P}(z)
\nonumber\\
&-& \frac{C_A}{2 \rho} \,
\frac{z^2 (5+z)}{(1-z)^2_{+,1}}
+ \frac{C_1}{3 \rho}
\Big(48 + 37 z + z^2 (5 + 12 L_\rho)\Big)  
\, \frac{z}{(1-z)^2_{+,1}}
\nonumber\\
&-& \frac{2 C_1}{\rho} \, f_3(z) \, z^2 
\,,
\nonumber\\
  P_{q g, 1; 2}^{2; \nabla}(z) &=&
2 P_{q g, 1; 2}^{2; \Delta_P}(z)
+\frac{2}{\rho} \,
\Big(C_A z - 2 C_1 (2+z)\Big) \, 
 \frac{z^2}{(1-z)^2_{+,1}}
\,,
\nonumber\\
  P_{q g, 2; 1}^{2; \nabla}(z) &=&
2 P_{q g, 1; 2}^{2; \Delta_P}(z)
+ \frac{C_A}{4 \rho} \,
\frac{1 - 5 z - 11 z^2 - z^3}{(1-z)^2_{+,1}} 
\nonumber\\
&+& \frac{C_1}{6 \rho}  \,
\frac{3 - 12 z + 63 z^2 + 92 z^3 - 50 z^4}{(1-z)^2_{+,1}} 
- \frac{C_1}{2 \rho}  \,
\frac{z (1 + 3 z) (3 + z)}{(1-z)_+} \, L_\rho 
\,,
\nonumber\\
  P_{q g, 2; 2}^{2; \nabla}(z) &=&
2 P_{q g, 2; 2}^{2; \Delta_P}(z)
+ \frac{1}{\rho} \,
\Big(C_A - 6 C_1 z)
\, \frac{z^2 (1+z)}{(1-z)^2_{+,1}} 
+ \frac{2 C_1}{\rho} 
\, \frac{z^2 (1+z)}{(1-z)_+} \, L_\rho
\,,
\nonumber\\
  P_{q g, 2; 3}^{2; \nabla}(z) &=&
2 P_{q g, 2; 3}^{2; \Delta_P}(z)
\,.
\en

The partial derivatives
$x_1^m x_2^n \, \partial^{m}_{x_1} \partial^{n}_{x_2} \ W^{a b}_{0; J}(x_1,x_2)$
and
$x_1^m x_2^n \, \partial^{m}_{x_1} \partial^{n}_{x_2} \ W^{a b}_{1; J}(x_1,x_2)$
are evaluated using the technique discussed
in Appendix~C. 
To the accuracy we are working at, we need the derivatives 
$\partial_{x_i}$, $\partial_{x_1}^2$ with $i=1,2$
and $\partial_{x_1x_2}^2$ acting on the $W_0$ structure functions and
$\partial_{x_i}$ with $i=1,2$ acting on the $W_1$ structure functions.

These derivatives can be written as
\eq
& &x_i   \, \partial_{x_i}W_k   = - W_k + V_k^{(i)}
\,, \nonumber\\
& &x_i^2 \, \partial_{x_i}^2W_k =
- (1 + 3 x_i \, \partial_{x_i})  \, W_k + V_k^{(ii)}
\,, \nonumber\\
& &x_1 x_2 \, \partial_{x_1x_2}^2W_k =
- (1 + x_1 \, \partial_{x_1} + x_2 \, \partial_{x_2}) \, W_k
+ V_k^{(12)} \,, 
\en 
where $V_k^{(L)}$ with $L=1,2,11,22,12$ are the specific expressions
for the corresponding derivatives, subprocess types, and type of
structure function. Following Eq.~(\ref{W_exp}) we write the $V_k^{(L)}$
as
\eq
V^{a b; (L)}_{k; J}(x_1,x_2,L_\rho)
&=& - \frac{g_{a b}}{4 x_1 x_2} \biggl[ 
R_{ab,k}^{(L); J}(x_1,x_2,L_\rho)  \, f_{a/H_1}(x_1) \,  f_{b/H_2}(x_2) 
\nonumber\\
&+& \Big(P_{ba,k}^{(L); J} \otimes f_{b/H_2}\Big)(x_2,x_1,L_\rho) \ f_{a/H_1}(x_1)
\nonumber\\
&+&  \Big(P_{ab,k}^{(L); J} \otimes f_{a/H_1}\Big)(x_1,x_2,L_\rho) \ f_{b/H_2}(x_2)
\biggr]  \,,
\label{V_exp}
\en
and expand the perturbative functions
$R_{ab,k}^{(L); J}(x_1,x_2,L_\rho)$,
$P_{ab,k}^{(L); J}(z_1,x_2,L_\rho)$, and
$P_{ba,k}^{(L); J}(z_2,x_1,L_\rho)$ following Eq.~(\ref{RP_exp}): 
\eq\label{RPV_exp}
R_{ab,k}^{(L); J}(x_1,x_2,L_\rho) &=& \sum\limits_{N=1}^9
\, R_{a b; N}^{k (L); J}(L_\rho) \, T^R_N(x_1,x_2) \,,
\nonumber\\
P_{ab,k}^{(L); J}(z_1,x_2,L_\rho) &=&
\sum\limits_{N=1}^3
\, P_{ab, 1; N}^{k (L); J}(z_1,L_\rho) \, T^P_N(x_2) \,,
\nonumber\\
P_{ba,k}^{(L); J}(z_2,x_1,L_\rho) &=&
\sum\limits_{N=1}^3
\, P_{ab, 2; N}^{k (L); J}(z_2,L_\rho) \, T^P_N(x_1) \,,
\en
Below we again list only the nonvanishing $R_{a b; N}^{k (L); J}(L_\rho)$,
$P_{ab, 1; N}^{k (L); J}(z_1,L_\rho)$, $P_{ab, 2; N}^{k (L); J}(z_2,L_\rho)$. 

For the $\Delta\Delta_P$ structure function we have,
in the $q \bar q$ annihilation channel:
\eq
R_{q q; 1}^{0 (1); \Delta\Delta_P} &=&
R_{q q; 1}^{0 (2); \Delta\Delta_P}   \, = \, 
R_{q q; 2}^{0 (12); \Delta\Delta_P}  \, = \, 
R_{q q; 3}^{0 (12); \Delta\Delta_P} 
= - \frac{C_1}{3}
\,, \nonumber\\
R_{q q; 2}^{0 (1); \Delta\Delta_P}   &=&
R_{q q; 3}^{0 (2); \Delta\Delta_P} \, = \,
- C_A \, (1 + L_\rho)
\,, \nonumber\\
R_{q q; 1}^{0 (11); \Delta\Delta_P} &=&
R_{q q; 1}^{0 (22); \Delta\Delta_P} \, = \,
\frac{C_A}{2} + \frac{C_1}{6} 
\,, \nonumber\\
R_{q q; 2}^{0 (11); \Delta\Delta_P} &=&
R_{q q; 3}^{0 (22); \Delta\Delta_P}  \, = \,
- C_A \, \Big(2 + L_\rho\Big) -  \frac{C_1}{3} 
\,, \nonumber\\
R_{q q; 4}^{0 (12); \Delta\Delta_P} 
&=& 
- C_A \, \Big(2 + L_\rho\Big) 
\,, \nonumber\\
R_{q q; 5}^{0 (11); \Delta\Delta_P} &=&
R_{q q; 6}^{0 (22); \Delta\Delta_P} \, = \,
- \frac{C_A}{2}  \, (3 + 2 L_\rho)
\,.
\en

\eq
P_{q q, 1; 1}^{0 (1); \Delta\Delta_P}(z)  &=&
P_{q q, 2; 1}^{0 (2); \Delta\Delta_P}(z)  \ = \
P_{q q, 1; 2}^{0 (12); \Delta\Delta_P}(z) \ = \
P_{q q, 2; 2}^{0 (12); \Delta\Delta_P}(z) 
\nonumber\\
&=&
\frac{C_A}{2}
\frac{z (1 + 2 z - z^2)}{(1-z)^2_{+,1}}
+ C_1 z \partial_{z}f_1(z)
\,, \nonumber\\
P_{q q, 2; 2}^{0 (1); \Delta\Delta_P}(z) &=&
P_{q q, 1; 2}^{0 (2); \Delta\Delta_P}(z)  \, = \,
P_{q q, 2; 2}^{0 (11); \Delta\Delta_P}(z) \, = \, 
P_{q q, 2; 3}^{0 (11); \Delta\Delta_P}(z) \, = \,
\nonumber\\
&=& 
    P_{q q, 1; 2}^{0 (22); \Delta\Delta_P}(z) \, = \,
    P_{q q, 1; 3}^{0 (22); \Delta\Delta_P}(z) \, = \,
\frac{C_A}{2} \frac{1 + z^2}{(1-z)_+} + C_1 f_1(z)
\,, \nonumber\\
P_{q q, 1; 1}^{0 (11); \Delta\Delta_P}(z) &=&
P_{q q, 2; 1}^{0 (22); \Delta\Delta_P}(z)
\nonumber\\
&=&
\frac{C_A}{2} \frac{z (1 + 5 z - 3 z^2 + z^3)}{(1-z)^3_{+,2}}
+ C_1 \, \Big(z \partial_{z} + z^2 \partial_{z}^2\Big)f_1(z) 
\,.
\en

\eq
R_{q q; 1}^{1 (1); \Delta\Delta_P} &=& R_{q q; 1}^{1 (2); \Delta\Delta_P}
= \frac{C_1}{6}
 \,, \nonumber\\
R_{q q; 2}^{1 (1); \Delta\Delta_P} &=&
R_{q q; 3}^{1 (2); \Delta\Delta_P} \, = \,
\frac{5}{2} C_A  + C_1 + L_\rho \Big(\frac{3}{2} C_A + C_1\Big)
 \,, \nonumber\\
R_{q q; 3}^{1 (1); \Delta\Delta_P} &=&
R_{q q; 2}^{1 (2); \Delta\Delta_P} \, = \,
- \Big(\frac{C_A}{2} + \frac{C_1}{6}\Big)  
\,, \nonumber\\
R_{q q; 4}^{1 (1); \Delta\Delta_P} &=&
R_{q q; 4}^{1 (2); \Delta\Delta_P} \, = \,
- \frac{C_1}{3} \Big(1 + L_\rho\Big)
\,, \nonumber\\
R_{q q; 5}^{1 (1); \Delta\Delta_P} &=&
R_{q q; 6}^{1 (2); \Delta\Delta_P} \, = \,
\Big(C_A - \frac{C_1}{3}\Big)  
\Big(\frac{1}{2} + L_\rho\Big)  
 \,, \nonumber\\
R_{q q; 7}^{1 (1); \Delta\Delta_P} &=&
R_{q q; 8}^{1 (2); \Delta\Delta_P} \, = \,
- C_A \Big(\frac{1}{2} + L_\rho\Big)  
\,. \nonumber\\
\en

\eq
P_{q q, 1; 1}^{1 (1); \Delta\Delta_P}(z) &=&
P_{q q, 2; 1}^{1 (2); \Delta\Delta_P}(z)
\nonumber\\
&=&
-  \frac{C_A}{4} \frac{z (5 + 5 z - 3 z^2 + z^3)}{(1-z)^3_{+,2}}
-  \frac{C_1}{3} \frac{z (1+z) (2 - 4 z + z^2)}{(1-z)^3_{+,2}}
\nonumber\\
&-&
\frac{C_1}{2} z \partial_{z}\Big(f_3(z) \, (1 + 3 z)\Big)
\,, \nonumber\\
P_{q q, 1; 2}^{1 (1); \Delta\Delta_P}(z) &=&
P_{q q, 2; 2}^{1 (2); \Delta\Delta_P}(z)
\nonumber\\
&=&
\frac{C_A}{2} \frac{z (1 + z + 3 z^2 - z^3)}{(1-z)^3_{+,2}}
+  \frac{C_1}{3} \frac{z}{(1-z)^2_{+,1}}
\nonumber\\
&+& C_1 z \partial_{z}\Big(f_3(z) \, z\Big)
\,, \nonumber\\
P_{q q, 2; 2}^{1 (1); \Delta\Delta_P}(z) &=&
P_{q q, 1; 2}^{1 (2); \Delta\Delta_P}(z)
\nonumber\\
&=&
- \frac{C_A}{4}
\frac{1 + z + z^2 - 3 z^3}{(1-z)^2_{+,1}}
- \frac{C_1}{6} \frac{1 - z^2 - 2 z^3}{(1-z)^2_{+,1}}
\nonumber\\
&-& \frac{C_1}{2} f_3(z) \, (1+z) 
\,, \nonumber\\
P_{q q, 2; 3}^{1 (1); \Delta\Delta_P}(z) &=&
P_{q q, 1; 3}^{1 (2); \Delta\Delta_P}(z)
\nonumber\\
&=&
\frac{C_A}{2} 
\frac{z (1 + z^2)}{(1-z)^2_{+,1}}
+ \frac{C_1}{3} \frac{z}{(1-z)_+}
\nonumber\\
&+& C_1 f_3(z) \, z 
\,.
\en

For the $\Delta_P$ and $\nabla$ structure functions we have,
in the $q \bar q$ annihilation channel: 
\eq
R_{q q; 1}^{0 (1); \nabla}
&=&
  2 R_{q q; 1}^{0 (1); \Delta_p} \, = \,
  - R_{q q; 1}^{0 (2); \nabla}   \, = \,
- 2 R_{q q; 1}^{0 (2); \Delta_p}
\nonumber\\
    R_{q q; 2}^{0 (11); \nabla} &=&
  2 R_{q q; 2}^{0 (11); \Delta_P} \, = \,
-   R_{q q; 3}^{0 (22); \nabla}  \, = \,
- 2 R_{q q; 3}^{0 (22); \Delta_P}
\nonumber\\
    R_{q q; 3}^{0 (12); \nabla} &=&
  2 R_{q q; 3}^{0 (12); \Delta_P} \, = \,
  - R_{q q; 2}^{0 (12); \nabla} \, = \,
- 2 R_{q q; 2}^{0 (12); \Delta_P}
\nonumber\\
&=&
\frac{2}{\rho} \, \Big(2 C_A + C_1\Big)
\,,
\nonumber\\
    R_{q q; 1}^{0 (11); \nabla} &=&
  2 R_{q q; 1}^{0 (11); \Delta_P} \, = \,
-   R_{q q; 1}^{0 (22); \nabla}  \, = \,
- 2 R_{q q; 1}^{0 (22); \Delta_P}
\nonumber\\
&=&
\frac{2}{\rho} \, \Big(C_A -  \frac{C_1}{3}\Big)
\,. 
\en

\eq
    P_{q q, 1; 1}^{0 (1); \nabla}(z) &=&
  2 P_{q q, 1; 1}^{0 (1); \Delta_P}(z) \, = \,
-   P_{q q, 2; 1}^{0 (2); \nabla}(z)  \, = \,
- 2 P_{q q; 2; 1}^{0 (2); \Delta_P}(z)
\nonumber\\
    P_{q q, 1; 2}^{0 (12); \nabla}(z) &=&
  2 P_{q q, 1; 2}^{0 (12); \Delta_P}(z) \, = \,
-   P_{q q, 2; 2}^{0 (12); \nabla}(z)  \, = \,
- 2 P_{q q; 2; 2}^{0 (12); \Delta_P}(z)
\nonumber\\
&=&
- \frac{2}{\rho} \, z \, \Big(C_A + 2 C_1 \partial_zf_2(z)\Big)
\,, \nonumber\\
    P_{q q, 2; 1}^{0 (1); \nabla}(z) &=&
  2 P_{q q, 2; 1}^{0 (1); \Delta_P}(z) \, = \,
-   P_{q q, 1; 2}^{0 (2); \nabla}(z)  \, = \,
- 2 P_{q q; 1; 2}^{0 (2); \Delta_P}(z)
\nonumber\\
    P_{q q, 2; 2}^{0 (11); \nabla}(z) &=&
  2 P_{q q, 2; 2}^{0 (11); \Delta_P}(z) \, = \,
-   P_{q q, 1; 2}^{0 (22); \nabla}(z)  \, = \,
- 2 P_{q q; 1; 2}^{0 (22); \Delta_P}(z)
\nonumber\\
    P_{q q, 2; 3}^{0 (11); \nabla}(z) &=&
  2 P_{q q, 2; 3}^{0 (11); \Delta_P}(z) \, = \,
-   P_{q q, 1; 3}^{0 (22); \nabla}(z)  \, = \,
- 2 P_{q q; 1; 3}^{0 (22); \Delta_P}(z)
\nonumber\\
&=&
\frac{2}{\rho} \, \Big(C_A \, (1+z) + 2 C_1 f(z)\Big)
\,, \nonumber\\
    P_{q q, 1; 1}^{0 (11); \nabla}(z) &=&
  2 P_{q q, 1; 1}^{0 (11); \Delta_P}(z) \, = \,
-   P_{q q, 2; 1}^{0 (22); \nabla}(z)  \, = \,
- 2 P_{q q; 2; 1}^{0 (22); \Delta_P}(z)
\nonumber\\
&=&
- \frac{2}{\rho} \, z \,
\Big(C_A + 2 C_1 (\partial_z + z \partial_z^2)f_2(z)\Big)
\,.
\en 

\eq
R_{q q; 1}^{1 (1); \nabla} &=& - R_{q q; 1}^{1 (2); \nabla}
\, =  \, 
2 R_{q q; 1}^{1 (1); \Delta_P} - \frac{C_1}{3\rho}
\, = \,
- \Big(2 R_{q q; 1}^{1 (2); \Delta_P} - \frac{C_1}{3\rho}\Big)
= - \frac{2}{\rho}
\, \Big(C_A + \frac{7 C_1}{3}\Big)
\,,
\nonumber\\    
 R_{q q; 2}^{1 (1); \nabla} &=& - R_{q q; 3}^{1 (2); \nabla} \, = \,
2 R_{q q; 2}^{1 (1); \Delta_P} \, = \, - 2 R_{q q; 3}^{1 (2); \Delta_P} 
\nonumber\\
&=& - R_{q q; 4}^{1 (1); \nabla}  \, = \, R_{q q; 4}^{1 (2); \nabla} \, = \,
- 2 R_{q q; 4}^{1 (1); \Delta_P} \, = \, 2 R_{q q; 4}^{1 (2); \Delta_P} 
= - \frac{2 (1 + L_\rho)}{\rho}
\, \Big(2 C_A + C_1\Big)
\,,
\nonumber\\
  R_{q q; 3}^{1 (1); \nabla} &=& - R_{q q; 2}^{1 (2); \nabla} \, = \,
2 R_{q q; 3}^{1 (1); \Delta_P} \, = \, - 2 R_{q q; 2}^{1 (2); \Delta_P} \, = \,
- \frac{2}{\rho}
\, \Big(C_A + \frac{C_1}{3}\Big)
\,,
\nonumber\\
  R_{q q; 5}^{1 (1); \nabla} &=&
- R_{q q; 6}^{1 (2); \nabla}  \, = \,
  2 R_{q q; 5}^{1 (1); \Delta_P} \, = \,  2 R_{q q; 6}^{1 (2); \Delta_P}  
\, = \,
- \frac{1 + 2 L_\rho}{\rho}
\, \Big(2 C_A + C_1\Big)
\,.
\en 

\eq
P_{q q, 1; 1}^{1 (1); \nabla}(z) &=&
- P_{q q; 1}^{1 (2); \nabla}(z)  \, = \, 
    2 P_{q q, 1; 1}^{1 (1); \Delta_P}(z) - \frac{C_1}{\rho}
    \, \partial_z\Big(f_1(z) \, (1+z)\Big)
\nonumber\\
&=&  
    - \Big(2 P_{q q, 1; 1}^{1 (2); \Delta_P}(z) - \frac{C_1}{\rho}
    \, \partial_z\Big(f_1(z) \, (1+z)\Big)\Big)
\nonumber\\
&=&
\frac{C_A}{\rho} \, \frac{z (5 + z + 3 z^2 - z^3)}{(1-z)^3_{+,2}}
+ \frac{2 C_1}{\rho} \, \frac{z (2 - 2 z + 3 z^2 - z^3)}{(1-z)^3_{+,2}}     
\nonumber\\
&+&  \frac{C_1}{\rho}
\, \Big(2 z \, \partial_z\Big(f_1(z) \, (1+3z)\Big)
- \partial_z\Big(f_1(z) \, (1+z)\Big)\Big)
\,,
\nonumber\\
   P_{q q, 1; 2}^{1 (1); \nabla}(z) &=& - P_{q q, 2; 2}^{1 (2); \nabla}(z) 
   \, = \,   2 P_{q q, 1; 2}^{1 (1); \Delta_P}(z)
   \, = \, - 2 P_{q q, 12 2}^{1 (2); \Delta_P}(z) 
  \nonumber\\
&=&
- \frac{2 C_A}{\rho} \, \frac{z (1 + 2 z - z^2)}{(1-z)^2_{+,1}}
- \frac{2 C_1}{\rho} \, \frac{z}{(1-z)^2_{+,1}}     
-  \frac{2 C_1}{\rho} \, z \, \partial_z\Big(f_1(z) \, z\Big)
\,,
\nonumber\\
      P_{q q, 2; 2}^{1 (1); \nabla}(z) &=&
   -  P_{q q, 1; 2}^{1 (2); \nabla}(z) \, = \,
    2 P_{q q, 2; 2}^{1 (1); \Delta_P}(z) - 
    \frac{C_1}{\rho}
    \, \Big(1 - f_1(z) \, (1+z)\Big)
\nonumber\\
&=&
    - \Big(2 P_{q q, 1; 2}^{1 (2); \Delta_P}(z) - 
    \frac{C_1}{\rho}
\, \Big(1 - f_1(z) \, (1+z)\Big)\Big)
  \nonumber\\
&=&
- \frac{C_A}{\rho} \, \frac{z (1 + 3 z - z^2 + z^3)}{(1-z)^2_{+,1}}
- \frac{C_1}{\rho} \, \frac{1 - z + 3 z^2 - 3 z^3 + 2 z^4}{(1-z)^2_{+,1}}     
\nonumber\\
&+&  \frac{C_1}{\rho}
  \, f_1(z) \, (1 - 3 z^2) 
\,,
\nonumber\\
   P_{q q, 2; 3}^{1 (1); \nabla}(z) &=&
-  P_{q q, 1; 3}^{1 (2); \nabla}(z)    \, = \, 
   2 P_{q q, 2; 3}^{1 (1); \Delta_P}(z) \, = \,
-  2 P_{q q, 1; 3}^{1 (2); \Delta_P}(z) \, = \,
  \nonumber\\
&=&
\frac{2 C_A}{\rho} \, \frac{z (1 + z)}{(1-z)_+}
+ \frac{2 C_1}{\rho} \, \frac{z}{(1-z)_+}     
+  \frac{2 C_1}{\rho} \,  z \, f_1(z)
\,.
\en

For the $\Delta\Delta_P$ structure function we have, in the $qg$ channel:
\eq
R_{q g; 1}^{0 (2); \Delta\Delta_P} &=&
R_{q g; 3}^{0 (22); \Delta\Delta_P} \, = \,
R_{q g; 2}^{0 (12); \Delta\Delta_P} \, = \,
- \frac{3 C_A}{2} - 2 C_1 (1 + L_\rho) 
\,, \nonumber\\
R_{q g; 1}^{0 (22); \Delta\Delta_P} &=&
- C_A - 2 C_1 L_\rho
\,.
\en

\eq
P_{q g, 2; 2}^{0 (1); \Delta\Delta_P}(z) &=&
P_{q g, 2; 2}^{0 (11); \Delta\Delta_P}(z) \, = \,
P_{q g, 2; 3}^{0 (11); \Delta\Delta_P}(z) 
\nonumber\\
&=&
\frac{C_A}{2} (1 + 2 z) 
+ C_1 (1 + 2 z - z^2 + 2 z L_\rho) 
\,, \nonumber\\
P_{q g, 2; 1}^{0 (2); \Delta\Delta_P}(z) &=&
P_{q g, 2; 2}^{0 (12); \Delta\Delta_P}(z)
\, = \, 
z \Big(C_A + 2 C_1 (1 - z + L_\rho)\Big) 
\,, \nonumber\\
P_{q g, 2; 1}^{0 (22); \Delta\Delta_P}(z) &=& 
z \Big(C_A + 2 C_1 (1 - 2 z + L_\rho)\Big) 
\,.
\en

\eq
R_{q g; 1}^{1 (1); \Delta\Delta_P} &=&
- C_A - C_1 \Big(\frac{11}{6} + 2 L_\rho\Big)
\,, \nonumber\\
R_{q g; 2}^{1 (1); \Delta\Delta_P} &=&
\Big(C_A + C_1 \, \Big(\frac{5}{3} + 2 L_\rho\Big)\Big)
\, (1 + L_\rho) 
\,,\nonumber\\
R_{q g; 5}^{1 (1); \Delta\Delta_P} &=& 
-  \Big(\frac{3 C_A}{4} + C_1 \, (1 + L_\rho)\Big)
\, (1 + 2 L_\rho) 
\,,\nonumber\\ 
R_{q g; 1}^{1 (2); \Delta\Delta_P} &=&
- \frac{C_A}{4}  
- C_1 \, \Big(\frac{5}{3} - 3 L_\rho\Big)
\,, \nonumber\\
R_{q g; 2}^{1 (2); \Delta\Delta_P} &=&
C_A + 2 C_1 \, L_\rho
\,, \nonumber\\
R_{q g; 3}^{1 (2); \Delta\Delta_P} &=&
\frac{5 C_A}{2} (1 +  L_\rho) - 4 C_1 \, (1 - L_\rho) \,
\Big(\frac{11}{12} + L_\rho\Big) 
\,, \nonumber\\
R_{q g; 4}^{1 (2); \Delta\Delta_P} &=&
- \Big(\frac{3 C_A}{2} + 2 C_1 \, (1 + L_\rho) \Big) \, (1 + L_\rho)
\,.
\en

\eq
P_{q g, 1; 1}^{1 (1); \Delta\Delta_P}(z) &=&
C_A \frac{z^2 (1 + 3 z - z^2)}{(1-z)^3_{+,2}}
+  \frac{C_1}{3} \frac{z^2 (2 + z (3 - z) (5 + 6 L_\rho))}{(1-z)^3_{+,2}}
\nonumber\\
&+& C_1 z \, \partial_z\Big(f_3(z) \, z^2\Big)
\,, \nonumber\\
P_{q g, 2; 2}^{1 (1); \Delta\Delta_P}(z) &=&
- \frac{C_A}{4} \frac{1 + 5 z - 2 z^2}{(1-z)_+}
\nonumber\\
&-& \frac{C_1}{6}
\frac{3 + 27 z - 33 z^2 + 13 z^3 + 6 (1 + z) L_\rho}{(1-z)_+}
\,, \nonumber\\
P_{q g, 2; 3}^{1 (1); \Delta\Delta_P}(z) &=&
\frac{C_A}{2} \frac{z (1 + 2 z)}{(1-z)_+}
+ C_1 \frac{z (1 + 2 z - z^2 + 2 z L_\rho)}{(1-z)_+}
\,, \nonumber\\
P_{q g, 2; 1}^{1 (2); \Delta\Delta_P}(z) &=&
- \frac{C_A}{2} 
\frac{z (4 + 2 z - z^2)}{(1-z)^2_{+,1}}
\nonumber\\
&+& \frac{C_1}{3}  \frac{z 
  (18 - 21 z + 21 z^2 - 7 z^3 + 3 (1 + 6 z - 3 z^2) L_\rho)}
{(1-z)^2_{+,1}}
\,, \nonumber\\
P_{q g, 2; 2}^{1 (2); \Delta\Delta_P}(z) &=&
\frac{C_A}{2} 
\frac{z (1 + 4 z - 2 z^2)}{(1-z)^2_{+,1}}
+ C_1  \frac{z (1 + 4 z - 5 z^2 + 2 z^3 + 2 z (2-z) L_\rho)}
{(1-z)^2_{+,1}}
\,, \nonumber\\
P_{q g, 1; 2}^{1 (2); \Delta\Delta_P}(z) &=&
\frac{C_A}{2} 
\frac{z^2 (1 + 2 z)}{(1-z)^2_{+,1}}
+ \frac{C_1}{3} 
\frac{z^2 (1 + 5 z + 6 z L_\rho)}
{(1-z)^2_{+,1}} + C_1 z^2 f_3(z)
\,.
\en

Finally, for the $\Delta_P$ and $\nabla$ structure functions we have,
in the $qg$ channel:
\eq
      R_{q g; 1}^{0 (2); \nabla}     &=&
    2 R_{q g; 1}^{0 (2); \Delta_p}  \, = \,
      R_{q g; 3}^{0 (22); \nabla}  \, = \,
      2 R_{q g; 3}^{0 (22); \Delta_p}
      \nonumber\\
&=&   R_{q g; 2}^{0 (12); \nabla}  \, = \,
    2 R_{q g; 2}^{0 (12); \Delta_p} \, = \,
\frac{4 C_1}{\rho} 
\,,
\nonumber\\
    R_{q g; 1}^{0 (22); \nabla} &=&
  2 R_{q g; 1}^{0 (22); \Delta_p} \, = \,
\frac{2}{\rho} \Big(C_A + 2 C_1 (4 + L_\rho)\Big)
\,.
\en

\eq
    P_{q g, 2; 2}^{0 (1); \nabla}(z)     &=&
  2 P_{q g, 2; 2}^{0 (1); \Delta_P}(z) \, = \,
    P_{q g, 2; 2}^{0 (11); \nabla}(z)  \, = \,
  2 P_{q g, 2; 2}^{0 (11); \Delta_P}(z)
\nonumber\\
  &=&  P_{q g, 2; 3}^{0 (11); \nabla}(z)  \, = \,
     2 P_{q g; 2; 3}^{0 (11); \Delta_P}(z)
\nonumber\\
&=& 
\frac{2 C_A}{\rho} (1 - z)
+ \frac{4 C_1}{\rho} \Big(1 - 2 z^2 + z (1-z) L_\rho\Big)
\,, \nonumber\\
    P_{q g, 2; 1}^{0 (2); \nabla}(z) &=&
  2 P_{q g, 2; 1}^{0 (2); \Delta_P}(z) \, = \,
    P_{q g, 2; 2}^{0 (12); \nabla}(z)  \, = \,
  2 P_{q q, 2; 2}^{0 (12); \Delta_P}(z)
\nonumber\\
&=&
- \frac{2 C_A}{\rho} z
- \frac{4 C_1}{\rho} z \Big(4 z - (1 - 2 z) L_\rho\Big)
\,, \nonumber\\
    P_{q g, 2; 1}^{0 (22); \nabla}(z) &=&
  2 P_{q g, 2; 1}^{0 (22); \Delta_P}(z) 
\, = \,
- \frac{2 C_A}{\rho} z
- \frac{4 C_1}{\rho} z \Big(8 z - (1 - 4 z) L_\rho\Big)
\,.
\en

\eq
  R_{q g; 1}^{1 (1); \Delta_P} &=& - \frac{C_A - 2 C_1}{2 \rho} 
\,,
\nonumber\\
    R_{q g; 2}^{1 (1); \nabla} &=&
  2 R_{q g; 2}^{1 (1); \Delta_P}
  - \frac{C_A - 6 C_1}{\rho} \, (1 + 2 L_\rho)
  \nonumber\\
&=&
- \frac{C_A}{\rho} \, L_\rho
+ \frac{2 C_1}{\rho} \, (2 + 5 L_\rho) 
\,,
\nonumber\\
    R_{q g; 5}^{1 (1); \nabla} &=&
  2 R_{q q; 5}^{1 (1); \Delta_P} 
= \frac{2 C_1}{\rho} \, (1 + 2 L_\rho)
\,,
\nonumber\\
  R_{q g; 1}^{1 (2); \nabla} &=&
  2 R_{q g; 1}^{1 (2); \Delta_P}
- \frac{2 C_1}{\rho}  \, (3 + 2 L_\rho) 
\nonumber\\
&=& \frac{2 C_A}{\rho} + \frac{4 C_1}{\rho} \, (2 + L_\rho)
\,,
\nonumber\\
    R_{q g; 2}^{1 (2); \nabla} &=&
  2 R_{q g; 2}^{1 (2); \Delta_P} 
  \nonumber\\
  &=&
- \frac{2 C_A}{\rho} - \frac{4 C_1}{\rho} \, (4 + L_\rho)
\,,
\nonumber\\
    R_{q g; 3}^{1 (2); \nabla} &=&
  2 R_{q g; 3}^{1 (2); \Delta_P}
  - \frac{C_A}{2\rho} \, (2 + 3 L_\rho)
  + \frac{3 C_1}{\rho} \, (4 + 3 L_\rho) 
  \nonumber\\
  &=&
- \frac{C_A}{2\rho} \, L_\rho
+ \frac{3 C_1}{\rho} \, (2 + L_\rho) 
\,,
\nonumber\\
    R_{q g; 4}^{1 (2); \nabla}  &=& 
  2 R_{q g; 4}^{1 (2); \Delta_P} - 
\frac{2 C_1}{\rho} L_\rho  
\ = \
 \frac{2 C_1}{\rho} \,
\Big(2 + L_\rho\Big)
\,.
\en 

\eq
      P_{q g, 1; 1}^{1 (1); \nabla}(z) &=&
    2 P_{q g, 1; 1}^{1 (1); \Delta_P}(z) + \frac{2 z}{\rho}
\Big(C_A \frac{z (2 - z)}{(1-z)^2_{+,1}}
-  4 C_1 \frac{1 + 2 z}{(1-z)^3_{+,2}}\Big)  
\nonumber\\
&=&
\frac{z}{\rho}
\, \Big(C_A \frac{z (2 - z)}{(1-z)^2_{+,1}}
- 2 C_1 \frac{4 + 10 z + 3 z^2 - z^3}{(1-z)^3_{+,2}}\Big) 
\,, 
\nonumber\\
      P_{q g, 2; 2}^{1 (1); \nabla}(z) &=&
    2 P_{q g, 2; 2}^{1 (1); \Delta_P}(z) + \frac{z}{\rho}
    \Big(C_A \frac{1 + 3 z}{(1-z)_+} - 
    6 C_1 \frac{1 + z}{(1-z)_+} +
    2 C_1 \, (1 + z) \,
    L_\rho\Big) 
\nonumber\\
&=&  - \frac{1}{\rho} \, 
\Big(C_A \frac{1 - 2 z - 2 z^2}{(1-z)_+} 
+ 2 C_1 \frac{1 + 5 z - 7 z^2 + 6 z^3}{(1-z)_+}
\Big) 
\,, 
\nonumber\\
      P_{q g, 2; 3}^{1 (1); \nabla}(z) &=&
      2 P_{q g, 2; 3}^{1 (1); \Delta_P}(z)
      + \frac{2 C_A}{\rho} \, \frac{z^2}{(1-z)_+}  
\nonumber\\
&=& \frac{2 z}{\rho} \, \Big(\frac{C_A + 2 C_1 (1 - 2 z^2)}{(1-z)_+}
+ 2 C_1 \, z \, L_\rho\Big)
\,, 
\nonumber\\
      P_{q g, 2; 1}^{1 (2); \nabla}(z) &=&
    2 P_{q g, 2; 1}^{1 (1); \Delta_P}(z) - \frac{2 z}{\rho} C_1 
    \, \Big(\frac{3 + 6 z - 3 z^2}{(1-z)^2_{+,1}}
    - (1 + 2 z) \, L_\rho\Big)
\nonumber\\
&=&
- \frac{z}{\rho} \, 
    \Big(C_A \frac{2 - 2 z + z^2}{(1-z)^2_{+,1}}
     + 2 C_1 \frac{(2 - z) ( 4 - 5 z + 4 z^2)}{(1-z)^2_{+,1}}
     + 8 C_1 z L_\rho\Big)
\,, 
\nonumber\\
    P_{q g, 1; 2}^{1 (2); \Delta_P}(z) &=& - \frac{z^2}{2 \rho}
    \, \Big(\frac{C_A}{(1-z)_+}
    + 2 C_1 \frac{1 + z}{(1-z)^2_{+,1}}\Big) 
\,,
\nonumber\\
      P_{q g, 2; 2}^{1 (2); \nabla}(z) &=&
      2 P_{q g, 2; 2}^{1 (1); \Delta_P}(z)
      + 4 C_1 z^2 (2 - 3 z) \, L_\rho
\nonumber\\
      &=&
      \frac{2 z}{\rho}
    \, \Big(C_A 
    + 2 C_1 \frac{1 - 6 z^2 + 4 z^3}{(1-z)^2_{+,1}}
    + 4 C_1 \, z \, L_\rho\Big) 
\,. 
\en

\end{document}